**United Nations University**
Institute in Macau

POLICY REPORT

# Interoperability in AI Safety Governance: Ethics, Regulations, and Standards

**THE UNITED KINGDOM, SOUTH KOREA, CHINA AND SINGAPORE**

YIK CHAN CHIN, DAVID A RAHO, HAG-MIN KIM, CHUNLI BI, JAMES ONG, JINGBO HUANG, SERGE STINCKWICH

## Contributors

PROJECT LEADS

Yik Chan Chin, Beijing Normal University and United Nations University Institute in Macau

Jingbo Huang, United Nations University Institute in Macau

Serge Stinckwich, United Nations University Institute in Macau

COUNTRY RESEARCH LEADS

**UK:** David A Raho, Sheffield Hallam University

**China:** Chunli Bi, China Academy of Information and Communications Technology

**South Korea:** Hag-Min Kim, Kyung Hee University

**Singapore:** James Ong, Artificial Intelligence International Institute

COUNTRY RESEARCH TEAM MEMBERS

**China:** Leilei Zhang, Tianyu Wang, Na Fu, China Academy of Information and Communications Technology

**South Korea:** Wenshuai Su, Kyungwon Kim, and Minyu Jiang, Kyung Hee University

**Singapore:** Sameer Gahlot, Artificial Intelligence International Institute

RESEARCH ASSISTANTS

Songruowen Ma, University of Oxford

Eduarda Mello, Beijing Normal University

## Acknowledgements

This report is developed by the United Nations University Institute in Macau. The publication has greatly benefited from the research assistance of Songruowen Ma and Eduarda Mello. We appreciate the comments and feedback from Antonella Maia Perini, Gurjit S. Sandhu, Christopher David Jones, Alan Duncan King, Raymond Forbes and Rolf H. Weber. This work is made possible through the generous funding of SenseTime.

ISBN: 978-92-808-9166-9

# Executive Summary

## 1. Purpose of the Report

This policy report draws on country studies from China, South Korea, Singapore, and the United Kingdom to identify effective tools and key barriers to interoperability in AI safety governance. It offers practical recommendations to support a globally informed yet locally grounded governance ecosystem.

Interoperability is a central goal of AI governance, vital for reducing risks, fostering innovation, enhancing competitiveness, promoting standardization, and building public trust. However, structural gaps such as fragmented regulations and lack of global coordination, and conceptual gaps, including limited Global South engagement, continue to hinder progress.

Focusing on three high-stakes domains-autonomous vehicles, education, and cross-border data flows-the report compares ethical, legal, and technical frameworks across the four countries. It identifies areas of convergence, divergence, and potential alignment, offering policy recommendations that support the development of interoperability mechanisms aligned with the Global Digital Compact and relevant UN resolutions. The analysis covers seven components: objectives, regulators, ethics, binding measures, targeted frameworks, technical standards, and key risks.

## 2. Methods of study

This research adopts a regulatory learning approach, comparing AI safety governance across China, South Korea, Singapore, and the United Kingdom. By engaging stakeholders from each jurisdiction, it captures local insights and fosters interjurisdictional learning. Coordinated by the United Nations University (UNU), the initiative introduces a collaborative model for transnational policymaking by identifying shared challenges and co-developing strategic responses.

## 3. Recommendations

We recommend leveraging effective interoperability instruments at both global and national levels. Additional recommendations span three key dimensions:

**Ethical Interoperability**
- Promote Ethical Self-Certification Reports
- Uphold the UN System as the Primary Forum for AI Ethics Deliberation
- Advance the Global AI Ethics Framework

**Regulatory Interoperability**
- Establish a Multilateral System for Coordinated AI and Data Governance
- Establish a Coherent National Entity for Global Engagement
- Support Inclusive Multi-Stakeholder Engagement for AI Safety Governance
- Enhance Public Engagement to Strengthen International AI Governance
- Launch a Global Benchmark for AI Safety and Security
- Enhance Transparency and Accountability in AI Safety Governance
- Promote Adaptation and Expansion of Data Interoperability Mechanisms
- Introduce Additional Protections to Mitigate Data Flow Risks
- Develop Greater Standardization and Harmonization of Liability Models for Autonomous Vehicles
- Support the Development of Interoperable Digital Public Infrastructure
- Invest in AI Safety, Frontier Risk Management, and Alignment Research Collaborations
- Promote AI Safety in Education

**Technical Interoperability**
- Promote Interoperability by Design
- Call for International Consensus-Driven AI Standards and Avoid Duplication
- Prioritize the Development of Dedicated AI-in-Education Standards
- Prioritize Standards Interoperability at the Security Layer
- Call for Scenario Planning for AI-Related Catastrophic Risks and Improved Regulatory Forecasting

## 4. Conclusion

The future of AI safety governance is evolving towards an evidence-based, outcomes-oriented model that complements principle-led frameworks. This shift reflects a growing international consensus a strengthening alignment with emerging global frameworks and standards. To sustain momentum, policymakers must prioritize deepening normative specificity, enhancing legal interoperability, expanding the adoption of interoperable standards, strengthening data governance mechanisms, and collaboratively investing in and conducting research into technical, institutional, and AI literacy capacity building.

# Interoperability in AI Safety Governance: Ethics, Regulations, and Standards

15 October 2025

## Introduction

The need for interoperable AI governance is widely recognized as essential for reducing risks, fostering innovation, enhancing competitiveness, promoting standardization, and building public trust. In May 2024, the European Union and ten countries—Australia, Canada, France, Germany, Italy, Japan, South Korea, Singapore, the United Kingdom, and the United States—signed the *Seoul Declaration* for safe, innovative, and inclusive AI, signalling a strong commitment to cooperation and interoperable governance frameworks.

Despite this momentum, technical incompatibilities hinder cross-border data sharing and collaboration. Moreover, a lack of shared understanding of AI's societal roles contributes to ethical inconsistencies, while diverging regulatory approaches to AI safety governance across countries, coupled with geopolitical tensions, underscore the urgent need for international coordination to prevent the fragmentation of ethical, technical, and legal standards. To address these challenges, all UN member states committed in 2024 to implementing the Global Digital Compact (GDC) at national, regional, and global levels, with interoperability repeatedly emphasized as a core objective of AI governance.

This research advances those commitments by conducting a comparative analysis of ethical, legal, and technical frameworks for AI safety governance across China, South Korea, Singapore, and the United Kingdom. It focuses on three critical domains - autonomous vehicles, education, and cross-border data flows - and systematically reviews existing national and regional AI safety frameworks, regulations, standards, and best practices. By identifying areas of convergence, divergence, complementarity, and potential alignment, the project offers policy recommendations to support the development of interoperability mechanisms aligned with the GDC and relevant UN resolutions on AI safety governance.

## 1. Background

Interoperability measures can enhance the health and competitiveness of the broader market, enable wider participation in AI innovation across system layers and data use. Conversely, a lack of interoperability may cause tangible harm or prevent users from accessing the full benefits of technological innovation.

Policy research published by the UN Internet Governance Forum (IGF)'s policy network on AI (2023-2024) and others shows growing efforts on interoperability and international cooperation to address the challenges posed by AI systems. These efforts focus on various areas, such as standardisation, safety, cross-border data transfers, and risk mitigation. However, significant challenges remain in achieving effective interoperability in AI governance. These include structural gaps, such as the absence of global coordination mechanisms and inconsistent regulatory frameworks as well as conceptual gaps, including limited understanding of regulatory principles and insufficient input from the Global South in shaping and interpreting the concept of interoperability in AI governance.

### **1.1** THE UN AND INTEROPERABILITY IN AI GOVERNANCE

In 2024, at the UN, all states adopted the GDC, which emphasises the importance of interoperability in AI governance across its various scopes, including:

- Interoperable cross-border data flows: Identifying interoperable mechanisms to enable secure and trusted cross-border data flows, particularly for micro, small, and medium enterprises, within and between countries.
- Interoperable data governance (Objective 4 of the GDC): Promoting interoperability between national, regional, and international data policy frameworks. Interoperability of data governance legislation is a crucial issue in AI governance, as there can be no AI without data.
- Coordination, interoperability, and compatibility of emerging AI governance frameworks (Objective 5 of the GDC) are promoted through
    - Establishing the International Scientific Panel on AI and the Global Dialogues on AI Governance.
    - Sharing best practices and promoting common understanding in AI.
    - Encouraging transparency, accountability, and strong human oversight of AI systems in line with international law.
    - Encouraging standards development organisations to collaborate on interoperable AI standards that uphold

safety, reliability, sustainability, and human rights.

- Establishing international partnerships to develop education and training programmes, increase access to open AI models and systems, share training data and computing resources, and support AI model training and development.
- Addressing local needs, fostering cross-regional partnerships, and connecting them globally to ensure AI interoperability frameworks are inclusive, adaptable, and capable of tackling local challenges.
- Establishing a dedicated working group on data governance under the Commission on Science and Technology for Development.

The UN General Assembly adopted two resolutions on AI in 2024, i.e. "Seizing the opportunities of safe, secure and trustworthy artificial intelligence systems for sustainable development" and "Enhancing international cooperation on capacity-building of artificial intelligence" mark a significant milestone of global multilateral collaboration on AI governance[1]. The first one concerns international cooperation on AI capacity-building. It encourages international collaboration to strengthen AI capacity in developing countries. The second landmark resolution calls for the establishment of regulatory and governance frameworks to ensure AI systems are safe, secure, and trustworthy. It underlines that governance measures must be interoperable, flexible, adaptable, inclusive, and grounded in international law, catering to the needs and capabilities of different countries and guaranteeing fair benefits worldwide.

It is recognized that the United Nations has an important role to play in shaping, enabling and supporting agile, multidisciplinary and adaptable multi-stakeholder AI governance. This UNU policy report builds on these efforts and commitments by conducting an interoperability analysis of ethical, legal, and technical standards frameworks across China, South Korea, Singapore, and the UK in AI safety governance within the education sector, cross-border data flows, and autonomous driving domains.

## **1.2** AI SAFETY GOVERNANCE AND INTEROPERABILITY

Safety is the top priority in managing artificial intelligence systems, acting as both a regulatory requirement and an ethical duty. AI Safety or Safety of an AI system refers to the understanding, prevention, mitigation, and management of potential harms arising from the design, development, and deployment of AI systems, ensuring that AI technologies protect human well-being throughout their lifecycle. These safety harms may be deliberate or accidental, and can affect individuals, groups, organizations, nations, or even global systems, taking various forms such as physical, psychological, or economic impacts (UK Government, 2023; Bengio, Y., et al, 2025)[2]. Some common examples of AI risks include algorithmic bias, privacy leakage, misinformation and deepfakes, unreliable decision-making.

AI safety is particularly critical in high-stakes domains, such as education, cross-border data management, and autonomous vehicles where risks to rights, public trust, and security are heightened. The salient risks they pose include threats to the right to privacy, the right to life, and the right to equitable access to knowledge and digital literacy, along with concerns related to data security and vehicle cybersecurity. Additionally, building public trust in the safe and confident use of digital tools with robust safety and privacy protections, while fostering consensus among government, industry, and citizens, is critical across all three sectors. These risks lie at the heart of the UN's AI Safety framework, which prioritizes the development of safe, secure, and trustworthy artificial intelligence systems. Such systems should uphold human rights, promote sustainable development, and be governed by ethical principles throughout their lifecycle. Key elements of the framework include preventing harm and misuse, promoting transparency and accountability, ensuring inclusive participation especially from the Global South, and supporting the peaceful use of AI in alignment with the Sustainable Development Goals (SDGs).

AI safety governance encompasses frameworks, policies, and operational practices that ensure AI is developed, deployed, and maintained in a safe, reliable, and ethical way, reducing risks and avoiding harm to individuals and society (Jobin et al., 2019[3]; Lee et al., 2021[4]; Tabassi, 2023[5]; OECD, 2019/2024[6]).

The term interoperability is commonly understood as the ability of different systems, tools, and components to work together seamlessly both technically through enabling data sharing and normatively through aligning laws and standards, it can include substantive measures such as international norms, shared protocols, interfaces, and data models, enabling communication, data exchange, standardizations etc. (PNAI 2023& 2024[7]; Zeng, 2019[8]; Berg, 2024[9]; Onikepe, 2024[10]).

Interoperability of AI safety governance in this report is understood as essential substantive methods that enable two or more different jurisdictions to collaborate in order to support a common understanding, interpretation, and implementation of transborder AI safety governance. Interoperability functions can encompass four broad layers of complex systems (table 1). Greater interoperability can reduce risks, foster innovation, enhance competitiveness, promote standardization, and build public trust.

| | |
|---|---|
| **Technology** | The capacity to transfer and interpret data and other information across AI systems, applications, or components. |
| **Data** | The ability to read and understand the data. |
| **Human elements** | The capacity for communication |
| **Institutional aspects** | The ability to collaborate effectively. |

**Table 1:** Interoperability Layers

Three key aspects of interoperability in AI safety governance studied in this report include ethical, legal and technical interoperability (see *Glossary of Terms)* and their effects on AI regulations and standards (Table 2).

| | |
|---|---|
| **Functions of Ethical interoperability** | The ability of institutions, systems, or actors to collaborate across different moral frameworks to support the development of AI regulations and technical standards, as well as international cooperation. |
| **Functions of Legal interoperability** | Involves the coordination of regulatory frameworks and establishing international cooperative mechanisms. The development of legal interoperability's substantive and structural dimensions as a "third way" between fragmentation and harmonisation merit increased attention. |
| **Functions of Technical interoperability** | Ensures the compatibility of AI technical standards in addressing governance issues related to technical interconnectivity, transactional interconnectivity, physical externalities, and policy externalities. |

**Table 2:** Interoperability Aspects

## 2. The Scope, Definitions and Methods of Investigation

### **2.1** THE SCOPE OF INVESTIGATION

The research investigates existing ethical and legal frameworks, standards, and best practices in AI safety governance across four jurisdictions - China, South Korea, Singapore, and the United Kingdom - focusing on three critical sectors: education, cross-border data flows, and autonomous vehicles. It identifies ethical, legal, and technical standards gaps, overlaps, and divergences. It develops policy recommendations to support the creation of interoperability mechanisms aligned with the GDC and relevant UN resolutions on AI safety governance.

The study is guided by the following key questions:

- What significance and dimensions does interoperability hold in the context of AI safety governance?
- How should a matrix of regulatory, ethical, and technical models be designed within the AI safety governance framework?
- How can substantive interoperability serve as a tool to address fragmentation in AI safety governance?

### **2.2** FORMULATING INTEROPERABILITY

Interoperability manifests in various forms, and different mechanisms can be employed to enable the implementation of **ethical**, **legal**, and **technical interoperability** (see *Glossary of Terms*).

Ethical interoperability involves below key mechanisms:

- **Promotion of shared or common terminology.** One of the most evident ways to address this challenge is to identify shared or universal terminologies capable of articulating diverse ethical principles. For example, NATO members commit to ensuring that the AI applications they develop and consider for deployment will comply with the following six principles: lawfulness, responsibility and accountability, explainability and traceability, reliability, governability, and bias mitigation. This obligation provides a coherent common basis for both NATO and Allies to design and develop AI applications while also supporting interoperability goals. However, even with shared or common terminology in place, ethical interoperability remains challenging. Terms such as "privacy" or "security" lack universally accepted definitions, and their interpretations often vary across legal, cultural, and institutional contexts.

- **Compatible cross-institutional and international ethical assessment and accountability mechanisms.** A practical approach to achieving ethical interoperability involves structured processes for disclosure and evaluation. Developers of AI systems should provide transparent information about the system's operational characteristics - covering its design, development, deployment, and usage. Adopting institutions can then assess whether these characteristics align with their own ethical principles. This mechanism is particularly valuable when it is unclear whether two ethical standards are substantively different. It enables diverse normative commitments to be translated into comparable, outcome-oriented assessments, fostering alignment on ethical expectations and building mutual trust across institutional and national boundaries.
- **Multi-stakeholder engagement.** The successful implementation of ethical interoperability depends on collaborative engagement across key stakeholders such as AI developers, ethicists, healthcare professionals, legal experts, policymakers, and community representatives, and establishing platforms to support continuous dialogue. By institutionalizing inclusive engagement, we strengthen the ethical foundations of interoperable AI systems and foster a culture rooted in shared values, transparency, and mutual responsibility.

Legal interoperability involves several mechanisms with **harmonisation**, **standardisation**, **mutual reciprocity** and **cooperation** (see *Glossary of Terms*) being seen as key ones.

Technical interoperability can also be achieved through various mechanisms, each providing different advantages and compromises depending on the technological setting.

- Mutual consensus through the development and adoption of open standards
- Creation of infrastructure for integrations[11]
- Open source code[12]
- Policy intervention to advance interoperability in a number of specific technology contexts[13]
- Common protocols
- Couplings between hardware and software
- Sharing data between services
- "Adversarial interoperability" - engineering interoperability without its maker's consent or involvement

Notably, governments may directly cite standards in legislation, effectively turning voluntary standards into legal requirements.[14]

## 2.3 METHODS OF STUDY

This research explicitly adopts regulatory learning in its design by comparing four jurisdictions to assess AI safety governance. Regulatory learning promotes learning from domestic experiments and experiments, ideas, experiences, and insights of other jurisdictions or communities. To determine the most suitable regulatory response to the challenges posed by AI, interjurisdictional learning and regulatory innovation are necessary through the exchange of ideas and experiences.

Learning to regulate AI effectively requires robust international cooperation and coordination. This project brings together stakeholders from four jurisdictions to harness local insights and promote interjurisdictional learning. Led and coordinated by the United Nations University (UNU), the initiative pioneers a new approach to transnational policy-making by defining shared challenges, and developing collaborative strategies. Through this framework, the project aims to generate practical solutions to AI safety governance's complex and pressing policy challenges, fostering a globally informed yet locally grounded regulatory ecosystem.

This research employs a framework to compare AI safety governance across four jurisdictions. Developed from recurring patterns observed in various governance initiatives, the framework systematically identifies differences, similarities, complementarities, and potential areas for alignment among them.[15] It comprises seven core elements:

- **Objectives**: The primary aims of each governance initiative, such as promoting cross-border data flows, improving regulatory coordination, or ensuring the ethical alignment of AI systems.
- **Principles and values:** The foundational norms guiding the design and implementation of the framework. These may include transparency, accountability, inclusivity, fairness, and the protection of human rights.
- **Governance approach (Top-Down vs. Bottom-Up):** Bottom-up approaches typically emerge organically through multi-stakeholder collaborations, often emphasizing inclusivity and adaptability. Conversely, top-down approaches are generally driven by deliberate decisions from governments or international institutions, such as mandatory disclosure regimes, regulatory licensing, or the establishment of standards. Private-sector-led initiatives - such as open standards, reverse engineering, and technical cooperation - may also exhibit top-down characteristics.
- **Binding nature:** The legal enforceability of the framework. This can range from non-binding instruments, such as declarations, principles, taxonomies, or mutual recognition agreements, to binding mechanisms, including treaties, legislation, or enforceable standards.
- **Level of integrations:** The degree of detail regarding alignment

with international governance frameworks. Frameworks may feature highly prescriptive guidelines or adopt more flexible, principle-based approaches that allow for contextual adaptation.

- **Components:** The institutional and structural elements of each governance framework, which may encompass legal, organisational, semantic, and technical dimensions.
- **Regulator**: a regulator is typically tasked with maintaining order, ensuring compliance, or stabilizing a system through rules, feedback, or enforcement mechanisms.

To propose interoperability recommendations, we follow the principles below: Interoperability is not solely a technical issue; it encompasses multiple layers of implications - ethical, legal, and technical. Addressing these challenges requires a holistic approach that spans all layers and operates across different administrative levels, from local to global. Achieving this remains a complex and ongoing challenge.

**Ethical interoperability**

- **Assessment of interoperability levels.** This method involves a structured evaluation of the extent to which moral standards and principles are aligned across different actors.
  - Step 1: Comparative review of ethical commitments. Collect and examine publicly available documents outlining the ethical principles adopted by various countries, international organisations, or corporations (e.g., UNESCO's Recommendation on the Ethics of AI, OECD AI principles). Use qualitative analysis or text comparison methods to identify common areas of convergence and divergence in ethical values.
  - Step 2: Semantic alignment and interpretation analysis. Investigate how different actors interpret the same ethical terms within their cultural, legal, or operational contexts. This step helps identify underlying cultural or normative barriers to interoperability.
  - Step 3: Identification of barriers and areas of consensus. Synthesize findings from the first two steps into an "ethical convergence-divergence matrix" that categorises: Highly interoperable areas; Partially aligned areas with divergent interpretations; Structurally incompatible areas. This analytical process clarifies both common ground and points of divergence in ethical commitments and interpretations.
- **Case-based analysis.** Considering the context-dependent nature of ethical understanding, analysing case studies is crucial for revealing how ethical standards are applied and debated in real-world situations. Select impactful or illustrative cases where ethical tensions have arisen across different jurisdictions or cultural settings, then determine where and how ethical interpretations differ, and whether ethical compromises, negotiations, or circumventions occur.

**Legal interoperability**

- **"Interoperability checks"** by policymakers and regulators are essential in developing regulatory interoperability frameworks. They involve several steps: First, review existing legislation to identify interoperability barriers. Second, ensure coherence between legislations by evaluating the compatibility of enabling laws across different countries to enhance interoperability. This will promote interoperability between AI systems at lower (technical) levels and decrease costs and implementation time.
- **Applying either a top-down or a bottom-up process.** A top-down approach necessarily involves establishing a global agency, such as the UN or one of its specialised organisations. Typically, this approach results in the creation of large bureaucracies. A bottom-up process must follow a step-by-step model that includes the main relevant entities and individuals involved in the substantive issue. The bottom-up approach requires significant coordination but does not involve harmonisation or management by central bodies.

**Technical interoperability**

- **International collaboration is crucial.** It involves alignment between global standardisation organisations such as ISO, IEC, IEEE, and ITU. The UN AI resolutions encourage Member States to promote developing and deploying internationally interoperable technical tools, standards, or practices to harness AI opportunities for sustainable development. The GDC and the UN High-Level AI Advisory Body also emphasise the importance of inclusive international collaboration and ensuring AI standards are adaptable and globally applicable.
- **Technical standardisation is a key enabler of technical interoperability.** The main goal is to adopt common standards across jurisdictions, software, hardware components, and platforms. Standards are essential for facilitating interoperability. International standards play a vital role in the governance ecosystem by clarifying regulations, supporting international interoperability, and offering companies best practices. International standards are technical documents developed by international Standards Developing Organisations (SDOs) that follow formal procedures for reaching consensus and are widely recognised by national governments and international organisations. Standards can be categorised according to their functionalities (Roberts & Ziosi, 2025)[16].
  - Technological interconnectivity standards for advanced AI systems have comparatively low demand, as relatively little technical interoperability is required between different companies.

- Transactional interconnectivity standards enable transactions and interoperability between organisations - such as through standardised contracts and procedures - primarily found in the open-source community (McDuff et al., 2024)[17].
- Standards are used to directly mitigate physical externalities, where one actor's behaviour directly affects another. Examples include safety standards related to information security (ISO/IEC 27001:2013) and the functional safety of AI systems (ISO/IEC TR 5469:2024). International standards addressing physical externalities generally require regulatory, reputational, and commercial incentives to ensure adoption.
- Standards are also used to reduce policy externalities, where laws or policies in one jurisdiction affect actors in another. They enable regulatory interoperability and encourage uniform compliance (Abbott & Snidal, 2001)[18].

- **Interoperability by design** means that an AI system or its components must be built following the proposed model and certain interoperability requirements. Open standards usually form a key part of technologies designed for interoperability and can be licensed by the developing entity at reasonable and non-discriminatory costs.
- **Instruments:** Common or compatible models, agreements on shared infrastructure, regular third-party testing, use of formal technical specifications, certification, validation processes, and international trade agreements such as WTO's *Technical Barriers to Trade (TBT) Agreement* (WTO, 1995)[19] are necessary to ensure that AI systems from different providers meet these technical interoperability standards.
- **SDOs should develop "regulatory intermediary"** (Abbott et al., 2017)[20] partnerships with newer and more agile institutions that are creating "leading-edge" technical specifications. This will promote a clearer institutional division of labour, enhance harmonisation, and reduce opportunities for "forum shopping" by states and private actors (Roberts & Ziosi, 2025).

## 3. AI Safety Governance and Interoperability Overview

This section provides an overview of AI safety governance and interoperability approaches across four jurisdictions. It also introduces Table 3 to Table 5, which compare AI safety governance in three sectors across these jurisdictions. These tables are included in the appendix at the end of the report. The comparison covers seven components: Objectives, Initiatives, Ethics, Binding measures, Targeted regulation/framework, Technical standards and Risks and challenges. Additionally, drawing on the country reports for China, South Korea, Singapore, and the United Kingdom (attached in the appendix), we identify both the effective interoperability instruments in AI safety governance and the barriers to interoperability.

### **3.1** NATIONAL OVERVIEWS

#### China

China's approach to AI safety governance employs a multi-layered model that balances innovation with risk mitigation. AI safety seeks to ensure that the research and development (R&D), deployment, and operation of AI systems - particularly high-risk ones such as autonomous driving, generative AI, and AI in education - are safe, controllable, and responsible, without compromising individual rights, public interests, or national security. It emphasises people-centric design, safety controllability, fairness, transparency, accountability, privacy protection, and alignment with national development goals (e.g., "Digital China"). The governance approach combines a top-down legal and regulatory framework - including the Cybersecurity Law, Data Security Law, Personal Information Protection Law, and sector-specific rules (e.g., automotive data, algorithm recommendation) - supplemented by bottom-up pilot programmes such as local demonstration zones (e.g., Beijing AV pilot zone) and "sandboxes" (e.g., 20-city V2X pilots). China also participates in global standardisation efforts (ISO/IEC JTC1, UNECE WP.29), regional digital cooperation, aligns with OECD AI Principles and UNESCO AI Ethics Recommendations, establishes bilateral digital agreements (e.g., China-Singapore Digital Economy Agreement), and engages in multilateral forums (GPAI).

**Autonomous Vehicles:** Considers safety in four key dimensions. Functional safety is guaranteed through requirements for closed-field testing and open-road validation. Product access evaluations are integrated into international standards such as ISO 26262 (GB/T 34590) and SOTIF principles. Data safety mandates that all AVs be equipped with event data recorders ("black boxes") to log operational data, while sensitive information (e.g., facial recognition, high-precision maps) must adhere to anonymisation and "minimum collection" requirements, i.e. only the least amount of personal or sensitive data necessary to achieve a specific purpose should be collected and the retention period for data must not exceed the minimum duration necessary to achieve the intended processing purpose. Cybersecurity obligations specify that manufacturers must obtain certification, perform real-time system monitoring, and report vulnerabilities to regulators. Vehicle-to-everything (V2X) communication systems are required to utilise encrypted transmission protocols. Ethical risk considerations focus on "unavoidable accident scenarios" (e.g., prioritising pedestrians vs. passengers), with AV decision-making mandated to align with public moral consensus and to be publicly explainable. At this stage, the "presumed liability" model under the Road Traffic Safety Law remains in effect. When a vehicle is operating in autonomous driving mode, the driver bears full responsibility unless it can be proven that the liability raises from human error or force majeure. Environmental, social, and long-term sustainability are incorporated into safety governance, with approaches involving international cooperation through standardisation, bilateral collaboration, and participation in multilateral forums.

**Education:** AI safety is based on a multi-dimensional framework that treats data, content, algorithms, and ethics as core pillars. Student data (e.g., grades, attendance) is classified as "sensitive personal information" requiring pseudonymisation and encryption during data transit and storage. The Minors Internet Protection Regulations prohibit third-party sharing of student data without guardian consent. Content safety is ensured through a mandatory "double review" - AI plus human oversight - to verify the accuracy of AI-generated educational material (e.g., textbooks, virtual experiments). Algorithmic safety involves record-keeping with regulators and providing "one-click adjustment" tools for teachers to modify AI suggestions, as well as transparency of algorithms to parents. Ethical safeguards prevent AI from replacing fundamental teaching tasks, with policies requiring "human-in-the-loop" oversight for high-stakes AI decisions (e.g., exam grading, student placement).

**Cross-Border Data Flows:** Multiple aspects of AI safety are addressed. The first concerns data classification risk, as AI training data may include "important data" (e.g., industrial AI uses manufacturing data) that must be assessed before cross-border transfer. Another critical aspect is model training safety, since AI models trained on cross-border data must comply with Chinese and foreign laws. The security of transmission requires cross-border data transfers to use encrypted channels (e.g., TLS 1.3) and secure APIs. Finally, enterprises must conduct due diligence on foreign recipients (e.g., verifying their data security capabilities) and include "onward transfer" restrictions in contracts, preventing recipients from transferring data to third parties without explicit approval.

### Singapore

Singapore's approach to AI safety governance is strategic and pragmatic. It is characterised by agile and voluntary frameworks rather than a single rigid AI law. This flexibility is balanced with the introduction of targeted legislation and mandatory legal requirements aligned with the risk level of AI systems. Its AI governance emphasises trust to ensure citizens can confidently use digital tools with safety and privacy protections and consensus-building between government, industry, and citizens. Such an AI safety governance approach is reflected across three critical sectors.

**Autonomous Vehicles:** Safety is overseen via a "physical-first" approach, with the Land Transport Authority (LTA) enforcing strict, pre-deployment testing at the Centre of Excellence for Testing and Research of AVs (CETRAN). Liability is managed through existing common law, supported by mandatory insurance and data recorders.

**Education:** The Ministry of Education (MOE) incorporates AI into its EdTech Masterplan 2030 to improve learning and teacher efficiency. Data privacy and ethical use are regulated by the Personal Data Protection Act (PDPA) and specific guidelines issued by the Personal Data Protection Commission (PDPC).

**Cross-Border Data Flows:** This approach prioritises "data free flow with trust", achieved by leveraging the Personal Data Protection Act (PDPA) alongside international, voluntary frameworks such as the Global Cross-Border Privacy Rules (Global CBPR) System and the ASEAN Model Contractual Clauses (MCCs) to ensure regulatory interoperability. At a broader level, ongoing grassroots efforts have independently proposed the establishment of "AI for Humanity" as United Nations Sustainable Development Goal #18 (UN SDG 18)—a call for united and collective global alignment that balances technology, commercialisation, and safety governance.

### South Korea

Korea's AI safety governance adopts an innovation-friendly, principles-plus-risk-based framework that tailors requirements to risk, prioritizes guidelines and voluntary norms, and legislates where needed. MSIT leads overall strategy while sectoral ministries set domain-specific standards, with strong alignment to OECD, G7, and the AI Seoul Summit. Testbeds and sandboxes are expanding, and safeguards centre recommendations outcomes on PIPA-driven data and privacy protection, alongside measures for deepfake labelling and election integrity. High-risk areas such as healthcare, transport, finance, and public safety face reinforced standards, certification, risk management, human oversight, logging/traceability, and explainability.

Key bodies include MSIT, PIPC, KCC, financial regulators, MFDS, MOLIT, KATS, and cybersecurity agencies—split responsibilities,

using a multilayered mix of laws, guidelines, standards, and certification to strengthen transparency, labelling, auditability, and ex-post liability and redress. Compared internationally, Korea shares the EU's risk-based orientation but emphasizes innovation, co-regulation, and sectoral oversight more; versus the US, it features a comprehensive privacy law and stronger platform/content rules with a centralized privacy regulator; and versus Japan, it stands out for testing, certification, sandboxes, and global engagement. Emerging issues include governance of general-purpose/frontier models, defining high-risk thresholds, allocating responsibility across the value chain, balancing trust and competitiveness, and ensuring interoperability with the EU AI Act, NIST AI RMF, and ISO/IEC standards.

**Autonomous Vehicles:** Korea has the Act on the Promotion of and Support for Commercialization of Autonomous Vehicles in force, providing the legal basis for introduction, sale, and safe operation of AVs. Korea has harmonized its Motor Vehicle Safety Standards with UN R157 (ALKS), and industry compliance emphasizes UN R155 (CSMS)/R156 (SUMS) requirements tied to type approval. The national K-City test bed provides multi-scenario, 5G-based validation environments for AVs (highway, urban, suburban, parking, community facilities). C-ITS and V2X pilots: Government-led connected-ITS pilots (notably Daejeon–Sejong and expressway corridors) verify vehicle–infrastructure interoperability under MOLIT coordination. MOLIT announced step-by-step verification within actual operational areas to manage safety for (incl. unmanned) automated driving services.

**Education:** Ensuring content integrity and accuracy requires safeguards that prevent biased, discriminatory, or harmful outputs in tutoring, assessment, and feedback, aligned with the MOE Ethical Principles of AI in Education (2022) and international frameworks (the UNESCO 2021 Recommendation and OECD AI Principles), and supported domestically by the AI Basic Act (enacted 2025; in force 2026). Protecting student data entails strict minimization, pseudonymization, secure storage, controlled access, and auditability under PIPA, with public-sector cloud deployments following CSAP baselines; lawful cross-border processing is further enabled by EU–Korea GDPR adequacy. Teacher oversight is anchored in KERIS/MOE classroom guidance and the 7th Master Plan for Educational Informatization (2024–2028), ensuring AI augments not replaces professional judgment. Equitable access follows KWCAG 2.1 and the Act on the Prohibition of Discrimination against Persons with Disabilities, and AI digital textbooks were reclassified from official textbooks to supplementary materials, reflecting a precautionary posture.

**Cross-Border Data Flows:** Protecting data security and confidentiality requires strong encryption, continuous transfer monitoring, and verification of controls when data moves across jurisdictions, under PIPA overseas-transfer rules and PIPC powers to suspend transfers where protection is inadequate. Regulatory alignment reduces uncertainty via EU–Korea GDPR adequacy and participation in the Global CBPR Forum; organizational controls are evidenced by K-ISMS/ISMS-P and ISO/IEC 27001/27701/27018 certifications. Transparency and accountability demand traceable processing, explainability for automated decisions, and clear allocation of harm responsibilities (DPIA/contractual measures). Interoperability safeguards combine standardized data formats/APIs and legal tools (SCCs, codes of conduct, CBPR) consistent with DFFT (Data Free Flow with Trust).

Interoperability is achieved through standardized data models and APIs carrying consent and purpose metadata, SSO and federated identity with role-based access, and auditable event logs for safe, traceable cross-border processing. Vendor assurance relies on transfer impact assessments, strict onward-transfer controls, geo-specific residency options, and recognized certifications, while provenance and watermarking standards help propagate content authenticity signals across jurisdictions, with a practical pattern of retaining data in-region where possible, exporting only minimized or pseudonymized datasets, encrypting with Korea-controlled keys, and ensuring portability and deletion.

## United Kingdom

The UK's AI governance balances innovation and safety, emphasising flexibility, principles, and sector-specific regulators. Instead of a single comprehensive AI law, the UK sets out five guiding principles for AI development and use across all sectors: safety, security, and robustness; appropriate transparency and explainability; fairness; accountability and governance; and contestability and redress. It recognises that inconsistent global regulations can hinder innovation and safety. It broadly aligns its framework with international efforts while remaining dedicated to engaging globally to support interoperability across different regulatory regimes. It aims to reduce business compliance burdens and embed the UK's values into emerging global AI governance strategies. The UK also used its G7 presidency to promote the idea of "Data Free Flow with Trust" (DFFT) for digital governance. The UK aligns its standards and laws with international benchmarks to ensure that AI systems and safety practices can operate seamlessly across borders.

The UK's principles-based AI safety governance approach is complemented by sector-specific regulations across all three focus areas. The common principles are interpreted in each sector to protect AI systems and data from harm, such as accidents, misuse, and cyber threats.

**Autonomous Vehicles:** Formal legislation, such as the Automated Vehicles Act, and rules explicitly incorporate these principles, including safety equivalence to human drivers and operator accountability. Security encompasses functional safety and vehicle cybersecurity; for example, the UK mandates compliance with international vehicle cyber standards like UNECE R155 and has testing protocols for fail-safe behaviour. It fosters a culture of safety and continuous improvement, similar to how aviation security is managed globally - an interoperable practice across borders.

**Education:** The UK Department for Education's Generative AI: Product Safety Expectations (DfE, 2025), schools and colleges should ensure that generative AI tools meet robust safety standards, including effective content filtering, access controls, and data protection measures. These expectations are designed to support compliance with safeguarding duties outlined in Keeping Children Safe in Education (DfE, 2025) and the Public Sector Equality Duty. While advisory rather than legally binding, the guidance emphasises that AI deployment in educational settings must align with statutory frameworks such as the Online Safety Act 2023, which introduces phased obligations for platform accountability to protect children online (Ofcom, 2025). Safeguarding and equality remain central principles, requiring schools to integrate AI use into their safeguarding policies and conduct risk assessments to uphold the ethical principle of "do no harm." Collectively, these measures aim to ensure that generative AI systems operate within a secure, responsible, and child-centred digital environment (DfE, 2025; Ofcom, 2025).

**Cross-Border Data Flows:** The UK's strategy is based on data protection laws and trade policies, aiming to enable free data movement for innovation while upholding high standards and public trust. The UK's framework ensures that cross-border data transfers are lawful, secure, and straightforward. The key ethical principle is that fundamental rights travel with the data, reflecting a belief in universal privacy and dignity. The UK relies on existing regulations such as the UK GDPR and Data Protection Act 2018 and frameworks from the AI Security Institute and the Alan Turing Institute (ATI). Post-Brexit, the UK makes its own adequacy decisions. The International Data Transfer Agreement (IDTA) offers templates and guidance for a risk-based approach. The Data Protection and Digital Information Bill (2023–2024) aims to facilitate flexible international data transfers. The UK incorporates data flow provisions in trade deals aligned with the G7 DFFT concept, removing barriers while building trust through shared principles. Security is crucial: encryption, secure APIs, and compliance with frameworks like ISO 27001 are expected for any data transfer. The UK's international agreements often include commitments to security practices, ensuring that its security measures align with those of its allies. Cybersecurity and resilience are viewed as sector-specific issues and national priorities covering all digital systems, including AI. The National Cyber Strategy and NCSC guidance provide a common foundation across sectors.

## **3.2** INTEROPERABILITY OVERVIEW

The major integration measures each jurisdiction has adopted to align with global or regional AI governance frameworks are identified in **Table 6**, **Table 7** and **Table 8.** These measures provide the foundations for interoperability in AI safety governance in the three crucial sectors studied.

## **3.3** EFFECTIVE INTEROPERABILITY INSTRUMENTS

Effective interoperability in this report is defined as two components: 1) Convergency: The instruments adopted by most of the four jurisdictions in their global or regional integrations; 2) Complementarity: Different governance instruments that reinforce each other to achieve complementary performance, leading to more robust, ethical, and effective integration and oversight of AI safety (e.g., AI regulations can be complemented by liability rules to address AI harms).

#### Effective ethical interoperability instruments

**Shared or common terminology:** Across the four case studies presented in this report, countries demonstrate similar ethical concerns in AI safety governance. This alignment lays a foundational basis for ethical interoperability.

- South Korea has issued the *Ethical Guidelines for Autonomous Vehicles*, emphasizing AI safety, accountability, and transparency. Its *Personal Information Protection Act* aims to establish a robust privacy regime.
- China promotes a human-centric ethical framework for AI, integrating values such as fairness, justice, transparency, accountability, and respect for human dignity into AI design and governance.
- The UK's AI regulatory framework is shaped by the *National AI Strategy* (2021) and the *AI Regulation White Paper* (2023), which promote five core principles: safety, fairness, accountability, contestability, and adaptability, ensuring AI decision-making aligns with ethical norms.
- Singapore's national AI strategy is built on key values including safety, accountability, transparency, and protection.

| Jurisdiction | Level of Integration |
|---|---|
| **China** | • **Legal and Principle Alignment:** Its core regulatory framework (anchored in the Data Security Law, Personal Information Protection Law (PIPL), and Provisions on Promoting and Regulating Cross-Border Data Flows (2024)) aligns with OECD Data Governance Principles, emphasizing "data free flow with trust" and balancing security with legitimate cross-border cooperation. It adopts global technical standards such as TLS 1.3 for encrypted transmission and ISO/IEC 27701 for privacy management. At the same time, GB/T 43697-2024 (Data Classification and Grading Rules) aligns with international data risk assessment methodologies. China has established a structured framework for integrating its cross-border data flow governance with international norms, and is striving to achieve alignment with international standards.<br>• **International Engagement and Mutual Recognition:** China participates in multilateral forums including the OECD Working Party on Data Governance and GPAI's Data Governance Working Group, and acts as an observer in the Global CBPR Forum to promote standard mutual recognition. Via the "Digital Silk Road," it provides technical assistance to developing countries to build data security frameworks, expanding global trust networks. |
| **South Korea** | • **Legal and Principle Alignment:** Robust foundation for AI interoperability, especially in personal data transfer regulation; Strong alignment with global data standards (GDPR adequacy, membership in the Global CBPR, etc.); Integrated legal framework with emerging international standards on data protection. Mirroring GDPR's "continuity of protection" principle. Data Protection Impact Assessment (DPIA) aligned with GDPR methodology -interoperability of risk management approaches.<br>• **International Engagement and Mutual Recognition:** Dual-track engagement working with regulatory and industry-driven models to cover more ground. G7's "Data Free Flow with Trust (DFFT)" – cross-border innovation while preserving rigorous privacy and safety standards. Seoul Declaration for Safe, Innovative and Inclusive AI, a high-level pledge to cooperate on interoperable AI governance standards across borders. Exploring mutual recognition of AI audits and certifications with partner countries.<br>• **Standardization and Capacity Building:** Contributing to ISO/IEC and ITU working groups on AI and data standards, OECD on AI system risk classification , ISO/IEC JTC 1/SC 42 on AI. |
| **Singapore** | • **Legal and Principle Alignment:** Leveraging PDPA with international voluntary frameworks to ensure regulatory interoperability and facilitate commerce. Singapore has developed "crosswalks" that map its AI Verify framework to international standards like NIST AI Risk Management Framework and ISO/IEC 42001 to enhance interoperability and reduce business compliance costs. Fostering the central norm is "data free flow with trust" to balance the economic benefits of data flows with the need to protect individual privacy and national interests, and to align AI with local values.<br>• **International Engagement and Mutual Recognition:** Founding member of the Global Cross-Border Privacy Rules (Global CBPR) System. The Infocomm Media Development Authority (IMDA) develops certifications like the Global CBPR to promote digital trust.<br>• **Capacity Building and Global Cooperation:** Becoming trusted and capable international partner in AI innovation & governance. |
| **UK** | • Reduce business compliance burdens and embed the UK's values into emerging global AI governance strategies; encourage mutual recognition and regulatory cooperation.<br>• **UK Adequacy Decisions:** recognized the EU/EEA, Japan, Canada, and others as adequate partners' UK–US Data Bridge.<br>• Incorporate data flow provisions in trade deals aligns with the G7 DFFT concept, removing barriers while building trust through shared principles. Participates in data flow discussions at the G20.<br>• Joining the Global CBPR Forum to develop an international certification system, making data exchange easier among member countries. Participating in the Global Privacy Assembly. The UK co-drafted the OECD AI Principles and the Declaration on Government Access to Data, establishing common ground with many other countries.<br>• **Bilateral Cooperation:** conducts data policy dialogues with key partners to share best practices and align regulatory approaches for AI governance. Collaborating with groups like the Global Partnership on AI (GPAI) to promote high-quality, representative data sharing.<br>• **Capacity Building:** The UK supports developing countries in creating their own data protection laws. |

**Table 6:** Comparison of AI safety governance's interoperability of four jurisdictions China, South Korea, Singapore and UK (Cross border Data Flow)

| Jurisdiction | Level of Integration |
|---|---|
| **China** | • China's AI-in-education governance integrates with international ethics and standards, and is striving to achieve dedicated standardization and platform compatibility.<br>• **Ethical and Principle Alignment:** It aligns domestic policies with UNESCO's Recommendation on the Ethics of Artificial Intelligence, prioritizing student well-being, educational equity, and the preservation of teachers' roles—consistent with global norms. The Minors Internet Protection Regulations (2024) and PIPL's minor protection provisions align with international child data privacy standards, requiring guardian consent for student data collection and limiting data processing to "minimum necessity."<br>• **Standardization and Collaboration:** China contributes to ISO/IEC JTC1 SC36 (Learning Technologies) to develop international standards for AI educational tools, and references global benchmarks such as WCAG 2.1 (accessibility standards) in technical requirements (e.g., text-to-speech, screen reader compatibility for students with disabilities). The National Smart Education Platform uses standardized APIs to ensure domestic interoperability, and Sino-foreign university partnerships conduct joint research on mitigating algorithmic bias and protecting minor data—aligning with global AI-in-education research agendas. |
| **South Korea** | • **Ethical and Principle Alignment:** Aiming to align with—and ensure interoperability with—international data-protection norms (e.g., the EU's GDPR).<br>• Ensuring any third-party AI service (possibly cloud-based or foreign-developed) adheres to Korea's child data protection standards.<br>• UNESCO (2023) "safe and effective" AI in education can be an international reference point for adoption decisions.<br>• Laying groundwork for **technical interoperability and standards**, exploring international standards (e.g., IEEE/ISO) for AI in education; referencing emerging IEEE/ISO AI-in-education standards; securing approval of the ISO/IEC 5259-1:2024 data-quality standard for AI; KATS is deepening cross-border standards coordination via the 2024 U.S.–Korea Standards Forum; public bodies have started adopting ISO/IEC 42001 AI management systems. |
| **Singapore** | • **Ethical and Principle Alignment:** MOE's EdTech are benchmarked against International practices and are guided by ethical AI use.<br>• The broader Model AI Governance Framework, while not specific to education, is consistent with frameworks from the EU and OECD. |
| **UK** | • **Ethical and Principle Alignment:** Schools refer to broader frameworks like the UNESCO Recommendations on AI in Education or the IEEE's ethical design guidance.<br>• A key member of UNESCO's Global Education Coalition on AI, aligning its domestic policies with international principles of fairness and inclusion.<br>• Supports digital and AI skills initiatives through the OECD and shares its projects via the OECD AI Policy Observatory.<br>• Carrying out cross-border research pedagogical innovations and AI progress. Learning from and sharing global best practices.<br>• **Technical Standard Alignment:** UK EdTech often adopts standards like Learning Tools Interoperability (LTI).<br>• UK schools must follow standard IT security practices, often referencing ISO/IEC 27001. This ensures that AI tools do not create security vulnerabilities or expose sensitive student data. |

**Table 7:** Comparison of AI safety governance's interoperability of four jurisdictions China, South Korea, Singapore and UK (Education)

| Jurisdiction | Level of Integration |
|---|---|
| **China** | • China's AV governance achieves strong alignment with international technical standards and engages in targeted bilateral cooperation, and is committed to ensuring the consistency between domestic standards and international standards.<br>• **Technical Standard Alignment:** Key domestic standards align with global benchmarks: GB/T 40429-2021 (AV Grading) adopts UNECE WP.29's automation classification; the GB/T 34590 series is equivalent to ISO 26262 (functional safety); and GB 44495-2024 (Vehicle Cybersecurity) references ISO/SAE 21434. It also aligns with UNECE WP.29 regulations on automated driving, ensuring compatibility with global AV terminology and safety validation frameworks.<br>• **Bilateral/Multilateral Cooperation:** China and Germany have signed the Joint Statement of Intent on Cooperation in the Field of Autonomous and Connected Driving, and will jointly develop vehicle-to-everything (V2X) technology. Domestically, 34 AV pilot zones (e.g., Beijing) adopt practices (e.g., unified operation data platforms, "black boxes" for event logging) consistent with international AV safety monitoring norms. |
| **South Korea** | • **Technical Standards Alignment:** Actively engages in joint V2X/C-ITS R&D and cross-border pilots via MoUs and standardization working groups. Actively participates in UNECE WP.29 (incl. UN R155 cybersecurity, UN R156 software updates) and ISO/TC 204, aligning national rules with evolving UN/ISO standards.<br>• **International Alignment:** Korea adapted KMVSS to UN Regulation No. 157 (ALKS). Implemented via MOLIT Notice No. 2022-670, enhancing consistency with UNECE rules.<br>• **National Testability & Verification:** Operates "K-City" as a multi-scenario proving ground (highway, urban, suburban, parking, community facilities). Validation platform is implemented for 5G/C-ITS integration. The vehicle–infrastructure interoperability is verified in real traffic environments. |
| **Singapore** | • Participates in the Asia-Pacific Economic Cooperation (APEC) to broaden technical coordination and support harmonized standards and regulatory approaches.<br>• Aligns with UN regulation on Cybersecurity Management Systems that refers to standards like ISO 26262 for Functional Safety and ISO/SAE 21434 for Cyber-security of Road Vehicles. |
| **UK** | • **Interoperability by design**, enabling UK and foreign AVs to operate safely across different markets.<br>• **Harmonise AV regulations** through the UNECE and bilateral agreements. · Based on international safety standards, such as UNECE regulations and ISO standards, by incorporating specific measures designed for autonomous functionality. Standardisation initiatives cover terminology and scenario descriptions that are essential for interoperability and safety validation.<br>• Horizon Europe and partnership accords with countries. |

**Table 8:** Comparison of AI safety governance's interoperability of four jurisdictions China, South Korea, Singapore and UK (Autonomous Vehicles)

**Compatible cross-institutional and international ethical assessment and accountability mechanisms:** Countries are adopting multi-tiered approaches to establish compatible ethical governance systems at both domestic and international levels.

- Domestically, the four governments combine foundational legislation, industry-specific rules, and adaptive ethical guidelines to build comprehensive AI safety governance ecosystems. These layered systems enable AI development under relatively clearly defined ethical boundaries.
- Internationally, countries align domestic initiatives with global frameworks. Specifically, South Korea, China, the UK, and Singapore all participate in the UN's Global Digital Compact, the OECD AI Principles, and UNESCO's Recommendation on the Ethics of AI. These global norms identify and integrate diverse national values, providing a common reference for assessing AI practices, facilitating cross-border regulatory coordination, and promoting mutual trust and interoperability in AI among countries. Moreover, the emphasis these four countries place on sustainability in AI aligns closely with the UN Sustainable Development Goals (SDGs), serving as a valuable ethical anchor point. The SDGs highlight measurable data such as energy consumption, AI coverage, and usage rates, which offer greater comparability and standardization potential across jurisdictions. This, in turn, facilitates the mutual recognition of ethical assessment tools and strengthens efforts toward ethical interoperability.

**Multi-stakeholder engagement mechanism:**

- Countries have adopted inclusive governance models incorporating diverse stakeholders' voices. For example, in South Korea, the government convenes AI forums and working groups in sectors such as education, fostering dialogue among educators, industry, and citizens.
- The UK supports institutions such as the Alan Turing Institute and the AI Safety Institute to provide independent advice and

policy input, as well as to conduct public consultation and co-production on its GenAI in education policy, ensuring broad stakeholder engagement.

- Singapore has established an Advisory Council on the Ethical Use of AI and Data, comprising representatives from technology firms, AI users, and other stakeholders.
- Internationally, these countries actively participate in global forums and initiatives. For example, South Korea engages with the OECD, UNESCO, the Global Privacy Assembly, the G20, and the Global CBPR Forum, advocating for interoperable AI safety standards. The UK co-leads efforts like the OECD Declaration on Government Access to Personal Data Held by Private Sector Entities, facilitating shared learning and best practices in AI. China contributes to the OECD and GPAI working groups on data governance and privacy, and is active in the UNESCO Global AI in Education Alliance. Singapore is a trusted and capable international partner in AI innovation and governance, supporting practical and interoperable frameworks such as the Global CBPR system and ASEAN MCCs.

Through the Singapore Consensus in 2025, Singapore stives to facilitate meaningful conversations between AI scientists and AI policymakers globally for maximally beneficial outcomes by bringing together AI scientists across geographies to identify and synthesise research priorities in AI safety.

### Effective regulatory interoperability instruments

**Standardisation:** States are aligning their regulatory frameworks with globally or regionally recognised normative benchmarks (guidelines, rules or practices etc) to ensure consistency and trust across borders. These benchmarks are:

**a. Cross-border Data Flows**

EU's General Data Protection Regulations (GDPR) and its mechanisms including adequacy decision, "continuity of protection" principles, data classification etc; G7's "Data Free Flow with Trust" (DFFT) framework; OECD's Data Governance Principles; UNESCO Recommendation on the Ethics of AI; The new Global CBPR (Cross-Border Privacy Rules) evolved from the APEC's Cross-Border Privacy Rules system, is developing towards a global data transfer framework and certification processes for personal data transfer. All four jurisdictions are either its members or observers (China). South Korea also launched a domestic CBPR certification system.

**b. Education**

UNESCO's Recommendation on the Ethics of AI and Global Education Coalition on AI, prioritise student well-being, educational equity, and the preservation of teachers' roles as well as principles of fairness and inclusion.

**c. Autonomous Vehicles**

United Nations Economic Commission for Europe (UNECE) vehicle regulations

**Harmonization:** States have established unified regulations through bilateral or multilateral agreements.

**a. Cross-border Data Flows**

China has applied to join digital trade agreements with Singapore, Chile, and New Zealand. South Korea forges digital trade agreements and partnerships with countries like Singapore, Vietnam, and the UK, including commitments on cross-border data flows, digital trust, and cooperation on AI ethics.

**b. Autonomous Vehicles**

UK harmonises AV regulations through the UNECE WP.29

**c. Education**

The UK is negotiating bilateral and multilateral agreements (such as the CPTPP) that promote EdTech exchange etc.

**Mutual recognition:** Jurisdictions are increasingly accepting each other's regulatory mechanisms to reduce compliance burdens.

**a. Cross-border Data flows**

GDPR-style adequacy mechanisms have been accepted and actively adopted in countries like South Korea, Singapore and the UK. UK Adequacy Decision recognises the EU/EEA, Japan, South Korea, Canada, and others as adequate partners. South Korea aims to set up adequacy decisions with the UK, Singapore, and other jurisdictions. South Korea and its partners are exploring mutual recognition of AI audits and certifications. China and Germany have signed a Memorandum of Understanding (MoU) on China-Germany Cooperation in Cross-Border Data Flow, facilitating cross-border data exchange for enterprises

**Cooperation:** Regulators from different states cooperate to overcome the disparity of different regulatory regimes via multilateral dialogue forums, dedicated working groups, joint research and sharing best practices or pilot experience.

**a. Cross-border data flows**

AI Summit dialogues held in the UK, South Korea and France as a high-level forum to advance discussion on AI governance, facilitating AI safety, innovation and inclusivity. China, South Korea and UK participated in multilateral forums including OECD Working Party on Data Governance and Global Partnership on AI (GPAI)'s Data Governance Working Group and Global Privacy Assembly

**b. Autonomous Vehicles**

China and Germany have signed the Joint Statement of Intent on Cooperation in the Field of Autonomous and Connected Driving, and will jointly develop vehicle-to-everything (V2X) technology. UK participates in Horizon Europe and partnership initiatives and initiatives like Horizon Europe and partnership accords and the G7 Transport Ministers' declarations to develop joint research, regulatory alignment and shape global best practices.

**c. Education**

China participates in UNESCO's Global Education Coalition on AI, and has Sino-foreign university joint research on AI safety in education. These collaborations focus on key issues such as mitigating algorithmic bias and protection of minors' data. UK collaborates in digital and AI skills through the OECD and OECD AI Policy Observatory, and cross-border research in pedagogical innovations and AI progress, and shares global best practices to meet its strict safety and ethical standards.

**Effective technical interoperability instruments**

**Technical standardisation**: by adopting common standards across jurisdictions, across software, hardware components, and platforms

**a. Cross-border Data Flows**

China adopts global technical standards such as TLS 1.3 for encrypted transmission and ISO/IEC 27701 for privacy management, its GB/T 43697-2024 standard for Data Classification and Grading Rules aligns with international data risk assessment methodologies. South Korea adopts international data format standards such as IEEE Learning Data standards for ed-tech, or ADAS/AD sensor data formats for vehicles so that when data crosses borders it remains interpretable and usable by foreign AI systems without needing error-prone conversion. UK complies with frameworks like ISO 27001 for information Security, UNECE regulations for safety, ISO standards' specific measures designed for autonomous functionality and ISO/IEC 27701 for privacy management

**b. Autonomous Vehicles**

China's grading standards GB/T 40429-2021 adopts the UNECE WP.29's automation classification to ensure compatibility with global AV terminology. Functional safety standards GB/T 34590 series are equivalent to ISO 26262, and GB/T 38667-2020 (SOTIF) and GB 44495-2024 (Vehicle Cybersecurity) references ISO/SAE 21434. Together they set the requirements for system design, testing, and validation. UK follows global standards such as ISO 26262 (electronic systems), ISO 21448 (intended functionality safety), and ISO/SAE 21434 for cybersecurity. As a member of UNECE WP.29, the UK enforces its regulations including Regulation 157 (Automated Lane Keeping Systems) and Regulation 155 (cybersecurity). BSI's programme promotes adopting emerging standards like ISO 34503 on operational design domains, with PAS 1883:2025 guiding local implementation. South Korea works with UNECE WP.29 on vehicle regulations.

**c. Education**

South Korea's reference emerging IEEE/ISO AI-in-education standards. China contributes to ISO/IEC JTC1 SC36 (Learning Technologies) and sharing experiences from the National Smart Education Platform.

**Collaborations:** Joint efforts in research and standard-setting to enhance global interoperability.

**a. Cross border data flows**

South Korean experts participate in ISO/IEC and ITU working groups on AI and data standards, developing common data schemas, metadata standards, and ontologies. They also contribute to the OECD's work on AI system risk classification and ISO/IEC JTC 1/SC 42 on AI which develops standards for the AI lifecycle. These technical standards facilitate cross-border acceptance of AI products.

**b. Autonomous Vehicles**

China participates in UNECE WP.29 on global AV regulations and ISO/IEC JTC1 on AI safety standards, with standards such as GB/T 40429-2021 being referenced in international AV standardisation discussions. China collaborates with Germany on joint Vehicle-to-Everything (V2X) technology R&D aligning with global efforts to test cross-infrastructure AV interoperability.

**Creation of Infrastructure for Integrations**

South Korea invests in technical infrastructure for cross-border data flows to secure international network links and cloud arrangements under the principle of "secure data corridors" – a hardened pipeline for data exchange with designated certified cloud centres in Korea and other countries with encrypted VPN connections and mutual audits. This infrastructure, combined with common technical standards and certifications, aims to provide a backbone for trustworthy AI collaboration internationally. South Korea can enforce its rules and lower practical barriers for companies to comply. It also fosters innovation since companies can integrate into global data ecosystems using standardised APIs and certifications rather than negotiating one-off arrangements. Singapore's AI Verify Toolkit helps companies assess the responsible implementation of their AI system against 11 internationally recognised AI governance principles. The framework is aligned with other international frameworks such as those from EU, G7, OECD, and the US.

## 3.4 INTEROPERABILITY BARRIERS

Despite growing international attention to AI interoperability, several key barriers continue to hinder effective interoperability across jurisdictions and sectors.

**Ethical interoperability barriers**

**Challenges in Algorithmic Transparency and Accountability for Safety-critical Systems:** Insufficient algorithmic transparency and the absence of standardized mechanisms limit the ability of different systems and stakeholders to interpret, audit, and align decisions according to shared ethical standards, preventing consistent ethical evaluation and information exchange. Moreover, accountability in complex AI ecosystems remains inherently fragmented. Determining moral responsibility and ensuring ethical alignment across complex AI ecosystems is difficult. Together, these factors create significant interoperability barriers.  For example, in autonomous driving sector, this leads to ambiguity in assigning liability in the event of an accident across drivers, owners, manufacturers, and software developers. Similarly, in educational settings, interoperability challenges emerge when AI systems designed under different pedagogical and ethical assumptions cannot integrate or ensure equitable learning outcomes across institutions.

**Limited Universality of International Ethical Frameworks:** Existing frameworks, such as the OECD AI Principles, while significant, are not universally adopted. Efforts by geopolitical blocs like the G7, G20, or Global CBPR Forum aim to align AI safety standards, but often exclude perspectives from the Global South and lack mechanisms for truly global interoperability.

**Voluntary Nature of Ethical Principles:** Unlike laws and binding industry regulations, ethical AI principles are often voluntary and advisory. Without translating ethics to legal enforcement, their impact on countries' or organizations' AI practices may be limited, especially when safety concerns conflict with commercial or political interests.

**Uneven Maturity and Implementation of AI Standards:** Significant disparities in AI development across regions result not only in uneven technical and institutional capacity to design and deploy effective AI safety mechanisms, but also in divergent ethical priorities and governance approaches. These ethical and structural differences create barriers to transferring and adapting best safety practices, particularly from regions with established AI safety governance to those with limited resources and differing value frameworks.

**Regulatory interoperability barriers**

**Lack of Global Regulatory Standardization:** Apart from a few UN frameworks - such as the United Nations Economic Commission for Europe (UNECE) vehicle regulations and UNESCO's Recommendation on the Ethics of AI - most normative benchmarks adopted by the four countries are either formulated by regional trading blocs (e.g., the EU) or by self-selected multilateral organizations and forums (e.g., the OECD, G7, and Global CBPR). There is a notable absence of globally inclusive AI safety and security benchmarks that can guide regulatory alignment among states. Furthermore, a universally accepted mechanism to coordinate regional and multilateral efforts is lacking. This gap risks fragmenting AI safety governance and undermining the cohesion of the global digital economy.

**Uneven Regulatory Benchmark Contributions:** Most benchmarks originate from trading blocs, organizations, or forums based in the Global North, driven by concentrated regulatory leadership, soft power, economic dominance, and diplomatic influence. This asymmetry in shaping international norms risks widening AI development and governance disparities, leading to uneven global benefits. Effective AI safety governance requires broader participation, yet many countries lack the economic, political, and technical leverage to influence global standards meaningfully.

**Geopolitical Tensions Undermine Frontier AI Safety Collaboration:** Competition for technological and economic leadership, data securitization, and eroding trust are obstructing global cooperation on AI safety - particularly in AGI research and cross-border data governance. Data flows risk politicization amid strained diplomatic ties. Mutual recognition of data protection regimes remains limited, with EU GDPR adequacy decisions confined to a few countries.

**Regulatory Harmonization through Digital Trade Agreements:** Bilateral agreements offer advantages like faster resolution, greater control over outcomes, and private negotiations. Digital trade agreements support harmonisation in cross-border data flows and educational exchanges, mutual recognitions are used in data protection standards, AI audits and certifications, and commercial vehicle testing. However, many cannot effectively adjudicate modern data and AI safety issues. While dispute mechanisms and mutual recognition efforts are valuable, they require extensive coordination. A multilateral system with a dedicated institution is needed to guide a common approach to the safe exchange and use of data and AI technologies in different sectors.

**Prioritising Political and Commercial interest of Interoperability Over Legal Harmonisation:** Relying on voluntary mechanisms like the Global CBPR System or ASEAN MCCs to promote trusted data flows effectively prioritises political and commercial interests of interoperability over legal harmonisation. These frameworks, by design, do not ensure compliance with the domestic laws of receiving countries and require additional safeguards to prevent data exposure in jurisdictions with weaker protection standards.

**Diverse Liability Frameworks for Autonomous Driving:** Autonomous vehicle liability models are varied across jurisdictions. China applies a "presumed liability" approach, where the operator is liable when the vehicle is in autonomous mode unless fault is proven to stem from human error or force majeure. Singapore relies on common law principles, addressing accidents under the tort of negligence, supported by mandatory third-party insurance and black box data recorders. However, negligence-based models struggle to assign fault in complex AI systems, where errors often arise from software or algorithmic decision-making. South Korea's 2020 Guarantee of Automobile Accident Compensation Act adopts a shared liability model, assigning primary responsibility to the operator while allowing claims against manufacturers or software providers if defects contributed to the incident. The UK's AV Act 2024 shifts liability to the system operator or manufacturer when a vehicle operates autonomously, recognising the AI system as the accountable entity. While a global liability model remains unlikely in the near term, international standardisation is increasingly necessary. Bodies like UNECE WP.29 are working towards harmonised principles. The UK also monitors EU AI and product liability rules to ensure cross-border compatibility and market access. Regulatory harmonisation would ease public concerns and support AV deployment across jurisdictions, especially given the international nature of software updates and map data. In addition, China, Singapore, the UK and South Korea lack specific statutory liability rules for fully autonomous (Level 5) systems, posing significant regulatory risks.

**Fragmented Liability and Data Protection in Cross-Border Data Flows:** Cross-border data flows raise complex liability and safety risks. When AI services cause harm internationally, liability may be unclear and subject to multiple legal systems. Fragmented protection frameworks increase compliance costs, safety risks, and erode trust. UK law stresses due diligence, holding companies liable for breaches even if caused by foreign partners. Its ethical stance - fundamental rights to travel with data - reflects a commitment to universal privacy. South Korea enforces robust safeguards, including pre-transfer risk checks, onward transfer controls, and compliance monitoring. It applies GDPR's "continuity of protection" and OECD's "data free flow with trust" principles, requiring each recipient in a data chain to meet protection standards such as adequate level of protection. China's model, rooted in data sovereignty with trust, mandates risk assessments before transfer, evaluating recipient security, misuse risks (e.g., foreign government unauthorised access), and national security and individual rights impacts. Post-transfer, enterprises must monitor data use and use of audit logs. The government encourages companies to purchase "cross-border data liability insurance" to mitigate risks. Singapore adopts voluntary international frameworks like the Global CBPR System and ASEAN MCCs to balance economic benefits with privacy protection and national interests, and align AI with local values.

**Limited International Alignment in AI Safety in Education:** AI safety alignment in education remains largely focused on child-specific data protection and joint research, partly due to the limited internationalisation of education systems. However, schools often rely on broader frameworks such as the UNESCO Recommendations on AI in Education and IEEE's ethical design principles.

**Need to Invest in Digital Public Infrastructure:** Investing in digital infrastructure—such as standardised digital mapping of road regulations for AVs—is essential for global interoperability. These efforts support SDG 9 and reduce barriers to international tech deployment, promoting a more competitive and collaborative global market.

**Technical Standard Interoperability Barriers**

**Overlapping Efforts Among International AI Standard Bodies:** Multiple technical and intergovernmental organisations - including ISO, IEC, IEEE, ITU, CEN, CENELEC, ETSI, OECD, UNESCO, UNECE, APEC, and ASEAN - are actively developing AI standards. These efforts aim to guide responsible AI development and deployment by establishing terminology, principles, standards and technical frameworks. Market-based standards and technological solutions can complement government policies, foster multi-stakeholder consensus, and promote best practices. Technological solutions and market-based standards can supplement government measures and form part of a broader framework to achieve AI safety policy objectives. They also encourage consensus-building among stakeholders, strengthen cooperation on AI safety, and promote best practices. However, overlapping standardisation efforts, and inconsistent terminology across frameworks create complexity. Our country reports show both convergence (e.g., UNECE WP.29 VE regulation, ISO/SAE 21434 cybersecurity) and divergence (e.g., data formats and semantic standards). Without better coordination, fragmented standardization risks increasing compliance burdens - especially for SMEs - and hindering cross-border collaboration and innovation.

**Focus on Deployment Safety and Assurance:** Current national AI safety initiatives, such as the AI Verify toolkit and AI Assurance Sandbox, prioritize testing and assurance at the application or deployment level. While essential, this downstream focus leaves gaps in model-level controls - such as oversight of foundational training data or the intrinsic safety of General-Purpose AI (GPAI). Without addressing these upstream risks, downstream assurance may be ineffective if the underlying models are unsafe or opaque.

**Gaps in Catastrophic/Frontier Risk Mandates:** Governance standards largely concentrate on immediate and societal risks like fairness and data privacy. However, technical research into low-probability, high-impact threats - such as loss of control or dual-use risks involving bioweapons or advanced cyberattacks - remains limited, creating a blind spot in regulatory foresight.

## 4. Recommendations

Our recommendations are based on Section 2.3: Methods of Study and draw from effective interoperability measures and identified barriers, with the goal of aligning AI safety governance with the Global Digital Compact and other UN initiatives. A combination of ethical, regulatory, and technical interoperability mechanisms is essential to support global AI safety governance. Alignment in AI governance ensures that AI systems and their governing frameworks reflect human values, societal needs, and ethical principles - such as safety, fairness, transparency, and accountability - embedded throughout the development process. Robust regulatory frameworks must uphold these principles and meet compliance standards. Our recommendations offer ethical, regulatory, and technical measures to address fragmentation in AI safety governance, grounded in empirical evidence from China, South Korea, Singapore, and the United Kingdom.

We recommend **leveraging the effective interoperability instruments** outlined in Section 3.3 at both global and national levels. Additional recommendations span three key dimensions.

### Ethical Interoperability Recommendations

**Promotion of Ethical Self-Certification Report:** the UN takes the lead in convening, coordinating, and overseeing the development of the AI Ethical Self-Certification Report mechanism, while national governments, regional bodies, or organisations would be responsible for submitting their respective self-certification reports. This mechanism would serve as a structured tool for countries and organizations to disclose how their AI systems align with core ethical principles. The report could include (but is not limited to) transparency, accountability, social & environmental impact, public engagement, fairness & inclusiveness, and third-party auditors. By systematising the disclosure of ethical interpretation and practices, this self-certification report can: Provide a shared ethical language across jurisdictions and sectors, reducing ambiguity in cross-border collaboration; Enhance regulatory clarity and reduce compliance friction for developers operating in multiple countries; Empower adopters, regulators, and the public to make more informed and trust-based decisions regarding AI deployment. Such a mechanism could be promoted through multilateral platforms (e.g., OECD, GPAI, Global CBPR Forum) to promote global AI ethical convergence. This approach helps alleviate the challenge posed by the diversity of ethical narratives and interpretations across countries, which has made ethical interoperability difficult to achieve. To ensure that the self-certification mechanism remains credible and adaptive to evolving AI practices and evidence, an annual review cycle is recommended, allowing organizations to update their reports based on new data, user feedback, or ethical developments.

**Upholding the UN System as the Primary Forum for AI Ethics Deliberation:** We urge the UN to propose more concrete measures for global AI ethical governance, for example, introducing AI ethics indicators or establishing a global AI ethical monitoring dashboard. According to our research, most countries tend to rely on domestic initiatives or frameworks established by groupings such as the G7 and OECD when addressing AI safety and ethical issues. While these initiatives are valuable, they remain limited in scope, representativeness, and may reinforce fragmented approaches to global AI ethics governance. In contrast, the UN platforms, through their inclusive mechanisms for dialogue among both developed and developing countries, offer an opportunity to promote equitable participation and ensure that the Global South meaningfully contributes to shaping global AI ethics agendas. Building on the UN's long-standing experience in facilitating Global North - Global South collaboration across sectors such as finance, health, and climate, as well as its multilateral principles, specialized agencies, expert networks, and institutional mechanisms dedicated to supporting the Global South, the organization is uniquely positioned to coordinate global AI ethical governance. However, the current UN texts and frameworks on AI ethics have not yet received the level of global attention that they merit - likely because the content has thus far been too general and lacks actionable specificity.

**Advancing the Global AI Ethics Framework:** A major obstacle to AI ethical interoperability is the lack of strong, enforceable policy tools. Current efforts often rely on broad principles, non-binding guidelines, and voluntary commitments. Based on the country report, this research highlights how four countries reflect distinct priorities in and approaches to AI safety: China emphasizes ensuring that the R&D, deployment, and operation processes of AI systems are safe, controllable, and responsible, without compromising individual rights, public interests, or national security. South Korea focuses on practices, standards, and governance mechanisms, ensuring that AI systems operate reliably, securely, and in alignment with human values, minimizing unintended harm, misuse, and systemic risks. Singapore emphasizes ensuring AI systems are trustworthy, reliable, and secure throughout their entire lifecycle; and the UK focuses on addressing sector-specific physical, psychological, and privacy risks. These differences underscore the challenge of harmonizing diverse national interests and governance models on a global scale. We urge platforms such as the Global Dialogue on AI Governance (GDAIG) and the Independent International Scientific Panel on AI (SPAI) to lead in propose universally applicable ethical standards that can serve as the basis for concrete, operational, and potentially binding international policy instruments, supported by a set of clear, small-scale Key Performance Indicators (KPIs) to track progress, identify gaps, and enable regular cross-country and cross-organizational AI safety review.

### Regulatory Interoperability Recommendations

**>** GOVERNANCE ARCHITECTURE AND COORDINATION

These recommendations focus on building institutional structures and international cooperation mechanisms to support regulatory interoperability.

**Establish a Multilateral System for Coordinated AI and Data Governance:** To ensure the safe trade and safe use of data and AI

technologies across sectors, we recommend that states negotiate the creation of a multilateral system with a designated institution or mechanism to coordinate a common approach. This system should uphold the United Nations (UN) Charter's principle of sovereign equality among all Member States, as applied to national actions in cyberspace. This interpretation is endorsed by the UN's Group of Governmental Experts on Developments in the Field of Information and Telecommunications in the Context of International Security (UNGGE) and the subsequent Open-ended Working Group (OEWG).

**Establishment of a Coherent National Entity for Global Engagement:** To effectively engage with UN bodies and global partners, countries should formalise a coherent national entity - such as the UK's *AI Coordination Council* - to design a unified national approach. This ensures that contributions to global AI governance are consistent, strategic, and impactful. Countries may also consider establishing a *National AI Safety Coordination Council* to address regulatory inconsistencies and represent national interests in global AI governance forums. By centralizing the coordination of cross-sector AI safety policies, such entities can enhance international collaboration and ensure that national strategies align with emerging global norms.

**>** STAKEHOLDER ENGAGEMENT AND PUBLIC PARTICIPATION

These emphasize inclusive governance through collaboration with diverse actors and public involvement.

**Inclusive Multi-Stakeholder Engagement for AI Safety Governance:** Effective AI safety governance depends on inclusive mechanisms that bring together a wide range of perspectives. Governments should take the lead in convening AI safety forums and working groups that foster dialogue among civil society, technical experts, industry stakeholders, and citizens. They should also establish advisory bodies with the authority to provide policy input and oversee the safe use of AI technologies and data. Supporting independent AI safety institutes is equally important, as these organizations can offer expert, unbiased advice to guide decision-making. Independent International Scientific Panel on AI (SPAI) and the Global Dialogue on AI Governance (GDAIG) provide scalable models for inclusive governance. Similar structures should be developed at regional and national levels to ensure both bottom-up participation and top-down coordination in shaping AI safety governance.

**Enhancing Public Engagement to Strengthen International AI Governance:** The long-term success of AI governance depends on public trust and the democratic values emphasized in UN guidelines. To cultivate this trust, countries should launch national dialogues on AI's societal impacts, engaging educators, industry leaders, technical experts, and civil society early through schools, public forums, and dedicated task forces. Building trust in AI systems through media communication and fostering a culture of safety and ethics within industry are also essential steps. Establishing a societal consensus is critical for the responsible management of this transformative technology. Domestic public engagement is not only vital for national resilience - it is also foundational for effective international cooperation. A country's ability to participate credibly in global technology governance depends on the strength and integrity of its domestic ecosystem. A robust national framework grounded in research, education, and public trust provides the foundation for meaningful and impactful global engagement.

**>** STANDARDS, BENCHMARKS, AND TRANSPARENCY

These aim to improve interoperability through shared benchmark, transparency, and accountability mechanisms.

**Launch a Global Benchmark for AI Safety and Security:** To support regulatory alignment among states, there is an urgent need to establish a global benchmark for AI safety and security. While existing UN frameworks - such as the UNECE WP.29 and the UNESCO Recommendation on the Ethics of AI - offer foundational guidance, they are not sufficient to implement the Global Digital Compact (GDC) and the two recent UN resolutions on AI governance. This new global benchmark must be developed with the participation of a critical mass of countries to address the current imbalance in global norm-setting and policy influence. It can be formulated through the newly established mechanisms like the Independent International Scientific Panel on AI (SPAI) and the Global Dialogue on AI Governance (GDAIG), or by leveraging and coordinating existing regional efforts.

**Enhancing Transparency and Accountability in AI Safety Governance:** To strengthen public trust and improve oversight, countries are encouraged to develop clear AI safety key performance indicators - such as autonomous vehicle accident rates, data breach incidents, and public trust levels. Publishing an annual national *AI Safety Report* can help track progress, inform the public, and guide the refinement of policies and governance frameworks. Governments should also enhance transparency in policymaking by publicly sharing the reasoning, standards, and evidence behind AI-related regulations. Regular AI policy evaluations, including annual white papers or audits of key programs, should be conducted and made publicly available to ensure accountability and continuous improvement.

**>** TECHNICAL AND LEGAL INFRASTRUCTURE FOR INTEROPERABILITY

These address the development of interoperable systems, legal harmonization, and data governance mechanisms.

**Promote Adaptation and Expansion of Data Interoperability Mechanisms:** In the absence of a unified global framework for cross-border data transfers, it is essential to encourage the adoption of effective mechanisms for standardization, harmonization, mutual recognition, and international cooperation. Potential instruments include: Alignment with global or regional benchmarks; bilateral and multilateral agreements (e.g., digital trade agreements, UNECE WP.29); GDPR-style adequacy mechanisms; mutual recognition of AI audits, data protection certifications, and test results; multilateral dialogue forums and dedicated working groups, joint research initiatives and sharing of best practices or pilot experiences. These mechanisms can foster interoperability, reduce redundant compliance burdens for enterprises, mitigate risks, and lower cross-border frictions. They also enhance portability of compliance, supporting data free flow with trust. As a result, data and AI-enabled services can move across borders with greater legal certainty and shared expectations around risk management.

**Additional Protections to Mitigate Data Flow Risks:** In addition to voluntary international mechanisms such as the Global CBPR System and ASEAN Model Contractual Clauses (MCCs), more robust protections are urgently needed to safeguard cross-border data flows. These enhanced safeguards include: Adoption of ethical principles such as "fundamental rights travel with the data", "universal privacy and dignity", and the GDPR's "continuity of protection" principle; implementation of thorough due diligence as a key risk mitigation measure, ensuring that domestic companies remain accountable for data breaches occurring overseas; and introduction of cross-border data liability insurance to help companies manage financial risks associated with international data transfers. These legal and policy tools support the vision of the UN Global Digital Compact (2024) by promoting open data flows with trust, embedding security by design, and integrating safety, privacy, and human rights into digital cooperation.

**Developing Greater Standardization and Harmonization of Liability Models for Autonomous Vehicles:** While a globally uniform liability model for autonomous vehicles may not be foreseeable in the near future, there is a growing recognition of the need for greater international standardization and harmonization. Efforts are underway through international bodies such as the ISO 39003:2023 standard - *Road Traffic Safety (RTS): Guidance on Ethical Considerations Relating to Safety for Autonomous Vehicles* - and the United Nations Economic Commission for Europe (UNECE) WP.29 liability framework. UNECE WP.29's technical regulations and guidance documents help member countries address liability in the event of an accident within their own legal systems. However, no internationally binding legislation currently exists. Given that traffic regulation is embedded in system design, industry stakeholders have called for regulatory harmonization at the international level. Legislation needs to evolve alongside technological advancements, without hindering progress at national borders. In particular, statutory liability rules for fully autonomous (Level 5) vehicle systems must be collectively researched and addressed by states to ensure legal clarity, safety, and accountability.

**Development of Interoperable Digital Public Infrastructure:** States can collaborate and invest in digital infrastructure to enhance global interoperability - particularly through initiatives like standardized digital mapping of road regulations for autonomous vehicles (AVs). Such efforts are essential for harmonizing technological deployment across borders and reducing regulatory fragmentation. They also support Sustainable Development Goal 9 (Industry, Innovation, and Infrastructure) by promoting inclusive infrastructure for technology deployment, fostering innovation, and enabling a more competitive and cooperative global market.

**>** RESEARCH AND CAPACITY BUILDING

These support long-term interoperability through education, research, and risk management.

**Investment in AI Safety, Frontier Risk Management, and Alignment Research Collaborations:** Significantly increasing investment and international collaboration in AI safety research is essential for managing the risks associated with advanced AI systems. Key focus areas include AGI safety, privacy-enhancing technologies, infrastructure evaluation for frontier models, testing for dangerous capabilities and robustness, algorithmic bias mitigation, and cross-border AI risk assessment. Advancing research in these domains can generate critical public knowledge to help the global community understand, manage, and reduce risks linked to advanced AI. A coordinated network of AI Safety Institutes could play a leading role in driving this effort. Such collaboration supports the broader goal of ensuring that AI development remains safe and beneficial to humanity.

**Promoting AI Safety in Education:** To ensure the safe and responsible use of AI, the general public and workforce need a stronger understanding of its capabilities, risks, and ethical implications. This calls for a holistic approach to AI safety in education. AI literacy should be integrated into school curricula - not only within computer science classes but also through broader digital citizenship education. Teacher training must be expanded to include AI safety and ethics, supported by potential accreditation schemes. Evaluation mechanisms are needed to monitor bias and efficacy in AI-related educational content. Education systems should also broaden reskilling programmes for workers likely to be impacted by AI-driven job displacement, helping to prevent social harm. Safety and ethics training should be embedded in both initial teacher preparation and ongoing professional development. UNESCO Recommendations on AI in Education and the IEEE 3527.1™ Standard for Digital Intelligence (DQ) offer valuable guidance for countries seeking to align their educational strategies with International benchmarks.

## Technical Interoperability Recommendations

**Promoting Interoperability by Design:** Countries are encouraged to align their national standards with established and emerging international benchmarks, such as the ISO 42001 AI Management System Standard and the IEEE 3527.1™ Standard for Digital Intelligence (DQ). Regulators can support this effort by developing sector-specific checklists or AI audit requirements aligned with international frameworks or relevant UN guidance. Such alignment enables AI industries to build systems that are compliant by design, facilitating smoother integration into global marketplaces and supply chains. It also helps ensure that AI technologies are developed and deployed safely, ethically, and interoperable across borders, supporting a more cohesive and trustworthy international AI ecosystem.

**Call for International Consensus-Driven AI Standards and Avoiding Duplication:** To ensure safe, ethical, and interoperable AI development, countries should support the creation of international consensus-driven standards and avoid unnecessary duplication of standardization efforts. The Global Digital Compact (GDC) calls on standards-developing organizations "to collaborate to promote the development and adoption of interoperable AI standards that uphold safety, reliability, sustainability, and human rights." Such collaboration is essential to promote interoperability and assist policymakers in effectively using standards. Initiatives like the International Electrotechnical Commission (IEC), International Organization for Standardization (ISO), and International Telecommunication Union (ITU) have partnered through the World Standards Cooperation (WSC) to map AI and machine learning standardization activities, helping to coordinate efforts and reduce redundancy. Similarly, the UK's AI Standards Hub works to improve understanding of the AI standards landscape and foster international coordination. These collaborative efforts can help establish urgently needed common ground on protocols such as Vehicle-to-Everything (V2X) communication, which is critical for the safe deployment of autonomous vehicles.

**Prioritizing the Development of Dedicated AI-in-Education Standards:** Countries should prioritize the development of dedicated standards for AI in education. This could be achieved by launching *Technical Specifications for AI Educational Tools*, covering key areas such as content quality, algorithmic fairness, and data security. These specifications should reference global benchmarks like the IEEE 3527.1™ Standard for Digital Intelligence (DQ) to ensure consistency and quality across educational systems.

**Prioritizing Standards Interoperability at the Security Layer:** Interoperability efforts should focus more on the security layer than the technical layer. The security layer protects AI systems and data from harm - including accidents, misuse, and cyber threats - and requires common safety and security standards that can be applied across sectors and borders. Technical interoperability - covering data formats, model compatibility, and interfaces - has relatively lower demand, as advanced AI systems often operate independently. However, security-related standards such as ISO/IEC 27001 (information security) and ISO/IEC TR 5469 (functional safety of AI systems) should be further incentivized to mitigate physical and systemic risks.

**Call for Scenario Planning for AI-Related Catastrophic Risks and Improved Regulatory Forecasting:** In addition to focusing on AI deployment safety and assurance through tools like the AI Verify Toolkit, AI Assurance Sandbox, and mechanisms for testing and certification, governments and regulators must expand their attention to long-term and high-impact risks. While immediate societal concerns such as fairness and data privacy remain critical, scenario planning and regulatory forecasting for model-level controls are equally essential. This includes oversight of foundational training data, intrinsic safety features of general-purpose AI, and mitigation strategies for catastrophic risks such as loss of control or dual-use threats (e.g., bioweapons or advanced cyber offense). Most critically, soft ethical guidance for high-risk AI must be transformed into enforceable and auditable obligations. These should cover: Data governance; model evaluation; pre-deployment and post-deployment testing; incident reporting and corrective actions. This ensures that ethical commitments are translated into concrete engineering practices, operational processes, and verifiable evidence requirements, and laying the foundation for robust, forward-looking AI safety and security regulation.

| Category | Recommendation | Timeline | Priority |
|---|---|---|---|
| **Ethical Interoperability** | Promote the AI Ethical Self-Certification Report Mechanism | **2025–2027** | High |
| | Advance the Global AI Ethics framework with measurable KPIs (led by SPAI/GDAIG) | **2025–2027** | High |
| **Regulatory Interoperability** | Ensure Inclusive Multi-Stakeholder Engagement for AI Safety Governance | **Ongoing** | High |
| | Launch a Global Benchmark for AI Safety and Security | **2025–2026** | High |
| | Promote Adaptation and Expansion of Data Interoperability Mechanisms | **Ongoing** | High |
| | Introduce Additional Safeguards for Cross-Border Data Flows (due diligence, liability insurance) | **Ongoing** | High |
| | Develop Greater Standardization of Liability Models for Autonomous Vehicles (AVs) | **2025–2027** | High |
| | Invest in AI Safety, Frontier Risk Management, and Alignment Research Collaborations | **Ongoing** | High |
| | Promote AI Safety and Literacy in Education | **2025–2027** | High |
| | Establish a National AI Coordination Entity (e.g., National AI Safety Council) | **2025–2026** | High |
| | Enhance Transparency and Accountability through National AI Safety KPIs and Reports | **2025–2027** | High |
| **Technical Interoperability** | Support International Consensus-Driven AI Standards and Avoid Duplication | **2025–2027** | High |
| | Develop Dedicated AI-in-Education Standards | **2025–2027** | High |
| | Scenario Planning for AI-Related Catastrophic Risks and Regulatory Forecasting | **Ongoing** | High |

**Table 9:** Implementation Timeline of  Selective Recommendations

High Priority: Core global alignment + AI safety foundation

## Conclusion

The future of AI safety governance is evolving towards an evidence-based, outcomes-oriented model that complements principle-led frameworks. This shift reflects a growing international consensus and increasing alignment with emerging global standards. To sustain momentum, policymakers must prioritize deepening normative specificity, expanding the adoption of interoperable standards, strengthening data governance mechanisms, and collaboratively investing in and conducting research into technical, institutional, and AI literacy capacity building. Achieving AI safety and interoperability is a dynamic and iterative process. In this process, public trust defines the foundation, interoperable ethics guide strategic direction, regulations translate values into enforceable rules, standards and enforcement drive implementation, and international cooperation amplifies impact. Together, these elements form a resilient, inclusive, and globally aligned AI safety governance ecosystem.

## Glossary of Terms

COOPERATION
The process through which regulators or agencies from different legal regimes address disparities and establish clear mandates. Cooperation may involve collective regulatory rules or coordination in designing, implementing, and enforcing regulatory measures.

ETHICAL INTEROPERABILITY
The capacity of institutions, systems, or actors to cooperate across diverse moral frameworks. It focuses on when different AI systems should be integrated ethically. There is a notable consistency in core values (such as justice, dignity, and privacy) and universal principles guiding ethics (such as safety, fairness, transparency, and accountability) whose relative importance varies across regions. AI ethics discussions in various regions are often shaped by different narratives about how these values are challenged, how they can be protected, and why they matter. The improvement of ethical interoperability depends on a meaningful, effective, and concrete shared moral vision, practical strategies for implementation, compatible methodologies, and institutions capable of balancing conflicting principles when trade-offs are necessary.

HARMONISATION
The process of unifying law, often building on a prior approach of standardisation. Harmonisation can be applied to varying extents.

INTEROPERABILITY
The ability of different systems, tools, and components to work together seamlessly both technically and normatively. It can include substantive measures such as international norms, shared protocols, interfaces, and data models, enabling communication, data exchange, standardizations etc. (PNAI 2023& 2024 ; Zeng, 2019 ; Berg, 2024 ; Onikepe, 2024 ).

LEGAL INTEROPERABILITY
The ability and process of enabling different regulatory frameworks to cooperate and communicate across jurisdictions within a single state or between two or more states. It reduces regulatory friction, advances common policy goals, and balances global integration with domestic regulatory autonomy. Legal interoperability exists in a spectrum between harmonisation or fragmentation, where too much harmonisation may limit national flexibility, while too little may impede smooth economic or social interaction. An intermediate level of regulatory interoperability is often seen as good policy, since regulatory competition can be beneficial and productive as long as the best normative order prevails. Legal interoperability should align with substantive principles, with new or adjusted or reinterpreted laws tailored to the particular circumstances. When neither fundamental principles nor market forces promote interoperability, regulatory actions, such as laws or regulations specifically targeting interoperability, are requested.

LEGAL STANDARDISATION
A formal understanding approved by a recognized body that enables the consistent and repeated application of rules or guidelines. It provides a concrete and normatively relevant benchmark for the behavior of the concerned community.

## Glossary of Terms

MUTUAL RECOGNITION
A principle that assesses whether regulatory measures between countries are comparable or equivalent. It reflects an agreement in which one country may relinquish a degree of regulatory independence by accepting that another nation's regulations are sufficient or satisfactory. Mutual recognition acknowledges that different national standards can be considered interchangeable for domestic application.

RECIPROCITY
A traditional legal principle aimed at establishing equilibrium between two or more countries in specific legal domains. It typically involves negotiating a balanced exchange of concessions in cross-border agreements, ensuring that both parties benefit from a fair and equitable arrangement.

STANDARD
"A document, established by consensus and approved by a recognised body, that provides, for common and repeated use, rules, guidelines, or characteristics for activities or their results, aimed at the achievement of the optimum degree of order in a given context" (ISO, 2025) .

STANDARDISATION
A regulatory approach grounded in widely accepted principles, practices, or guidelines within a specific field. Standardisation serves as a foundational step toward eventual harmonisation, facilitating alignment across jurisdictions and sectors.

TECHNICAL INTEROPERABILITY
"The ability of two or more systems or components to exchange information and to use the information that has been exchanged", focusing on ensuring systems can communicate and work together. It implies the ability to exchange information in some form, and the ability to make use of it.

- Vertical interoperability refers to interoperability between applications or systems on different stack levels of the stack.

- Horizontal interoperability refers to interactions between applications or systems in the same stack layer of the stack that provide similar or complementary functionality. It is foundational for systems interacting with each other,especially across networked technologies like the internet.

**Table 3:** Comparison of AI safety governance of four jurisdictions China, South Korea, Singapore and UK (Transborder Data Flow)

| Jurisdiction | Level of Integration | **Components of framework:** 1) objectives; 2) ethics; 3) binding; 4) targeted regulations/frameworks; 5) technical standards; 6)regulators; 7)risks/challenges |
|---|---|---|
| **China** | China has established a structured framework for integrating its cross-border data flow governance with international norms, and is striving to achieve alignment with international standards.<br><br>• **Legal and Principle Alignment:** Its core regulatory framework (anchored in the Data Security Law, Personal Information Protection Law (PIPL), and Provisions on Promoting and Regulating Cross-Border Data Flows (2024)) aligns with OECD Data Governance Principles, emphasizing "data free flow with trust" and balancing security with legitimate cross-border cooperation. It adopts global technical standards such as TLS 1.3 for encrypted transmission and ISO/IEC 27701 for privacy management, while GB/T 43697-2024 (Data Classification and Grading Rules) aligns with international data risk assessment methodologies.<br><br>• **International Engagement and Mutual Recognition:** China participates in multilateral forums including the OECD Working Party on Data Governance and GPAI's Data Governance Working Group, and acts as an observer in the Global CBPR Forum to promote standard mutual recognition. China has established a comprehensive free trade relationship with Singapore and conducts overall planning and coordination of cross-border data rules. Via the "Digital Silk Road," it also provides technical assistance to developing countries to build data security frameworks, expanding global trust networks. | **1. Objectives**<br><br>• Establishing a "data sovereignty with trust" system to balance national security, individual privacy, and legitimate cross-border data needs.<br><br>• Promoting international cooperation via bilateral/multilateral agreements and participation in global forums.<br><br>• Supporting low-risk cross-border data activities through exemption mechanisms (per 2024 Provisions on Promoting and Regulating Cross-Border Data Flows).<br><br>• Advancing privacy-enhancing technologies such as federated learning and edge computing to reduce cross-border raw data transmission.<br><br>**2. Binding Nature: Legally binding requirements**<br><br>• Mandatory domestic storage of personal information and important data for Critical Information Infrastructure Operators (CIIOs) (Cybersecurity Law, 2017).<br><br>• Compulsory data export safety assessments for transfers of "important data" or personal information of 100,000+ individuals (Data Security Law, 2021; Measures for Data Export Safety Assessment, 2022).<br><br>• Mandatory filing of standard contracts for cross-border personal information transfers involving <100,000 individuals (Measures for the Administration of Personal Information Export Standard Contracts, 2023).<br><br>• Annual compliance audits for enterprises engaged in cross-border data transfers.<br><br>• Non-binding flexibility: Exemptions for low-risk activities per Provisions on Promoting and Regulating Cross-Border Data Flows (2024).<br><br>**3. Principles/Values**<br><br>• "Data sovereignty with trust": Ensuring cross-border data flows uphold national security and individual rights.<br><br>• "Fair and equitable governance": Discouraging excessive data localization requirements that hinder legitimate cross-border cooperation.<br><br>• Data security and privacy protection: Embedding "data minimization," "anonymization," and "pre-transfer risk assessment" into cross-border data rules.<br><br>• Sustainability: Promoting green data flows.<br><br>**4. Technical Standards**<br><br>• GB/T 35273-2020 Information Security Technology—Personal Information Security Specification: Provides technical criteria for identifying sensitive data and assessing cross-border risks.<br><br>• GB/T 43697-2024 Data Security Technology—Data Classification and Grading Rules: Guides technical classification of data to determine cross-border risk levels.<br><br>• Personal Information Protection Certification Implementation Rules (2022): Sets technical requirements for certification; recognized in over 10 countries via bilateral agreements.<br><br>• TC260 Artificial Intelligence Safety Governance Framework (2024): Includes guidelines for auditing AI models trained on cross-border data to ensure compliance with data origin laws.<br><br>• TLS 1.3 encryption: Mandated for secure cross-border data transmission.<br><br>• ISO/IEC 27701 (Privacy Management): Adopted to align with global privacy technical standards. |
| **China** | | **5. Targeted Legislation/Framework**<br>Foundational laws:<br>• Cybersecurity Law (2017): Mandates domestic storage of CIIOs' personal information and important data.<br>• Data Security Law (2021): Establishes data classification/categorization and requires assessments for cross-border "important data" transfers.<br>• Personal Information Protection Law (PIPL) (2021): Defines three cross-border personal information transfer pathways.<br>Sector-specific regulations:<br>• Measures for Data Export Safety Assessment (2022): Details assessment procedures for high-risk cross-border data transfers.<br>• Measures for the Administration of Personal Information Export Standard Contracts (2023): Provides template contracts for low-to-medium-risk personal information transfers.<br>• Provisions on Promoting and Regulating Cross-Border Data Flows (2024): Expands exemptions and clarifies "unidentified important data" rules.<br>Certification rules: Personal Information Protection Certification Implementation Rules (2022).<br>**6. Regulator**<br>Primary regulator: Cyberspace Administration of China (CAC), which leads cross-border data flow supervision, platform operation, and audit oversight.<br>Supporting bodies:<br>• Local IT associations: Provide subsidized consulting services for SMEs.<br>• National Information Security Standardization Technical Committee (TC260): Develops technical standards for cross-border data security.<br>**7. Main Challenges/Risk**<br>• AI model risk: Training AI models on biased or non-compliant cross-border data, leading to model unfairness or legal liability.<br>• Global definition inconsistency: Divergent international definitions of "important data" hinder cross-border alignment. |
| **South Korea** | • Dual-track engagement working with both regulatory-driven and industry-driven models to cover more ground<br>• Robust foundation for AI interoperability, especially in personal data transfer regulation; Strong alignment with global data standards (GDPR adequacy, new Global CBPR member, etc.); Integrated legal framework with emerging international standards on data protection;<br>• G7's "Data free flow with trust" – cross-border innovation with preserving rigorous privacy and safety standards<br>• Seoul Declaration for Safe, Innovative and Inclusive AI, a high-level pledge to cooperate on interoperable AI governance standards across borders.<br>• Exploring mutual recognition of AI audits and certifications with partner countries<br>• Mirroring GDPR's "continuity of protection" principle;<br>• Data Protection Impact Assessment (DPIA) aligned with GDPR methodology -interoperability of risk management approaches<br>• Contributing to ISO/IEC and ITU working groups on AI and data standards, OECD on AI system risk classification , ISO/IEC JTC 1/SC 42 on AI | **1. Objectives**<br>• Stablish AI safety and security-by-design: Safe AI deployment with robust data security – and vice versa – emphasize a holistic approach.<br>**2. Binding Nature**<br>Legally binding requirements:<br>• AI Framework Act (2024): establishes obligations for "high-impact AI" systems (those affecting safety or basic rights) in critical sectors, requiring risk assessments, impact evaluations, and transparency measures.<br>• Preventing personal data misuse abroad: protections "travel with the data" (through legal and technical bindings)<br>Non-binding flexibility:<br>• Preventing personal data misuse abroad: oversight "follows the data" (through cooperation and representative arrangements).<br>**3. Principles/Values**<br>• Structured, principle-based controls: prior assessment, documented safeguards, downstream restrictions, and individual empowerment. From pre-transfer risk assessments (to prevent unsafe or unethical data uses) to post-transfer monitoring and enforcement (to correct any harms). Active enforcement by PIPC (fines, model deletion orders).<br>**4. Technical Standards**<br>• MyData API for secure data portability with privacy by design, all major data controllers must implement standardized APIs.<br>• Technical standardization of data formats and semantics.<br>• "Secure data corridors." – Investing in secure international network links and clouds.<br>• Mandated pre-transfer risk assessment mandated with standardized risk assessment templates and software.<br>• AI Framework Act anticipates adopting international technical standards: it includes provisions that Korean AI assessment criteria should reference globally recognized standards. Korea's National AI Standards Council has already adopted dozens of ISO/IEC AI standards as KS (Korean Standards).<br>**5. Targeted Legislation/Framework**<br>Foundational laws:<br>• Personal Information Protection Act (PIPA): A "right to data portability"; A right to explanation of algorithmic decisions.<br>Sector-specific regulations:<br>• AI Framework Act (2024).<br>Data-specific rules:<br>• PIPC and Korea Internet & Security Agency (KISA) launched domestic CBPR certification system.<br>• In 2025, PIPC announced plans to adopt its own "whitelist" of countries with equivalent protection, starting with the EU and also evaluating others like the UK, U.S., and Japan for potential adequacy determinations. |
| **South Korea** | | **6. Regulator**<br>Primary regulator:<br>• PIPC (Active enforcement by PIPC: fines, model deletion orders. PIPC scaled up significantly, regulatory cooperation with counterparts in other countries to share information and assist in investigations; PIPC conducts periodic audits of companies' handling of cross-border data.)<br>Supporting bodies:<br>• Korea Internet & Security Agency (KISA) provides technical support and certification functions.<br>• Ministry of Science.<br>**7. Main Challenges/Risk**<br>• Gaps remain between jurisdictions that creates compliance challenges – for example, differences in what counts as "sensitive data"; the age of consent for children's data; or acceptable purposes for data processing.<br>• Growing data localization measures in some countries create compliance burdens for international businesses.<br>• Generative AI misuse continues to spread across borders, highlighting the need for international cooperation on detection technologies and common standards. |
| **Singapore** | • Becoming trusted and capable international partner in AI innovation & governance;<br>• Leveraging PDPA with international voluntary frameworks to ensure regulatory interoperability and facilitate commerce. | **1. Objectives**<br>• Highly dependent on international trade, free flow of data as strategic imperative.<br>• Promotion of data free flow to maximize the benefits.<br>**2. Binding Nature**<br>Legally binding requirements:<br>• Personal Data Protection Act (PDPA).<br>• ASEAN MCCs<br>• Agreement with the EU on digital trade.<br>Non-binding flexibility:<br>• Voluntary, potential future shift toward more specific mandatory laws.<br>**3. Principles/Values**<br>• Protecting consumer privacy and align AI systems with local languages, laws, and societal values; balancing free data flows with individual privacy and national interests.<br>**4. Technical Standards**<br>• Privacy Enhancing Technologies (PETs) to manage data privacy and security.<br>• "Crosswalks" maps.<br>• AI Verify framework maps with international standards like NIST AI Risk Management Framework and ISO/IEC 42001 to enhance interoperability and reduce compliance costs for businesses.<br>**5. Targeted Legislation/Framework**<br>Foundational laws:<br>• Personal Data Protection Act (PDPA).<br>Sector-specific regulations:<br>• National AI Strategy 2.0 (NAIS 2.0). |
| **Singapore** | | • Model AI Governance Framework.<br>Data-specific rules:<br>• AI Verify – validate AI systems against ethical principles.<br>• ASEAN MCCs<br>• Global CBPR System<br>• Agreement with the EU on digital trade<br>**6. Regulator**<br>Primary regulator:<br>• Personal Data Protection Commission (PDPC): Enforces data protection laws and provides specific guidance for the use of personal data in AI systems.<br>• The Infocomm Media Development Authority (IMDA): Provides voluntary, principles-based guidance and a technical testing framework for organizations.<br>Supporting bodies:<br>• Advisory Council on the Ethical Use of AI and Data.<br>**7. Main Challenges/Risk**<br>• Regulatory fragmentation negatively impact trade.<br>• The use of foreign technologies undermines the development of context-specific and culturally appropriate AI applications without targeted investment in local data collection and annotation. |
| **UK** | • Reduce business compliance burdens and embed the UK's values into emerging global AI governance strategies; encourage mutual recognition and regulatory cooperation<br>• UK Adequacy Decisions: recognized the EU/EEA, Japan, Canada, and others as adequate partners' UK–US Data Bridge,<br>• Incorporate data flow provisions in trade deals aligns with the G7's DFFT concept, removing barriers while building trust through shared principles. Participates in data flow discussions at the G20.<br>• Joining the Global CBPR Forum to develop an international certification system, making data exchange easier among member countries. Participating the Global Privacy Assembly. The UK co-drafted the OECD AI Principles and the Declaration on Government Access to Data, establishing common ground with many other countries<br>• Bilateral Cooperation: conducts data policy dialogues with key partners to share best practices and align regulatory approaches for AI governance. Collaborating with groups like the Global Partnership on AI (GPAI) to promote high-quality, representative data sharing<br>• Capacity Building: The UK supports developing countries in creating their own data protection laws. | **1. Objectives**<br>• Data protection laws and trade policies aim to enable free data movement for innovation while upholding high standards and public trust.<br>• Promoting international cooperation, through the participation in global forums, partnerships and councils.<br>• DFFT: digital governance, emphasising that data and AI innovations should be able to move freely across borders<br>• UK–US Data Bridge: facilitates data transfer to certified US companies that meet the UK's strict privacy and rights standards.<br>**2. Binding Nature**<br>Legally binding requirements:<br>• UK GDPR and Data Protection Act 2018: adequate level of data protection/ safeguards like SCCs.<br>• International Data Transfer Agreement (IDTA): templates and guidance for a risk-based approach.<br>Non-binding flexibility:<br>• The Data Protection and Digital Information Bill (2023–2024) aims to facilitate flexible international transfers of data.<br>**3. Principles/Values**<br>• Lawful, secure, and straightforward.<br>• Fundamental rights travel with the data, reflecting a belief in universal privacy and dignity.<br>• Ethical Data Use and Human Rights: initiatives like the OECD Declaration on Government Access to Personal Data ensure law enforcement access is necessary and proportionate.<br>• Transparency and accountability in AI supply chains: firms must ensure international partners adhere to safety protocols. |
| **UK** | | **4. Technical Standards**<br>• Security standards: TLS encryption, ISO 27001 for secure data transfer.<br>• Digital trade standards: ISO/IEC 27701 (privacy management) and BS 10012 (UK standard for data management).<br>• Global CBPR<br>**5. Targeted Legislation/Framework**<br>Foundational laws:<br>• UK GDPR; Data Protection Act 2018.<br>Sector-specific regulations:<br>• Guidance from National Cyber Security Centre (NCSC) on encryption and data protection.<br>Data-specific rules:<br>• International Data Transfer Agreement (IDTA): offers templates and guidance for a risk-based approach.<br>• Data Protection and Digital Information Bill (2023–2024): facilitates flexible international transfers of data.<br>10. Standard Contractual Clauses (SCCs)<br>**6. Regulator**<br>Primary regulator:<br>• Department for Science, Innovation & Technology (DSIT)<br>• Information Commissioner's Office (ICO).<br>Supporting bodies:<br>• AI Security Institute.<br>• Alan Turing Institute (ATI).<br>• National Cyber Security Centre (NCSC).<br>**7. Main Challenges/Risk**<br>• Data quality and bias.<br>• Security risks and cybersecurity: encryption, integrity.<br>• Concentration risk: the UK is diversifying its data partners and investing in local capabilities like a sovereign cloud.<br>• Geopolitical risks: concern in data flows.<br>• Ethical Data Use and Human Rights: lead initiatives like the OECD Declaration on Government Access to Personal Data to ensure that law enforcement access is necessary and proportionate.<br>• Transparency and accountability in AI supply chains: UK firms are increasingly required to ensure their international partners adhere to safety protocols through contracts and audits. |

**Table 4:** Comparison of AI safety governance of four jurisdictions China, South Korea, Singapore and UK (education)

| Jurisdiction | Level of Integration | **Components of framework:**<br>1) objectives; 2) ethics; 3) binding; 4) targeted regulations/frameworks; 5) technical standards; 6)regulators; 7)risks/challenges |
|---|---|---|
| **China** | • China's AI-in-education governance integrates with international ethics and standards, and is striving to achieve dedicated standardization and platform compatibility.<br><br>• Ethical and Principle Alignment: It aligns domestic policies with UNESCO's Recommendation on the Ethics of Artificial Intelligence, prioritizing student well-being, educational equity, and the preservation of teachers' roles—consistent with global norms. The Minors Internet Protection Regulations (2024) and PIPL's minor protection provisions align with international child data privacy standards, requiring guardian consent for student data collection and limiting data processing to "minimum necessity."<br><br>• Standardization and Collaboration: China contributes to ISO/IEC JTC1 SC36 (Learning Technologies) to develop international standards for AI educational tools, and references global benchmarks such as WCAG 2.1 (accessibility standards) in technical requirements (e.g., text-to-speech, screen reader compatibility for students with disabilities). The National Smart Education Platform uses standardized APIs to ensure domestic interoperability, and Sino-foreign university partnerships conduct joint research on mitigating algorithmic bias and protecting minor data—aligning with global AI-in-education research agendas. | **1. Objectives**<br><br>• Advancing the "Double Reduction Policy" (2021) to reduce students' homework burden and off-campus tutoring pressure, promoting balanced development of in-school education.<br><br>• Pushing forward "Educational Digital Transformation" (per the 14th Five-Year Plan for Educational Development) by building national smart education platforms to bridge urban-rural educational resource gaps.<br><br>• Strengthening vocational education development (via the 2022 revised Vocational Education Law) to cultivate technical talents, including establishing vocational education groups and promoting integration of industry and education.<br><br>• Promoting "Basic Education Quality Improvement Project" to optimize curriculum systems, enhance teacher professional capabilities, and ensure equitable access to quality basic education resources.<br><br>**2 Binding Nature**<br><br>Legally binding requirements:<br><br>• Compulsory nine-year education for children aged 6-15 (per Compulsory Education Law, 2021), with local governments required to ensure 100% enrolment rate.<br><br>• Mandatory safety management for schools per Measures for the Administration of School Safety, 2022; non-compliant schools face closure or fines.<br><br>• Obligation for educational institutions to protect student personal information per Regulations on the Protection of Minors' Personal Information in Schools, 2023.<br><br>• Mandatory reporting of off-campus tutoring violations by local education authorities (required by "Double Reduction" Policy implementation rules).<br><br>Non-binding flexibility:<br><br>• Autonomy for universities to design 30% of undergraduate courses based on disciplinary characteristics (per Higher Education Law, 2021).<br><br>• Flexible implementation of smart education tools in rural schools per Guidelines for Smart Education in Rural Areas, 2024.<br><br>**3. Principles/Values**<br><br>• Education equity: Narrowing gaps in educational resources between regions, urban/rural areas, and schools.<br><br>• Moral education first ("Lide Shuren"): Embedding ideological and ethical education in all subjects to cultivate students' social responsibility.<br><br>• Innovation-driven development: Promoting digital transformation of education and encouraging students' critical thinking and creativity.<br><br>• Student-cantered approach: Designing curricula and teaching methods based on students' physical and mental development needs.<br><br>• Cultural confidence and international vision: Integrating traditional Chinese culture into courses while promoting international exchanges to broaden students' horizons.<br><br>**4. Technical Standards** |
| **China** | | • Technical Standards for Online Education Platforms (2023): Requires platforms to meet data encryption and user authentication to protect student privacy.<br>• AI Application Ethics Guidelines in Primary and Secondary Education (2025): Sets technical limits for AI teaching tools to prevent ethical risks.<br>• GB/T 22239-2019 Information Security Technology—Security Level Protection of Information Systems (Level 2): Mandated for all educational institutions' information systems to prevent cyberattacks.<br>**5. Targeted Legislation/Framework**<br>Foundational laws:<br>• Education Law (2021): Defines national education goals, school establishment standards, and government education responsibilities.<br>• Compulsory Education Law (2021): Regulates compulsory nine-year education, including enrollment policies, teacher qualifications, and funding guarantees.<br>• Vocational Education Law (2022): Promotes industry-academia integration, enterprise participation in vocational education, and vocational skill certification.<br>Sector-specific regulations:<br>• "Double Reduction" Policy Implementation Rules (2021): Details measures to reduce homework and off-campus tutoring.<br>• National Medium- and Long-Term Education Reform and Development Plan (2021-2035): Outlines key goals.<br>• Regulations on the Protection of Minors' Personal Information in Schools (2023): Specifies rules for collecting, storing, and using student personal information.<br>• Guidelines for the Construction and Application of National Smart Education Platform (2022): Guides the operation and resource update of the national smart education platform.<br>Data-specific rules:<br>• Education Data Security Management Measures (2024): Regulates the collection, transmission, and sharing of education data.<br>**6. Regulator**<br>Primary regulator:<br>Ministry of Education, with key departments including:<br>• Basic Education Department: Oversees K-12 education.<br>• Vocational Education and Adult Education Department: Manages vocational education.<br>• Department of Educational Informatization: Leads smart education development.<br>Supporting bodies:<br>• Local Education Bureaus: Implement national policies at the provincial/municipal/county levels.<br>• National Technical Committee for Education Informatization Standardization (SAC/TC489): Develops technical standards for education.<br>• Cyberspace Administration of China: Co-supervises education data security.<br>• State Administration for Market Regulation: Regulates off-campus tutoring institutions' pricing and compliance. |
| **China** | | **7. Main Challenges/Risk**<br>• Risks of Education imbalance : Both disparities in the distribution of AI educational resources and technological barriers may exacerbate the educational divide between different groups, rather than narrowing it.<br>• Risk of inhibiting students' autonomy and creativity: AI's efficient assistance in repetitive tasks may lead students to over-rely on technology, thereby weakening the development of their core learning abilities. |
| **South Korea** | • Aiming to align with and ensure interoperability with international data-protection norms (e.g., EU GDPR adequacy applies to KR) ; Ensuring any third-party AI services used in schools should comply with PIPA and child-protection requirements.<br>• UNESCO (2023) guidance on generative AI in education can serve as a non-binding international reference<br>• laying groundwork for technical interoperability and standards, exploring international standards (e.g., IEEE/ISO) for AI in education; referencing emerging IEEE/ISO AI-in-education standards; securing approval of the ISO/IEC 5259-1:2024 data-quality standard for AI; KATS is deepening cross-border standards coordination via the 2024 U.S.–Korea Standards Forum; public bodies are piloting or aligning with ISO/IEC 42001 AI management systems | **1. Objectives**<br>• Introduce AI-enabled digital textbooks incorporate generative AI to personalize learning content and feedback for each student.<br>• Incorporation of generative AI to personalize learning content and feedback for each student, and they include features like real-time captions and translations to improve accessibility for students with different needs.<br>**2. Binding Nature**<br>Legally binding requirements:<br>• Framework Act on Education, art. 2, art. 12: AI that Supports Human Growth.<br>• Framework Act on Education, arts. 9, 12, 13, 14: Draw out the potential for<br>• human growth.<br>Non-binding flexibility:<br>• Ministry of Education Implementation Plan, 2024.<br>**3. Principles/Values**<br>• Protecting children's rights, ensuring equitable access to high-quality education, preventing algorithmic bias from disadvantaging specific groups of learners.<br>• Safety in teaching–learning processes.<br>• Transparency and explainability in data processing.<br>• Use data for legitimate purposes, privacy, data ownership, and the possibility of commercial misuse.<br>**4. Technical Standards**<br>• KERIS is establishing technical standards, publishing development guidelines, and operating an accreditation/review system for AI digital textbooks.<br>**5. Targeted Legislation/Framework**<br>Foundational laws:<br>• Personal Information Protection Act (PIPA).<br>Sector-specific regulations:<br>• Korea applies PIPA-based requirements including lawful basis, purpose limitation, data minimization, (where necessary) pseudonymization, strict access control, and child-protection rules to collectively oversee data protection compliance and bias/fairness reviews for the algorithms.<br>Data-specific rules:<br>• Guidelines to protect student privacy and well-being in digital learning: student data used by AI systems must be minimised and, where necessary, pseudonymised, and that access be limited to authorized educational personnel – in accordance with the national privacy law (PIPA).<br>• Child protection rules mandate that AI education tools avoid harmful content or bias.<br>• Technical standards for efficacy (ensuring the AI's recommendations are pedagogically sound and unbiased) are still in development. |
| | | **6. Regulator**<br>Primary regulator:<br>• Ministry of Education<br>Supporting bodies:<br>• KERIS.<br>**7. Main Challenges/Risk**<br>• Risk exposing students to biased, inaccurate, or potentially harmful content generated by AI, as well as over-reliance on automated feedback that could undermine pedagogical integrity.<br>• Child data protection.<br>• Algorithmic transparency and accountability.<br>• Sociotechnical challenge – stakeholder buy-in and change management.<br>• Achieving true interoperability (integration of AI tools into schooling) requires aligning the technology with the capacities and comfort levels of educators and students.<br>• Clear standards for accountability (who is liable if the AI makes a harmful error or if data is leaked).<br>• Striking the right balance between AI assistance and human instruction. |
| **Singapore** | • MOE's EdTech are benchmarked against International practices and are guided by ethical AI use. | **1. Objectives**<br>• "Customisation & personalisation" and emphasizing cyber wellness and ethical AI use.<br>• The Student Learning Space (SLS), a unified platform for all students and teachers to enhance personalized learning for students and improve the operational efficiency of teachers through automation.<br>**2. Binding Nature**<br>Legally binding requirements:<br>• Personal Data Protection Act (PDPA).<br>Non-binding flexibility:<br>• Voluntary, potential future shift toward more specific, mandatory laws.<br>**3. Principles/Values**<br>• Student well-being, unbiased and deliver equitable outcomes.<br>• Data privacy is a paramount concern, given the collection of sensitive student data for personalized learning.<br>**4. Technical Standards**<br>• EdTech Masterplan 2030 outlines the use of AI-enabled tools within the Student Learning Space (SLS).<br>**5. Targeted Legislation/Framework**<br>Foundational laws:<br>• Personal Data Protection Act (PDPA).<br>Sector-specific regulations:<br>• EdTech Masterplan 2030.<br>Data-specific rules:<br>• National AI Strategy 2.0 (NAIS 2.0).<br>• Model AI Governance Framework.<br>• AI Verify – validate AI systems against ethical principles.<br>• Specific advisory guidelines from the Personal Data Protection Commission.<br>**6. Regulator**<br>Primary regulator:<br>• Ministry of Education: Integrates AI into education to enhance learning outcomes while providing ethical guidelines for use within the school system.<br>Supporting bodies:<br>• Advisory Council on the Ethical Use of AI and Data.<br>• Personal Data Protection Commission (PDPC): Enforces data protection laws and provides specific guidance for the use of personal data in AI systems.<br>• The Infocomm Media Development Authority (IMDA): Provides voluntary, principles-based guidance and a technical testing framework for organizations. |
| | | **7. Main Challenges/Risk**<br>• Protecting sensitive student data.<br>• Ensuring ethical and unbiased algorithmic outcomes.<br>• Addressing teacher readiness to adopt new technologies. |
| **UK** | • Schools refer to broader frameworks like the UNESCO Recommendations on AI in Education or the IEEE's ethical design guidance.<br>• A key member of UNESCO's Global Education Coalition on AI, aligning its domestic policies with international principles of fairness and inclusion.<br>• Supports digital and AI skills initiatives through the OECD and shares its projects via the OECD AI Policy Observatory.<br>• Carrying out cross-border research pedagogical innovations and AI progress. Learning from and shares in global best practices<br>• Data Interoperability Standards: UK EdTech often adopts standards like Learning Tools Interoperability (LTI);<br>• Cybersecurity and IT Standards: UK schools must follow standard IT security practices, often referencing ISO/IEC 27001. This ensures that AI tools do not create security vulnerabilities or expose sensitive student data. | **1. Objectives**<br>• Education Law & Safeguarding: Schools have a legal duty to protect pupils' well-being.<br>• Data Protection and Pupil Privacy: Student data is safeguarded by the UK GDPR and the Data Protection Act.<br>• Equality and Non-Discrimination: The Equality Act 2010 obliges schools to prevent discrimination.<br>• Emerging AI Policies: Between 2023 and 2025, the UK government began issuing AI-specific guidance.<br>• Ethical Guidelines: Besides government efforts, organisations such as the UK Children's Commissioner's Office have called for stricter ethical standards.<br>**2. Binding Nature**<br>Legally binding requirements:<br>• Children Act 2004 and Education Act 2002.<br>• UK GDPR and the Data Protection Act.<br>• The Equality Act 2010; The Equality Act 2010 obliges schools to prevent discrimination.<br>Non-binding flexibility:<br>• 'Generative AI product safety expectations".<br>• Web Content Accessibility Guidelines (WCAG).<br>**3. Principles/Values**<br>• Safety remains the top priority.<br>• Doing no harm.<br>• Fairness and justice.<br>• Privacy.<br>**4. Technical Standards**<br>• Data Interoperability Standards: adoption of standards that ensure different systems can communicate securely.<br>• Product Safety Standards: Outline requirements for developers to include content filtering, age-appropriate design, and user controls.<br>• Cybersecurity and IT Standards: UK schools must follow standard IT security practices, often referencing ISO/IEC 27001.<br>**5. Targeted Legislation/Framework**<br>Foundational laws:<br>• Children Act 2004 and Education Act 2002.<br>Sector-specific regulations:<br>• Department for Education's (DfE) "Keeping Children Safe in Education".<br>• "Generative AI product safety expectations".<br>Data-specific rules:<br>• Data Interoperability Standards: UK EdTech often adopts standards like Learning Tools Interoperability (LTI).<br>**6. Regulator**<br>Primary regulator:<br>• Department for Education's (DfE).<br>Supporting bodies:<br>• Information Commissioner's Office's (ICO).<br>**7. Main Challenges/Risk**<br>• Professional Responsibility: If an AI tool is used negligently, the school could be liable, highlighting the need for due diligence and appropriate supervision.<br>• Product Liability: The provider may be responsible if a defective AI product causes injury.<br>• Data Breaches or Misuse: If student data is mishandled, the school and the vendor could face regulatory action from the ICO.<br>• Insurance: Schools are assessing whether their existing liability insurance covers AI-related incidents. The prevailing view is that AI does not exempt educators from their duty of care. |

**Table 5:** Comparison of AI safety governance of four jurisdictions China, South Korea, Singapore and UK (autonomous driving)

| Jurisdiction | Level of Integration | Components of framework:<br>1) objectives; 2) ethics; 3) binding; 4) targeted regulations/frameworks; 5) technical standards; 6)regulators; 7)risks/challenges |
|---|---|---|
| **China** | • China's AV governance achieves strong alignment with international technical standards and engages in targeted bilateral cooperation, and is committed to ensuring the consistency between domestic standards and international standards.<br><br>• Technical Standard Alignment: Key domestic standards align with global benchmarks: GB/T 40429-2021 (AV Grading) adopts UNECE WP.29's automation classification; the GB/T 34590 series is equivalent to ISO 26262 (functional safety); and GB 44495-2024 (Vehicle Cybersecurity) references ISO/SAE 21434. It also aligns with UNECE WP.29 regulations on automated driving, ensuring compatibility with global AV terminology and safety validation frameworks.<br><br>• Bilateral/Multilateral Cooperation: China and Germany have signed the Joint Statement of Intent on Cooperation in the Field of Autonomous and Connected Driving, and will jointly develop vehicle-to-everything (V2X) technology. Domestically, AV pilot zones (e.g., Beijing) adopt practices (e.g., unified operation data platforms, "black boxes" for event logging) consistent with international AV safety monitoring norms. | **1. Objectives**<br><br>• Advancing the "14th Five-Year Plan for the Development of Intelligent Connected Vehicles" (2021-2025) to accelerate L3/L4 autonomous driving technology iteration and promote large-scale demonstration applications.<br><br>• Launching national-level AV pilot zones in over 20 cities to test road access for L4 AVs and explore "vehicle-road-cloud integration" (V2X) scenarios.<br><br>• Promoting cross-industry collaboration by establishing the National Intelligent Connected Vehicle Innovation Centre (2019) and supporting alliances between automakers , tech firms, and telecom operators for V2X infrastructure construction.<br><br>• Pushing for international cooperation, such as participating in the UN WP.29 (World Forum for Harmonization of Vehicle Regulations) to align AV safety standards and exploring bilateral pilot projects with Germany and Singapore for cross-border AV data sharing.<br><br>**2. Binding Nature**<br><br>Legally binding requirements:<br><br>• Mandatory road test qualification for AVs (per Measures for the Administration of Intelligent Connected Vehicle Road Testing and Demonstration Application, 2022); unqualified vehicles face seizure and enterprise fines.<br><br>• Compulsory local storage of AV "important data" per Interim Provisions on the Management of Autonomous Driving Data Security, 2023; cross-border transmission requires CAC approval.<br><br>• Mandatory configuration of safety officers in L2-L3 level AVs during public road testing (required by AV Safety Testing Guidelines, 2022); safety officers must hold professional certifications.<br><br>• Obligation for AV enterprises to report traffic accidents within 1 hour of occurrence (per Road Traffic Safety Law Amendment, 2021); failure to report leads to license revocation.<br><br>Non-binding flexibility:<br><br>• Autonomy for enterprises to set test routes within national pilot zones (per Pilot Zone Management Rules, 2023); no additional approval needed for routine tests.<br><br>• Flexible testing permissions for L4-level AVs in closed scenarios (per Special Scenario AV Testing Guidelines, 2024); enterprises can adjust test parameters based on scenario needs.<br><br>**3. Principles/Values**<br><br>• Safety first: Prioritizing human life safety in AV R&D and testing to prevent casualties from technical failures.<br><br>• Innovation with regulation: Encouraging technological innovation while imposing basic safety and data compliance requirements to avoid unregulated development.<br><br>• Data security and privacy protection: Safeguarding AV-related data via classification storage and access control (aligned with Data Security Law).<br><br>• Infrastructure synergy: Promoting coordinated construction of smart roads, 5G networks, and cloud platforms to enhance AV operational efficiency and reliability.<br><br>• Green and low-carbon: Integrating AVs with NEVs to reduce carbon emissions and support national "dual carbon" goals.<br><br>• International compatibility: Aligning AV technical standards and test rules with global practices to facilitate cross-border cooperation and market access. |
| **China** | | **4. Technical Standards**<br>• GB/T 40429-2021 Classification of Driving Automation for Motor Vehicles - Part 1: Definitions and Classifications. This standard classifies automated vehicles (AV) into Levels L0 to L5, serving to guide the development of R&D and testing standards.<br>• GB T 43267-2023 Safety of Expected Functions for Road Vehicles: aims to standardize the technical requirements for safety of expected functions of L1-L5 autonomous driving systems and emergency intervention functions.<br>• GB 44495-2024 Technical Requirements for Information Security of Complete Vehicles: specifies the requirements for information security management system of vehicles.<br>**5. Targeted Legislation/Framework**<br>Foundational laws:<br>• Road Traffic Safety Law (Amendment) (2021): Adds provisions for AV road testing, accident liability preliminary identification, and safety officer requirements.<br>• Data Security Law (2021): Regulates the collection, storage, and cross-border transmission of AV “important data”.<br>• Personal Information Protection Law (PIPL) (2021): Protects user privacy in AVs.<br>Sector-specific regulations:<br>• Measures for the Administration of Intelligent Connected Vehicle Road Testing and Demonstration Application (2022): Details AV test application procedures, safety requirements, and demonstration scenario scope.<br>• Interim Provisions on the Management of Autonomous Driving Data Security (2023): Specifies data classification, local storage, and cross-border approval rules for AV enterprises.<br>• “14th Five-Year” Plan for the Development of Intelligent Connected Vehicles (2022): Outlines goals.<br>• Guidelines for AV Public Transportation Demonstration Applications (2023): Regulates autonomous bus operation.<br>Local pilot rules:<br>• Beijing Intelligent Connected Vehicle Road Testing Management Measures (2023): Allows L4-level AVs to test without safety officers in designated areas.<br>• Shanghai Autonomous Vehicle Demonstration Application Measures (2024): Permits AVs to provide commercial ride-hailing services in Pudong New Area.<br>**6. Regulator**<br>Primary regulators (tripartite coordination mechanism):<br>• Ministry of Industry and Information Technology (MIIT): Oversees AV production access technical standards formulation, and pilot zone management.<br>• Ministry of Public Security (MPS): Manages AV road access, traffic order supervision, and accident investigation/liability preliminary identification. |
| **China** | | • Ministry of Transport (MOT): Regulates AV application in transportation scenarios and infrastructure coordination.<br>Supporting bodies:<br>• China Association of Automobile Manufacturers (CAAM): Develops industry self-regulatory standards and organizes technical exchanges.<br>• Cyberspace Administration of China (CAC): Co-supervises AV data security and user privacy protection.<br>• Local Pilot Zone Management Committees: Implements local test policies, coordinates road infrastructure construction, and monitors enterprise compliance.<br>**7. Main Challenges/Risk**<br>• Technical safety risk: AV functional failures and cybersecurity breaches.<br>• Liability Division Risk: The legal liability determination for traffic accidents involving autonomous vehicles has not yet been introduced, and there is a risk of ambiguity regarding liability attribution. |
| **South Korea** | • Partial alignment with international vehicle safety standards (UNECE), still in progress | **1. Objectives**<br>• Autonomous driving technologies develop with public safety.<br>**2. Binding Nature**<br>Legally binding requirements:<br>• Road Traffic Act: drivers undergo safety training.<br>• Autonomous Vehicle Management Act: a temporary permit regime and a legal basis for paid autonomous transport services.<br>• Guarantee of Automobile Accident Compensation Act (2020) assigns primary liability<br>• for accidents to the vehicle's owner, even if the vehicle is operating autonomously, while also allowing recourse against manufacturers or software providers if a defect in the vehicle or its algorithms contributed to the incident.<br>Non-binding flexibility:<br>• Core laws exist (AI Act forthcoming), but ethical guidelines are non-binding.<br>• The 2019 Act on the Promotion and Support for the Commercialization of Autonomous Vehicles & 2020 Autonomous Vehicle Management: temporary permit regime and a legal basis for paid autonomous transport services.<br>**3. Principles/Values**<br>• Prioritize human life, fairness in accident scenarios, and transparency in decision-making processes.<br>• "Moving data hubs," privacy, data ownership, and stronger safeguard of potential misuse of biometric or location data. |
| **South Korea** | | **4. Technical Standards**<br>• All autonomous vehicles must be equipped with event data recorders to log driving data.<br>• Manufacturers and service operators are required to obtain cybersecurity certification and conduct real-time monitoring of vehicle systems to guard against cyberattacks – a practice aligning with emerging international standards (such as ISO 21434 for automotive cybersecurity).<br>• Actively pursuing technical standardization and enabling infrastructure.<br>• Aligns its vehicle standards with international benchmarks; for instance, Europe's WP.29.<br>• Designated pilot testing zones provide a controlled environment for technical validation and standard convergence.<br>**5. Targeted Legislation/Framework**<br>Foundational laws:<br>• Guarantee of Automobile Accident Compensation Act (2020).<br>Sector-specific regulations:<br>• Road Traffic Act.<br>• Third Automotive Policy Master Plan (2022–2026).<br>• Act on the Promotion and Support for Commercialization of Autonomous Vehicles (2019)<br>• Regulatory sandboxes and K-City test sites.<br>Data-specific rules:<br>• Rules for cross-border data transfer and use of sensitive data (location, sensor recordings, biometric data) need clarification, especially if data is processed on cloud servers outside Korea.<br>• AI Basic Act: mandates that foreign AI service providers appoint a local representative in Korea.<br>**6. Regulator**<br>Primary regulator:<br>• Ministry of Land, Infrastructure and Transport (MOLIT)<br>• Korea Transportation Safety Authority (KOTSA)<br>Supporting bodies:<br>• AI Safety Research Institute<br>**7. Main Challenges/Risk**<br>• Safety risks: sensor failures, algorithmic misjudgements, or malicious cyberattacks highlight the importance of redundant safety mechanisms, secure vehicle-to-cloud communication, and real-time monitoring systems.<br>• Remaining challenges include refining these rules for higher levels of autonomy (Levels 4–5 where a human driver may not be present) and ensuring insurance products and legal frameworks adapt accordingly.<br>• AI risks and challenges: system and perception failures under edge or degraded conditions that can produce unsafe behaviour; adversarial and cyber-physical threats against connected vehicles and over-the-air updates; human–machine interaction hazards including automation complacency and unclear handover; complex liability and data-governance exposure as continuous location and biometric collection crosses jurisdictions. |
| **South Korea** | | • Responsibility tracing becomes harder when software updates or map data originate abroad, which calls for clearer cross-border legal mappings and evidence chains.<br><br>• Privacy protections must keep pace as vehicles become "moving data hubs" across jurisdictions.<br><br>• Clear interoperability of legal frameworks (domestic and foreign) is needed when software updates or map data come from international sources. |
| **Singapore** | • Participates in the APEC to broaden technical coordination and support harmonized standards and regulatory approaches<br><br>• UN regulation on Cybersecurity Management Systems, which refers to standards like ISO 26262 for Functional Safety and ISO/SAE 21434 for Cyber-security of Road Vehicles | **1. Objectives**<br>• Prioritizes a "human-over-the-loop" approach, ensuring human oversight of AI's unexpected or undesirable outputs; prioritizes verifiable physical measures to mitigate immediate, high-consequence risks.<br>• Promotion of international cooperation through participation in forums, such as the ones from the APEC.<br>**2. Binding Nature**<br>Legally binding requirements:<br>• Road Traffic (Amendment) Act 2017 and its subsidiary legislation, the Road Traffic (Autonomous Motor Vehicles) Rules 2017.<br>• The law requires a "Blackbox" data recorder to be installed in every AV to store crucial vehicle telematics, which is essential for accident investigations and for facilitating liability claims.<br>• Mandatory comprehensive insurance against third-party liability and property damage, or a security deposit of at least S$1.5 million with the LTA.<br>• Mandatory safety assessment process before Autonomous Vehicles' (AVs) public deployment, at the Centre of Excellence for Testing and Research of AVs (CETRAN).<br>Non-binding flexibility:<br>• No specific liability rules: accidents are addressed under the tort of negligence and existing common law principles.<br>**3. Principles/Values**<br>• Human-over-the-loop oversight.<br>• Verifiable physical measures to mitigate immediate, high-consequence risks.<br>**4. Technical Standards**<br>• Blackbox data recorder required in every AV.<br>• Safety assessment process conducted at CETRAN before deployment.<br>**5. Targeted Legislation/Framework**<br>Foundational laws:<br>• Road Traffic (Amendment) Act 2017.<br>Sector-specific regulations:<br>• Road Traffic (Autonomous Motor Vehicles) Rules 2017.<br>Data-specific rules:<br>• Requirement for blackbox data recorder in every AV.<br>**6. Regulator**<br>Primary regulator:<br>• Land Transport Authority (LTA): Regulates and certifies Autonomous Vehicles to ensure physical safety and mitigate public liability risks.<br>Supporting bodies:<br>• Advisory Council on the Ethical Use of AI and Data.<br>• Personal Data Protection Commission (PDPC): Enforces data protection laws and provides specific guidance for the use of personal data in AI systems.<br>• The Infocomm Media Development Authority (IMDA): Provides voluntary, principles-based guidance and a technical testing framework for organizations.<br>Supporting bodies:<br>• Centre of Excellence for Testing and Research of AVs (CETRAN).<br>**7. Main Challenges/Risk**<br>• Absence of a clear legal pathway for assigning liability in a fully autonomous scenario represents a policy gap. |
| **UK** | • Interoperability by design, enabling UK and foreign AVs to operate safely across different markets.<br><br>• harmonise AV regulations through the UNECE and bilateral agreements.<br><br>• based on international safety standards, such as UNECE regulations and ISO standards, by incorporating specific measures designed for autonomous functionality. Standardisation initiatives cover terminology and scenario descriptions which are essential for interoperability and safety validation.<br><br>• Horizon Europe and partnership accords with countries | **1. Objectives**<br>• A leading nation in developing standards to support AV safety, guided by the British Standards Institution (BSI) in collaboration with the Centre for Connected and Autonomous Vehicles (CCAV).<br>• The Connected and Automated Mobility (CAM) Standards Programme.<br>**2. Binding Nature**<br>Legally binding requirements:<br>• Automated Vehicles Act 2024 (AVA 2024).<br>• UK GDPR and Data Protection Act 2018.<br>Non-binding flexibility:<br>• Law Commission's 2022 report<br>• Code of Practice for AV Trialling.<br>**3. Principles/Values**<br>• Fostering innovation while prioritising road safety and clear responsibility allocation.<br>• Safety-focused principles and ethical standards into law.<br>• AV technology should only be permitted on public roads if it performs as safely as, or safer than, a human driver, thus helping to reduce accidents.<br>• Supports national sustainability and net-zero goals, prioritising electric vehicles in most AV trials to encourage decarbonisation.<br>• Lawful processing, robust security measures, and respect for individual rights.<br>**4. Technical Standards**<br>• The "safety ambition" clause, which requires authorised automated vehicles to demonstrate a safety level at least equal to, if not exceeding, that of a skilled human driver.<br>• Strict approval and authorisation procedures, featuring a two-phase clearance system.<br>• Data Protection Impact Assessments and implementing privacy-by-design for relevant technologies.<br>• The Age-Appropriate Design Code.<br>**5. Targeted Legislation/Framework**<br>Foundational laws:<br>• Automated Vehicles Act 2024 (AVA 2024).<br>Sector-specific regulations:<br>• The Connected and Automated Mobility (CAM) Standards Programme.<br>• Regulation 157 (Automated Lane Keeping Systems)<br>Data-specific rules:<br>• UK GDPR and Data Protection Act 2018.<br>• Data Protection Impact Assessments.<br>• Privacy-by-design requirements.<br>• Age-Appropriate Design Code.<br>• Regulation 155 (cybersecurity).<br>**6. Regulator**<br>Primary regulator:<br>• Department for Transport / Centre for Connected and Autonomous Vehicles (CCAV).<br>Supporting bodies:<br>• British Standards Institution (BSI).<br><br>• Law Commission.<br><br>• Information Commissioner's Office (ICO).<br><br>**7. Main Challenges/Risk**<br><br>• The AI accountability system: by the AV Act 2024 addresses liability by shifting responsibility from the human user to the system operator or manufacturer/service provider when an AV operates in self-driving mode. |

## Footnotes

1 United Nations General Assembly, A/RES/78/311, Enhancing international cooperation on capacity-building of artificial intelligence, 1 July 2024; United Nations General Assembly, A/78/L.49, Seizing the opportunities of safe, secure and trustworthy artificial intelligence systems for sustainable development, 11 March 2024.

2 UK Government (2023) introduction to the AI Safety Summit. https://assets.publishing.service.gov.uk/media/65255655244f8e00138e7362/introduction_to_the_ai_safety_summit.pdf; Bengio, Y., Mindermann, S., & Privitera, D. (2025). International AI Safety Report 2025: Advanced AI Risks and Mitigation. AI Safety Institute, United Kingdom. https://www.gov.uk/government/publications/international-ai-safety-report-2025.

3 Jobin, A., Ienca, M., & Vayena, E. (2019). The global landscape of AI ethics guidelines. Nature machine intelligence, 1(9), 389-399.

4 Lee, M. S. A., Floridi, L., & Denev, A. (2021). Innovating with confidence: embedding AI governance and fairness in a financial services risk management framework. In Ethics, governance, and policies in artificial intelligence (pp. 353-371). Cham: Springer International Publishing.

5 Tabassi, E. (2023). Artificial intelligence risk management framework (AI RMF 1.0). The Straits Times. (2025, March 20). Around 30% of South Korean elementary schools use AI textbooks. The Straits Times. Retrieved from https://www.straitstimes.com/asia/east-asia/around-30-of-south-korean-elementary-schools-use-ai-textbooks

6 OECD. (2019/2024). Recommendation of the Council on Artificial Intelligence (OECD Legal Instruments, OECD/LEGAL/0449). Organisation for Economic Co-operation and Development.

7 PNAI (Policy Network on Artificial Intelligence) (2024) The AI Governance We Want, Call to action: Liability, Interoperability, Sustainability & Labour Sipinen, M. (Ed.) Internet Governance Forum. https://intgovforum.org/en/filedepot_download/282/28491; PNAI(Policy Network on Artificial Intelligence) (2023). Strengthening multistakeholder approach to global AI governance, protecting the environment and human rights in the era of generative AI. Sipinen, M. (Ed.) United Nations Internet Governance Forum, https://www.intgovforum.org/en/filedepot_download/282/26545

8 Zeng, M. L. (2019). Interoperability. KO Knowledge Organization, 46(2), 122-146.

9 Berg, C. (2024). Interoperability. Internet Policy Review, 13 (2).

10 Onikepe, A. (2024, July 17). Interoperability in AI governance: A work in progress. TechPolicy.Press. Retrieved from https://techpolicy.press/interoperability-in-ai-governance-a-work-in-progress/

11 Hamin, M. and Hanson, A. (2024) User in the middle: An interoperability and security guide for policymakers, Atlantic Council, https://www.atlanticcouncil.org/in-depth-research-reports/report/user-in-the-middle-an-interoperability-and-security-guide-for-policymakers/

12 Hamin, M. and Hanson, A. (2024) User in the middle: An interoperability and security guide for policymakers, Atlantic Council, https://www.atlanticcouncil.org/in-depth-research-reports/report/user-in-the-middle-an-interoperability-and-security-guide-for-policymakers/

13 Hamin, M. and Hanson, A. (2024) User in the middle: An interoperability and security guide for policymakers, Atlantic Council, https://www.atlanticcouncil.org/in-depth-research-reports/report/user-in-the-middle-an-interoperability-and-security-guide-for-policymakers/

14 Cary Coglianese, Standards and the Law (May 13 2023) Standards and the Law, Standardization: Journal of Research and Innovation, Vol. 2, no. 2, p. 15, , U of Penn Law School, Public Law Research Paper No. 23-18, Available at SSRN: https://ssrn.com/abstract=4452726

15 Cedric (Yehuda) Sabbah, (2024) Framework Interoperability: A New Hope for Global Digital Governance, Lawfare. https://www.lawfaremedia.org/article/framework-interoperability-a-new-hope-for-global-digital-governance

16 Roberts, H. and Ziosi, M. (2025) Can We Standardise the Frontier of AI?. Available at SSRN: https://ssrn.com/abstract=5271446 or http://dx.doi.org/10.2139/ssrn.5271446.

17 McDuff, D., Korjakow, T., Cambo, S., Benjamin, J. J., Lee, J., Jernite, Y., Ferrandis, C. M., Gokaslan, A., Tarkowski, A., Lindley, J., Cooper, A. F., & Contractor, D. (2024). On the Standardization of Behavioral Use Clauses and Their Adoption for Responsible Licensing of AI (No. arXiv:2402.05979). arXiv. https://doi.org/10.48550/arXiv.2402.05979

COUNTRY REPORTS

# UK Country-Level AI Safety Interoperability Report

DAVID A RAHO, SHEFFIELD HALLAM UNIVERSITY, UK

# UK Country-Level AI Safety Interoperability Report

15 October 2025

## 1. Introduction

Artificial Intelligence (AI) is a transformative technology reshaping many industries; however, ensuring its safe and responsible deployment requires strict governance and international cooperation. This report provides a country-level assessment of AI safety and interoperability within the United Kingdom (UK). It aims to evaluate how the UK's AI governance framework addresses safety issues across key sectors such as Autonomous Vehicles, Education, and Cross-Border Data Flow, and to examine how these measures are effectively implemented domestically and internationally.

**Scope:** This report reviews three sectors: autonomous Vehicles, Education, and Cross-Border Data Flows, chosen for their societal impact and need for safety assurance. A comparative assessment framework evaluates each sector's legal and ethical standards, technical criteria, operational capacity, sustainability, data governance, international collaboration, and liability/risk management.

**Methodology:** The analysis draws on recent UK government policy documents, legislation, standards, and authoritative reports from 2020 to 2025. Key sources include those produced by the Department for Science, Innovation & Technology (DSIT), sector-specific legislation (e.g., the Automated Vehicles Act 2024), guidance from the Department for Education (DfE) on AI, and international frameworks developed by OECD, UNESCO, and GPAI, which influence UK policy. Insights from UK institutions such as the Alan Turing Institute, Ada Lovelace Institute, and British Standards Institution (BSI), along with multilateral organisations (OECD, G7), are integrated to contextualise the UK's approach internationally.

**Research Questions:** This report is guided by questions such as:
*How does the UK's AI governance framework ensure safety and interoperability across various sectors?*
*What sector-specific regulations and standards exist for AI safety regarding Autonomous Vehicles, Education, and Cross-Border Data Flow?*
*How does the UK promote interoperability of AI safety measures across governance, security, and technical/data sectors domestically and in cross-border contexts?*
*The report also examines best practices emerging from the UK's experience and provides recommendations to strengthen AI safety governance across sectors and borders.*

## 2. Comparative Assessment Framework (UK AI Safety Governance Overview)

UK's AI Governance Approach: The UK intentionally balances innovation and safety in its AI governance, emphasising flexibility, principles, and sector-specific regulators. Instead of a single comprehensive AI law, the government's March 2023 AI Regulation White Paper sets out five guiding principles for AI development and use across all sectors.

1. Safety, security, and robustness.
2. Appropriate transparency and explainability.
3. Fairness
4. Accountability and governance.
5. Contestability and redress.

(DSIT, 2023b)

These principles, initially adopted on a non-statutory basis, serve as a unifying ethical and legal framework across various regulatory sectors. Regulators in transport, education, data, and others are expected to interpret and apply these AI principles within their respective remits, ensuring consistency while tailoring them to their specific contexts. This results-oriented approach addresses risks without unnecessarily hindering innovation and technological progress.

**Institutional Governance and Coordination:** The UK relies on existing regulators such as the Office for Safety and Standards in Transport, the Office for Students, the Information Commissioner's Office, and emerging support structures. The UK's AI Safety Institute (AISI), now renamed the AI Security Institute due to a change in government policy, was launched in 2023 as a government-backed research organisation within the Department for Science, Innovation & Technology.

Focused on researching advanced AI risks and advising policymakers, AISI and similar initiatives like the Alan Turing Institute's (ATI) public-sector AI advisory programmes aim to strengthen the scientific basis for governance decisions.

**Global Interoperability Commitment:** A key UK AI policy is to collaborate internationally towards interoperable governance, recognising that inconsistent global regulations can hinder innovation and safety. The UK broadly aligns its framework with international efforts and remains dedicated to engaging globally to support interoperability across different regulatory regimes. It aims to reduce business compliance burdens and embed the UK's values into emerging global AI governance strategies (DSIT, 2023b). This is demonstrated by the UK's active involvement in forums such as the OECD (which produced AI Principles in 2019 that the UK upholds), the Global Partnership on AI (GPAI), and the Council of Europe's AI governance discussions. The UK also used its G7 presidency to promote the idea of "Data Free Flow with Trust" (DFFT) for digital governance, emphasising that data and AI innovations should be able to move freely across borders, supported by strong protections (DSIT, 2023a).

In practice, the UK aligns its standards and laws with international benchmarks, such as following OECD guidelines on AI risk and collaborating on standards development, to ensure that AI systems and safety practices can operate seamlessly across borders.

**Assessment Framework Dimensions:** To systematically evaluate AI safety governance, this report uses a comparative framework with multiple dimensions:

- **Legal & Ethical Frameworks:** The laws, regulations, and ethical codes that govern AI.
- **AI Safety Considerations:** Addressing sector-specific physical, psychological, and privacy risks.
- **Technical Standards:** Formal standards (e.g., ISO) that ensure AI robustness, safety, and interoperability.
- **Operational Capacity:** The resources, skills, and tools available to implement and oversee AI safety.
- **Sustainability:** How governance tackles long-term environmental and social impacts, such as energy use and public trust.
- **Data Governance:** Policies for managing AI data, with emphasis on quality, privacy, and ownership.
- **International Cooperation:** Alignment with global partners on AI regulations and safety standards.
- **Liability & Risk Management:** The framework for assigning responsibility and managing harm from AI failures.

**This framework will be applied to each of the three sectors:** Autonomous Vehicles (AVs), Education, and Cross-Border Data Flow

## 3. Sectoral Analysis:

### **3.1** AUTONOMOUS VEHICLES (AVS)

**Legal & Ethical Framework:** The United Kingdom has built a robust legal system for autonomous vehicles, fostering innovation while prioritising road safety and clear responsibility allocation. Following the Law Commission's extensive law reform efforts, the Automated Vehicles Act 2024 (AVA 2024) was enacted, providing a strong legal basis for deploying self-driving vehicles (Tyers, 2025). This legislation incorporates recommendations from the Law Commission's 2022 report (LC, 2024), embedding safety-focused principles and ethical standards into law. A key feature is the "safety ambition" clause, which requires authorised automated vehicles to demonstrate a safety level at least equal to, if not exceeding, that of a skilled human driver, to improve overall road safety (Tyers, 2025). This sets a clear ethical standard: AV technology should only be permitted on public roads if it performs as safely as, or safer than, a human driver, thus helping to reduce accidents.

The AVA 2024 requires strict approval and authorisation procedures. It features a two-phase clearance system: traditional vehicle type approval to ensure basic safety and specialised self-driving authorisation for the automated driving system (LC, 2024). This framework is based on international safety standards, such as UNECE regulations and ISO standards, by incorporating specific measures designed for autonomous functionality. Importantly, it is a criminal offence to market a vehicle as "self-driving" without official authorisation as an automated vehicle (Tyers, 2025), highlighting a moral obligation to accurate communication and consumer protection.

**AI Safety Considerations:** Passenger and public safety remain central to the UK's approach, with strict requirements that an AV may take control only under safe conditions and must relinquish control or manage risk when system reliability is uncertain. Ongoing "in-use" safety assurance is emphasised, requiring continuous monitoring throughout the vehicle's lifecycle rather than relying solely on initial approval (LC, 2024). The AVA 2024 authorises regulators to suspend or revoke approval if an AV is deemed unsafe or non-compliant and establishes statutory AV inspectors to investigate incidents in a no-fault manner, like air accident investigation models.

From a technical perspective, robust safety engineering standards are recommended. Automotive AI systems must handle edge cases and provide demonstrable “safe fallback” behaviour, such as implementing minimal-risk manoeuvres when encountering critical faults. Transparency in safety information and alignment with human traffic expectations are essential for maintaining public trust.

**Technical Standards:** The United Kingdom has been a leading nation in developing standards to support AV safety, guided by the British Standards Institution (BSI) in collaboration with the Centre for Connected and Autonomous Vehicles (CCAV). The Connected and Automated Mobility (CAM) Standards Programme has produced numerous standards that facilitate safe trials and deployments (BSI, 2025). For example, PAS 1881:2022 specifies requirements for confirming the operational safety of automated vehicle trials (PAS 1881, 2022).

Internationally, the UK follows global standards such as ISO 26262 (electronic systems), ISO 21448 (intended functionality safety), and ISO/SAE 21434 for cybersecurity. As a member of UNECE WP.29, the UK enforces regulations including Regulation 157 (Automated Lane Keeping Systems) and Regulation 155 (cybersecurity). BSI’s programme also promotes adopting emerging standards like ISO 34503 on operational design domains, with PAS 1883:2025 guiding local implementation. Standardisation initiatives cover terminology and scenario descriptions (for example, CAM Vocabulary via BSI Flex 1890), which are essential for interoperability and safety validation.

**Operational Capacity:** The UK has invested heavily in infrastructure to support effective oversight. The Centre for Connected and Autonomous Vehicles (CCAV), a joint unit between the Department for Transport and DSIT, provides policy leadership and funds advanced testbeds for AV research. These platforms facilitate controlled and real-world testing, guided by the Code of Practice for AV Trialling (DfT, 2023), which outlines safety management protocols, including the use of a safety driver or remote operator, transparent data sharing with authorities, and community engagement, key elements for responsible development and risk mitigation.

**Sustainability:** AI in transportation provides both environmental and societal benefits and challenges. The UK’s AV strategy supports national sustainability and net-zero goals, prioritising electric vehicles in most AV trials to encourage decarbonisation. The Connected & Automated Mobility 2025 strategy links AV innovation to broader aims, such as reducing accidents and enhancing accessibility (CCAV, 2022).

**Data Governance:** Autonomous vehicles depend on and produce large amounts of data, which are governed by both general privacy laws and sector-specific regulations. Many AV data streams contain personal data; thus, processing must comply with the UK GDPR and Data Protection Act 2018 (UK GDPR & DPA, 2018), requiring lawful processing, robust security measures, and respect for individual rights. The Information Commissioner’s Office (ICO) provides guidance, including conducting Data Protection Impact Assessments and implementing privacy-by-design for relevant technologies. The Age-Appropriate Design Code also enforces high privacy settings and transparency when handling children’s data (DfEd, 2025). Regarding operational and safety data, AVA 2024 requires that local traffic regulations be provided in digital formats for AV integration. This open data requirement promotes up-to-date compliance and supports safety objectives. BSI’s PAS 1882:2021 sets out frameworks for incident-related data collection during AV trials (BSI, 2025).

**International Cooperation:** Due to the global nature of the automotive industry, the UK actively collaborates with international organisations, helping to harmonise AV regulations through the UNECE and bilateral agreements. Participation in initiatives like Horizon Europe and partnership accords with countries such as Japan and Singapore demonstrates its commitment to joint research and regulatory alignment. The UK’s involvement in the G7 Transport Ministers’ declarations and its contributions to ISO and IEEE standards further show its dedication to shaping global best practices.

The UK closely observes EU regulatory changes, especially concerning the EU AI Act and product liability rules, to ensure cross-jurisdictional compatibility and maintain market access. The primary strategy focuses on interoperability by design, enabling UK and foreign AVs to operate safely across different markets.

**Liability and Risks:** The AV Act 2024 addresses liability by shifting responsibility from the human user to the system operator or manufacturer/service provider when an AV operates in self-driving mode. If a regulated AV commits a traffic offence or is involved in an incident while autonomous, the licensed operator assumes liability, departing from traditional driving norms and recognising the AI system as the accountable entity.

## **3.2** EDUCATION

### Legal & Ethical Framework

AI in education encompasses everything from intelligent tutoring systems to automated exam marking and AI assistants. Although the UK lacks specific legislation for AI in education, a comprehensive legal framework already exists to safeguard children’s rights, safety, and privacy, and it fully applies to AI use. Education Law & Safeguarding: Schools have a legal duty to protect pupils' well-being under the Children Act 2004 and Education Act 2002. This responsibility extends into the digital world. The Department for Education's (DfE) statutory guidance, "Keeping Children Safe in Education," now explicitly includes AI tools, requiring schools to ensure online safety (Gen AI in Ed, 2025). Any AI tool used in a school must align with its safeguarding policies, reflecting the ethical principle of doing no harm. Reinforced by Generative AI: product safety expectations (Department for Education, 2025)

**Data Protection and Pupil Privacy:** Student data is safeguarded by the UK GDPR and the Data Protection Act. As data controllers, schools must ensure any EdTech or AI service they employ complies with these regulations. The Information Commissioner's Office's (ICO) Children's Code is essential (ICO Children's Code, 2024; Gen AI Prod Safety, 2025). This code establishes 15 standards for online services likely used by children, requiring high privacy settings by default and minimal data collection. For instance, an AI homework app cannot profile children for advertising and must manage personal data with strict controls.

**Equality and Non-Discrimination:** The Equality Act 2010 obliges schools to prevent discrimination. If an AI grading system were found to systematically disadvantage students from a certain background due to biased training data, it could be considered unlawful. This legal requirement upholds the ethical principles of fairness and justice.

**Emerging AI Policies:** Between 2023 and 2025, the UK government began issuing AI-specific guidance. The DfE's paper "Generative AI in education" (updated August 2025) sets out the official approach: embrace the opportunities but operate within a safe and ethical framework (Gen AI in Ed, 2025). It emphasises that safety remains the top priority.

**Ethical Guidelines:** In addition to government efforts, organisations such as the UK Children's Commissioner's Office have called for stricter ethical standards. While a specific Code of Practice for AI in education is not yet law, schools often refer to broader frameworks like the UNESCO Recommendations on AI in Education or the IEEE's ethical design guidance.

In brief, the UK's legal framework for AI in education is a patchwork of existing child protection, privacy, and equality laws, now complemented by sector-specific guidance prioritising student well-being.

## AI Safety Considerations

Safety in this context is multifaceted, encompassing physical harm as well as psychological, cognitive, and reputational risks.

**Accuracy and Reliability:** AI tools must be precise. The UK saw this firsthand during the 2020 A-level grading fiasco, where a flawed algorithm caused biased outcomes and sparked public outrage. This disaster highlighted the dangers of algorithmic opacity and bias. The main lesson was to adopt a "human in the loop" approach, where a person reviews all AI-generated decisions.

**Content Safety:** Generative AI might generate inappropriate, biased, or simply incorrect content. The DfE's guidance explicitly warns of this risk (Gen AI in Ed, 2025). Schools are advised to use AI tools with built-in filters and carefully supervise students to reduce this risk. Some schools have blocked general-purpose chatbots in favour of approved, educational AI platforms.

**Bias and Fairness:** AI systems can reinforce societal biases. An AI tutor trained on culturally narrow texts might not effectively support a diverse student body. Educators are advised to remain vigilant and report any disparities in performance among student groups.

**Psychological Impacts:** An AI tutor's unempathetic or discouraging feedback could damage a student's confidence. This is part of an educator's broad duty of care.

**Cheating and Academic Integrity:** AI makes essay writing easier, which risks undermining academic honesty. The DfE's recommended approach is not to ban AI but to promote AI literacy, educating students about its ethical use and making it a meaningful learning experience.

To manage these risks, schools are encouraged to conduct risk assessments before implementing any AI tool, examining its data usage, potential failure points, and auditability.

## Technical Standards

Formal technical standards for AI in education are still developing, but several relevant areas exist.
Data Interoperability Standards: UK EdTech often adopts standards like Learning Tools Interoperability (LTI) to ensure different systems can communicate securely. This allows AI tools to integrate seamlessly with existing school data systems.

**Product Safety Standards:** While not an official standard, the government has issued 'Generative AI product safety expectations" (Gen AI Prod Safety, 2025). These outline requirements for developers to include content filtering, age-appropriate design, and user controls.

**Cybersecurity and IT Standards:** UK schools must follow standard IT security practices, often referencing ISO/IEC 27001. This ensures that AI tools do not create security vulnerabilities or expose sensitive student data. Accessibility Standards: Digital content, whether AI-generated or not, must be accessible to students with disabilities, in line with guidelines like the Web Content Accessibility Guidelines (WCAG).

**Quality Benchmarks:** Without formal standards, quality is often evaluated using academic benchmarks and trusted programmes. For example, the EDUCATE programme at UCL supports mentoring EdTech startups on evidence-based development, serving as an informal mark of quality.

### Operational Capacity

Using AI safely and effectively depends on the capacity of educators and schools.

**Teacher Training and Guidance:** The DfE has launched online AI training modules for teachers to boost their confidence and skills. The goal is to empower educators to make informed decisions and supervise students effectively.

**Digital Infrastructure:** The nationwide rollout of high-speed internet in schools has established the basis for AI tools, which often depend on reliable connectivity.

**Platforms and Procurement:** Many schools supervise AI using established platforms like Microsoft 365 or Google Classroom. These platforms offer centralised administrative controls, allowing schools to manage AI features and ensure compliance.

**Oversight and Inspection:** The school's inspectorate, Ofsted, does not directly assess AI use but will evaluate how it is integrated into the curriculum and safeguarding policies. This oversight encourages schools to create clear internal policies for AI.

**Incident Response:** Schools update their safeguarding incident procedures to address AI-related concerns. Designated Safeguarding Leads (DSLs) are trained to handle situations where an AI tool might produce harmful content.

**Research and Collaboration:** An expanding ecosystem of research and collaboration supports schools in sharing best practices and learning from each other, ensuring they do not have to face the complexities of AI alone.

### Sustainability

Sustainability means ensuring AI's long-term pedagogical value, social equity, and environmental responsibility.

**Long-term Adoption:** To prevent the "pilot then fizzle" cycle, AI tools need to focus on real needs, such as lowering teacher workload or offering personalised support. The DfE's AI Opportunities Action Plan intends to fund research into sustainable deployment models for all schools (DfE, 2025).

**Digital Divide:** There is a risk that AI could widen the gap between well-resourced schools and those with fewer resources. Efforts to provide centrally funded AI tools and access through public libraries aim to promote social sustainability and equal opportunity.

**Environmental Footprint:** Training large AI models consumes substantial energy. The UK's National AI Strategy recognises the importance of "green compute" and energy-efficient data centres (UK's AI Future Depends on Decarbonising Data Centres, 2025). Schools can help by selecting cloud providers with strong sustainability commitments.

**Curriculum Sustainability:** Reliance on AI must not weaken students' core skills. The curriculum is being revised to emphasise digital literacy and critical thinking, teaching students to use AI to support rather than replace their own abilities.

### Data Governance

Data is the lifeblood of educational AI, and its governance is paramount.

**Student Data Privacy:** Schools must ensure that any AI provider complies with privacy laws, often through strict data-sharing agreements. The Children's Code "data minimisation" principle is essential; only necessary student data should be used.

**Parental Consent and Transparency:** Schools usually obtain consent for data processing on behalf of children under 13 for educational reasons under the "public task" legal basis. However, many schools proactively inform parents and might offer an opt-out for non-essential tools.

**Data Quality and Bias:** The quality of training data affects an AI's effectiveness and fairness. Initiatives like the Oak National Academy are exploring the release of curated, curriculum-aligned datasets to aid in developing better and less biased educational AI for UK schools.

**Learning Analytics Governance:** AI can generate detailed analytics on student performance. Schools must carefully govern this sensitive data, ensuring it is used to support learning rather than to label or penalise students. Retention and deletion policies are vital for deciding how long student data is kept. DfE guidance advises schools to be cautious with AI tools that learn from user inputs, ensuring no personal data is entered unless necessary (Gen AI in Ed, 2025).

**Security:** AI tools must be secure. A data breach involving children's information could result in severe legal repercussions under GDPR.

### International Cooperation
While education remains a national concern, the UK actively engages in the global discussion on AI. For example, the UK is a key member of UNESCO's Global Education Coalition on AI, aligning its domestic policies with international principles of fairness and inclusion. The UK also supports digital and AI skills initiatives through the OECD and shares its projects via the OECD

### AI Policy Observatory.
The UK conducts cross-border research, ensuring it stays at the forefront of pedagogical innovations and AI progress.

Post-Brexit, the UK is negotiating bilateral and multilateral agreements (such as the CPTPP) that promote digital trade and EdTech exchange, as long as its high standards for safety and privacy are maintained.

The UK learns from and shares in global best practices, aiming to adopt innovative and effective AI tools worldwide that meet its strict safety and ethical standards.

### Liability and Risks
If AI malfunctions in the classroom, accountability falls under existing legal frameworks.

**Professional Responsibility:** Teachers and schools are ultimately responsible for student welfare. If an AI tool is used negligently, the school could be liable, highlighting the need for due diligence and appropriate supervision.

**Product Liability:** The provider may be responsible if a defective AI product causes injury. Nevertheless, a student's right to request a human review of any significant automated decision (under UK GDPR) acts as an essential safeguard, prompting institutions to avoid relying solely on AI for critical decisions.

**Data Breaches or Misuse:** If student data is mishandled, the school and the vendor could face regulatory action from the ICO.

**Insurance:** Schools are assessing whether their existing liability insurance covers AI-related incidents. The prevailing view is that AI does not exempt educators from their duty of care.

To address these risks, the DfE's stance is clear: keep a human involved. Professionals must check AI outputs to prevent any student harm caused by AI failure without adult supervision.

### Summary
In the education sector, the UK is proceeding cautiously. It recognises AI's potential to personalise learning but emphasises that it must not compromise education's safety, privacy, or integrity. The approach focuses less on creating new, strict laws and more on applying existing child protection frameworks, professional standards, and clear guidance. This human-centric model, where AI serves as a tool for teachers rather than a replacement, is the foundation of the UK's strategy and shapes its international engagement.

## **3.3** CROSS-BORDER DATA FLOWS

### Legal & Ethical Framework
Cross-border data flows involve moving data across national borders, essential for AI development and services that depend on global information. The UK's strategy is based on data protection laws and trade policies, aiming to enable free data movement for innovation while upholding high standards and public trust.

The legal framework has several core components:
UK GDPR and Data Protection Act 2018: Specify how personal data can leave the UK. A transfer is permitted if the destination country provides an "adequate" level of data protection, or if safeguards like Standard Contractual Clauses (SCCs) are implemented (DSIT, 2021). Ethically, this ensures that the fundamental right to privacy isn't compromised simply because data is transferred.

**UK Adequacy Decisions:** Post-Brexit, the UK makes its own adequacy decisions. It has recognised the EU/EEA, Japan, Canada, and others as adequate partners. A notable example is the UK–US Data Bridge, launched in October 2023, which facilitates data transfer to certified US companies that meet the UK's strict privacy and rights standards.

**Alternative Transfer Mechanisms:** If a country isn't deemed "adequate," UK firms can rely on tools such as the International Data Transfer Agreement (IDTA). The government offers templates and guidance for a risk-based approach, expecting companies to conduct due diligence to prevent data misuse, like excessive government surveillance (ICO, 2025).

**Emerging Data Protection Reforms:** The Data Protection and Digital Information Bill (2023–2024) aims to facilitate flexible international transfers. The ethical debate concerns balancing economic growth with privacy rights, but the UK government insists it will not compromise its high standards.

**Trade Agreements and Digital Trade Rules:** The UK incorporates data flow provisions in trade deals, such as those with Japan and Singapore. These agreements commit to the free flow of data and ban forced data localisation, provided personal data is protected. This aligns with the G7 "Data Free Flow with Trust" (DFFT) concept, removing barriers while building trust through shared principles.

The UK's framework aims to ensure that cross-border data transfers are lawful, secure, and straightforward. The key ethical principle is that fundamental rights travel with the data, reflecting a belief in universal privacy and dignity.

## AI Safety Considerations

Cross-border data flows have several direct implications for AI safety.

**Data Quality and Bias:** AI models can inherit biases from global data. If data from another country is inaccurate or unrepresentative, an AI system could be unsafe (e.g., a medical AI trained on one ethnic group performing poorly on another). To address this, the UK collaborates with groups like the Global Partnership on AI (GPAI) to promote high-quality, representative data sharing.

**Security Risks and Cybersecurity:** Transferring data across borders expands the "attack surface" for cyber threats. AI systems are vulnerable to data poisoning attacks (manipulating training data) or data interception. UK policies and guidance from the National Cyber Security Centre (NCSC) highlight the importance of strong security measures like encryption to protect data integrity, which is vital for trusting the AI built on it.

**Concentration Risk:** A large portion of data is directed to a few major cloud providers, mainly in the US. This creates a strategic risk; if a political or legal change overseas restricts access to this data, it could affect UK AI services. To counter this, the UK is diversifying its data partners (such as through the CPTPP trade agreement) and investing in local capabilities like a sovereign cloud. Other countries are also adopting similar methods.

**Ethical Data Use and Human Rights:** Data transferred abroad could be misused for surveillance or discrimination, especially if it ends up in a country with a poor human rights record. The UK helps lead initiatives like the OECD Declaration on Government Access to Personal Data, which sets principles to ensure that law enforcement access is necessary and proportionate. This aims to protect individuals and, for example, ensure that UK data does not fuel an authoritarian state's AI-driven surveillance.

**Transparency and accountability in AI supply chains:** When parts of AI development, such as data labelling, are outsourced overseas, it is crucial to maintain safety standards. The idea of AI supply chain accountability is gaining traction, and UK firms are increasingly required to ensure their international partners adhere to safety protocols through contracts and audits.

## Technical Standards and Interoperability Layers

Making cross-border data flows work requires interoperability across governance, security, and technical standards. Governance/Policy Interoperability: The UK seeks to unify different countries' data protection regulations. By participating in forums like the Global Privacy Assembly, the UK encourages mutual recognition and regulatory cooperation, ensuring that AI systems across various countries adhere to regulations.

**Security Standards:** Technical standards like TLS encryption and ISO 27001 are crucial for secure data transfer. The UK's participation in frameworks such as the Global Cross-Border Privacy Rules (CBPR) Forum helps create a trusted technical environment for data to move securely.

**Data Format and Semantic Standards:** For AI to work effectively with global data, agreed-upon data formats are essential. The UK's NHS, for example, employs standards like HL7/FHIR, enabling international collaboration on medical AI research.

**Digital Trade Standards:** Certifications like ISO/IEC 27701 (for privacy management) and the UK's BS 10012 allow organisations to demonstrate internationally that they manage data securely, facilitating cross-border data exchanges.

### Operational Capacity

Effectively managing cross-border data flows requires capacity from several key players.

**Regulatory Capacity (ICO):** The UK's Information Commissioner's Office (ICO) is a highly regarded international regulator. It has cooperation agreements with counterparts in other countries, enabling joint enforcement on cross-border issues.

**Government Policy and Negotiation:** The Department for Science, Innovation and Technology (DSIT) has a dedicated team focused on international data policy. They rely on expert advice to develop forward-looking strategies and ensure the UK remains active in global discussions at the WTO, G7, and OECD.

**Industry Practices:** UK companies have built significant capacity for data compliance management. Many appoint a Data Protection Officer who is knowledgeable about the legal framework, and the government provides toolkits to help even small businesses navigate the regulations.

**Judicial and Legislative Oversight:** The UK has legal frameworks, such as the bilateral UK-US Data Access Agreement (2019), to regulate lawful data requests from other countries, like those under the US CLOUD Act. This helps prevent legal disputes and ensures a managed process.

**Public Sector and Cross-Border Data:** The UK public and research sectors depend on international data collaboration. UK universities and the NHS are experienced in multi-country data protocols and often use privacy-preserving techniques like federated learning to train AI models globally without transferring raw personal data.

### Sustainability

A sustainable system for cross-border data flows must be durable, trusted, and able to foster long-term innovation. Multilateral Solutions vs. Fragmentation: The UK views the fragmented global data regime as unsustainable and potentially problematic. It advocates for multilateral solutions, such as the Global CBPR Forum, to establish a stable, predictable environment that encourages responsible AI innovation.

**Environmental Footprint:** Data centres supporting global data flows have a significant carbon footprint. UK policy encourages aligning digital strategies with net-zero targets by promoting energy-efficient infrastructure and methods such as edge computing, which can reduce the need to transmit large data volumes over long distances.

**Economic and Social Sustainability:** Cross-border data flows are vital for the UK's service-based economy. Ethically, the UK also promotes sharing data for social good, such as fighting pandemics, which relies on a sustainable international trust framework.

**Risk of Data Nationalism:** The UK opposes widespread data localisation requirements, where countries force data to be stored within their borders. Such policies can obstruct innovation and even introduce security risks. The UK consistently advocates for open data flows, emphasising their significance for sustainable growth.

### Data Governance

Cross-border data flow mainly concerns data governance. The UK's National Data Strategy (2020) explicitly promotes "unlocking the value of data" across borders while safeguarding public trust (DSIT & DfDCMS, 2022). The government provides practical governance tools like the International Data Transfer Agreement (IDTA). These tools are crucial because the UK's data protection system is based on accountability, meaning a UK company remains responsible for data even after it is transferred abroad. The UK also supports Data Trusts and similar organisations that facilitate secure data sharing, such as the international collection of genomic data for AI-driven health research.

### International Cooperation

The UK's strategy depends heavily on active international cooperation.

**OECD:** The UK co-drafted the OECD AI Principles and the Declaration on Government Access to Data, establishing common ground with many other countries.

**G7 and G20:** The UK has promoted the Data Free Flow with Trust (DFFT) initiative within the G7 and participates in data flow discussions at the G20.

**Global CBPR Forum:** The UK is joining this forum to develop an international certification system for companies to demonstrate high data protection standards, making data exchange easier among member countries.

**Bilateral Cooperation:** The UK conducts data policy dialogues with key partners like Singapore to share best practices and align regulatory approaches for AI governance.

**Capacity Building:** The UK supports developing countries in creating their own data protection laws, helping to expand the global network of trust and enable safer future data exchanges.

### Liability and Risks

Cross-border data flows introduce unique liability questions and risks.

**Liability:** Under UK law, a UK-based company can be held responsible for a data breach, even if the mistake was made by its partner abroad. This emphasises the importance of thorough due diligence as a key risk mitigation measure.

**Enforcement Challenges:** Enforcing UK regulations in an uncooperative country is challenging. The UK addresses this risk by not permitting data transfers to such jurisdictions in the first place (i.e., no "adequacy" decision).

**Intellectual Property and Data:** Using global datasets to train AI introduces complex IP risks. The UK is collaborating internationally to clarify rules around text-and-data mining, aiming to reduce legal uncertainty for researchers and businesses.

**Geopolitical Risks:** Data flows can become politicised if diplomatic relations sour. The UK establishes a dispute mechanism between states by including data flow commitments in trade agreements, offering stability and protection for businesses against unwarranted data restrictions.

### Summary

The UK's approach to cross-border data flows carefully balances enabling innovation with upholding high standards. Through active diplomacy and robust domestic regulation, the UK seeks to create an interoperable global system where data can move freely and securely. Developers can access diverse international datasets for AI to develop fairer and more resilient systems. Simultaneously, the public can trust that their personal data remains protected to UK standards, no matter where it travels.

## 4. Interoperability

Overview of each sector's relative strengths in terms of Interoperability

Interoperability in AI safety refers to the capacity of different systems, frameworks, and stakeholders to work together seamlessly across sectors and borders. In the UK, achieving interoperability requires aligning governance (laws and policies), security protocols, and technical/data standards to uphold AI safety principles consistently in various fields.
The comparative matrix (see Interoperability Matrices) summarises how each layer is tackled in each sector, followed by an explanatory narrative:

**Governance Layer:** The UK employs a principles-based governance approach supplemented by sector-specific regulations across all three focus areas. The governance layer emphasises common principles (like safety, accountability, fairness) that relevant regulators interpret in each sector. For instance, governance concerning Autonomous Vehicles is conducted through formal legislation (Automated Vehicles Act) and rules that explicitly incorporate those principles (safety equivalence to human drivers, operator accountability).

| Sector | Legal/Organisational | Technical/Interface | Data/Semantic | Overall Assessment |
|---|---|---|---|---|
| **Autonomous Vehicles** | **Strong**<br>Comprehensive legal framework with clear liability structures | **Moderate**<br>Aligned with international standards, but implementation timelines diverge. | **Moderate**<br>UK GDPR compliance but adequacy uncertainties | **Good**<br>Strong foundation with some operational constraints |
| **Education** | **Moderate**<br>Guidance-based approach creates flexibility but potential inconsistencies. | **Weak**<br>Limited technical certification requirements | **Moderate**<br>Child-specific protections but limited international alignment | **Fair**<br>Developing framework requiring strengthening |
| **Cross-Border Data Flows** | **Moderate**<br>Independent adequacy processes, but EU alignment challenges | **Good**<br>International standards adoption with sector-specific guidance | **Moderate**<br>Modified protection standards may affect compatibility | **Good**<br>Operational effectiveness with strategic uncertainties |

**Table 1:** Overview of each sector's relative strengths in terms of Interoperability

In **Education**, governance relies less on specific AI legislation. Nevertheless, the same principles are upheld through existing legal obligations (safeguarding, equality) and official guidance emphasising safety and accountability in AI deployment.

For **Cross-Border Data**, governance is managed through data protection laws and international agreements, ensuring accountability and fairness are maintained across borders. Interoperability at this level means that a core value like safety is consistently prioritised, whether it's an autonomous car, a classroom AI, or an international data transfer regime. UK governance also aims to be compatible with international standards: for example, vehicle safety rules align with UNECE, education AI ethics conform to UNESCO/OECD guidelines, and data laws are sufficiently aligned with EU/OECD principles to facilitate mutual recognition in terms of adequacy.

**Security Layer:** This layer safeguards AI systems and data from harm (accidents, misuse, cyber threats). Interoperability here involves establishing common security standards and practices that can be implemented across sectors and internationally.
In the case of AVs, security includes functional safety and vehicle cybersecurity; the UK mandates compliance with international vehicle cyber standards (UNECE R155) and has testing protocols for fail-safe behaviour. No-blame incident investigators also speak to a culture of safety and continuous improvement, which is analogous to how aviation security is handled globally – an interoperable practice across borders.

In **Education**, security translates to safeguarding students in digital environments: content filtering, access controls, and data security measures are standard across all schools (many use the same accredited filtering software). The online safety principle is integrated with AI usage guidelines, meaning any AI application in schools must be consistent with the security measures schools already use for internet safety (interoperable with broader online safety regimes, including the UK's Online Safety Act once in force).

For **Cross-Border Data**, security is essential: encryption, secure APIs, and compliance with frameworks like ISO 27001 are expected for any data transfer. The UK's international agreements often include commitments to security practices, ensuring its security measures align with those of its allies. For example, if UK data is transferred to a US cloud under the Data Bridge, that cloud provider must have certified security credentials. Additionally, cooperation in incident response (e.g., if a cyber incident affecting personal data crosses borders, the UK's CSIRTs and the ICO collaborate with foreign counterparts) guarantees that security breaches are handled smoothly. A key point is that cybersecurity and resilience are seen not as isolated sector issues but as national priorities covering all digital systems, including AI. The National Cyber Strategy and NCSC guidance provide a common foundation across sectors.
Technical/Data Layer: Interoperability at the technical layer involves standards for data formats, model compatibility, and interfaces that ensure AI systems and datasets can be widely used and understood. The UK strongly supports technical standardisation for AI.

In **AVs**, this is reflected in developing standardised terminologies (such as BSI's CAM vocabulary) and data specifications (like PAS 1882 for incident data) that align with or feed into ISO standards, enabling UK-developed AV technology to connect with global automotive systems. The need for digital Traffic Regulation Order data essentially establishes a data standard for map and regulatory data that can be utilised by any AV worldwide – a move towards technical interoperability in smart infrastructure.

In **education**, although less formal, using common data standards for student information, such as standard file formats for e-learning content and SCORM/xAPI for tracking learning experiences, enables AI tools to integrate with school management systems and learning platforms. The UK's advocacy in international forums for open educational resources and data standards (through the IEEE and UNESCO) further endorses this alignment. Suppose a UK school uses an AI reading app from the US. In that case, it probably employs IMS Global standards, allowing it to integrate student data securely without custom adaptation, demonstrating technical interoperability in practice.

For **Cross-Border Data**, technical interoperability guarantees that data protection and usage preferences travel with the data. International projects on data provenance and machine-readable consent are in development; the UK's regulators are interested in such standards so that, for example, an individual's consent or privacy settings encoded in the UK can be recognised by an AI processing data in another country. The UK participates in ISO/IEC JTC1 SC27 (security techniques) and SC38 (cloud and distributed systems) to promote standards that facilitate safe cloud interoperability—essential for cross-border cloud-based AI services.

| Sector | Governance | Security | Technical/Data |
|---|---|---|---|
| **Autonomous Vehicles** | Automated Vehicles Act 2024; “Safety Ambition” clause; lifetime regulator oversight (DVSA/CCAV). | Functional safety (ISO 26262, ISO 21448); UNECE R155 cybersecurity; BSI PAS 1881/1884 trial safety standards. | BSI PAS 1882 (incident data); PAS 1883/ISO 34503 (ODD); digital Traffic Regulation Orders mandate. |
| **Education** | DfE Generative AI policy (2025) and guidance; statutory safeguarding duties; Equality Act 2010; ICO Children’s Code. | Safeguarding frameworks, content filtering and moderation, and risk assessments for AI adoption. | Learning data interoperability (xAPI, SCORM); accessibility standards (WCAG); EdTech vendor compliance frameworks. |
| **Cross-Border Data Flows** | UK GDPR and Data Protection Act 2018; adequacy agreements (EU, US, others); trade deals embedding DFFT. | ICO oversight; Global CBPR Forum; encryption, secure API, and international audit standards. | IDTA and SCCs; ISO/IEC 27701 (privacy management); OECD/GPAI interoperability guidelines. |

**Table 2:** Interoperability Matrix (Governance, Security, Technical/Data Layers by Sector)

| Layer | Key Instruments & Practices | Core Principles & Objectives | International Alignment & Interoperability |
|---|---|---|---|
| **Governance** | Automated Vehicles Act; Formal rules and regulations | Safety equivalence to a competent and careful human driver; Clear attribution of operator accountability | Alignment with United Nations Economic Commission for Europe (UNECE) vehicle regulations |
| **Security** | Mandated compliance with UNECE Regulation 155 (Cybersecurity); Formal testing protocols for fail-safe behaviour; "No-blame" incident investigation bodies | Functional safety; Cybersecurity by design; Continuous improvement and learning from incidents | Adoption of global aviation safety culture model; Adherence to international cybersecurity standards |
| **Technical/Data** | BSI PAS 1882 (Safety case data); BSI CAM Vocabulary; Standardised digital Traffic Regulation Orders (TROs) | Common terminology for communication; Standardised data formats for incident analysis and infrastructure interaction | Contribution to and alignment with International Organisation for Standardisation (ISO) standards |

**Table 3:** Interoperability Matrix for Autonomous Vehicles

| Layer | Key Instruments & Practices | Core Principles & Objectives | International Alignment & Interoperability |
|---|---|---|---|
| **Governance** | Official guidance from the Department for Education (DfE); Application of existing legal duties | Safeguarding and welfare of students; Non-discrimination under the Equality Act 2010; Fairness and transparency | Ethical alignment with UNESCO and OECD guidelines on AI in education |
| **Security** | Existing school online safety policies and procedures; DfE guidance on digital safety | Protection of student data; Safeguarding in digital environments; Access control and content filtering | Interoperable with broader national digital safety regimes (e.g., the UK's Online Safety Act) |
| **Technical/Data** | De facto industry standards (e.g., SCORM, xAPI, IMS Global); Common file formats for educational content | Data portability for student records; Integration with existing Learning Management Systems (LMS); Ease of use for educators | Advocacy for open educational resources and standards in international forums (e.g., IEEE, UNESCO) |

**Table 4:** Interoperability Matrix for AI in Education

| Layer | Key Instruments & Practices | Core Principles & Objectives | International Alignment & Interoperability |
|---|---|---|---|
| **Governance** | UK Data Protection Act 2018 / UK GDPR; International data transfer adequacy decisions; Bilateral agreements (e.g., UK-US Data Bridge) | Accountability and fairness follow the data; Lawful, fair, and transparent processing; Purpose limitation. | Sufficient alignment with EU GDPR; Adherence to OECD privacy principles for mutual recognition |
| **Security** | Legally binding contractual clauses; Binding Corporate Rules (BCRs); Adherence to ISO/IEC 27001 | Data protection by design and by default; Confidentiality, integrity, and availability of data; Verifiable security posture | International incident response cooperation (CSIRT-to-CSIRT); Mutual recognition of security certifications |
| **Technical/Data** | Participation in ISO/IEC JTC1 (SC27 & SC38); Development of privacy-enhancing technologies (PETs) | Machine-readable consent and privacy preferences; Data provenance and lineage tracking; Secure cloud interoperability | Shaping global standards for cloud computing, distributed systems, and security techniques |

**Table 5:** Interoperability Matrix for AI in Cross-Border Data Flow

In the Interoperability Matrix, each layer is mapped to the sectors, illustrating, for example, that: - Governance: UK AI Principles unify AV regulation, EdTech guidance, and Data transfer policy (with international alignment via OECD/GPAI).

**Security:** Risk management and incident reporting are mandated in AV (new regulator roles) and mirrored by safeguarding policies in education and cybersecurity requirements in data flows (with international cooperation via networks like GPEN and CERTs).

**Technical/Data:** Standards like PAS and ISO for AV, interoperability standards for learning tools in education, and the use of standard data transfer tools (SCCs/IDTA, encryption protocols) in data flows show consistency in the UK's approach to ensuring technical compatibility and data integrity across domains.

**Interoperability Strengths and Gaps:** The UK's multi-layered approach generally fosters coherence. Regulators across sectors talk to each other (for instance, the Digital Regulation Cooperation Forum, which includes ICO, Ofcom, and CMA, has started considering AI oversight holistically). This cross-regulator dialogue helps share best practices (education might learn from AV's safety case approach, while AV might learn from data protection's accountability mechanisms).

The UK's emphasis on global standards is a strength, making its domestic rules outward compatible. A potential gap is that not all sectors have mature technical standards (education AI is still somewhat nascent in standards compared to AV).

Another challenge is ensuring that cross-border governance (which is more intergovernmental) keeps pace with technical interoperability (which industry often drives faster). For example, AI models can be trained across federated networks beyond jurisdictional boundaries, and governance needs to catch up with these innovations.

By maintaining a strong presence in the technical standardisation bodies and the international policy arena, the UK increases the chances that its domestic safety measures in all layers will interoperate smoothly with those of other countries. This is a necessity for managing AI, which is a global phenomenon. The following section will consolidate the UK's profile, summarising these findings dimension-by-dimension and highlighting sector case studies, good practices, and remaining priorities.

## 5. Jurisdiction Profile – United Kingdom

### 5.1 SNAPSHOT

The UK is a high-income, technologically advanced country with around 67 million inhabitants. It allocates substantial funding to AI and is ranked third worldwide for research and innovation. It hosts prominent events such as the 2023 Global AI Safety Summit and focuses on nurturing talent and developing infrastructure.

**Governance System:**
A unified state with devolved administrations, the UK manages AI governance mainly through national bodies such as DSIT and sector regulators. The AI Council and AI Security Institute provide expert advice. Regulation generally promotes a pro-innovative, flexible, and principles-based approach, relying on guidance rather than legislation, except in key cases.

**Key AI and Data Legislation/Policies:**
The UK's regulatory framework for artificial intelligence is shaped by the National AI Strategy (2021) and the AI Regulation White Paper (2023), both of which promote a "pro-innovation" and sector-specific approach. Rather than enacting horizontal legislation, the UK has opted to strengthen the capacity of existing regulators, such as the Information Commissioner's Office (ICO), Ofsted, and the Office for Safety and Standards in Transport, to interpret and apply five core AI principles: safety, transparency, fairness, accountability, and contestability.

This principles-based model is implemented on a non-statutory basis but is increasingly supported by institutional infrastructure, including:

- **AI Security Institute:** Founded in 2023 to assess advanced AI risks and guide policymakers on safety and interoperability.
- **Alan Turing Institute:** Offers public-sector AI advisory services and plays a role in developing international standards.

While the UK continues to assess the need for a dedicated AI law, current governance is anchored in existing legislation tailored to specific domains:

- **Data Protection Act 2018:** This Act regulates the lawful handling of personal data in AI systems, including the right to human review in automated decision-making.
- **Automated Vehicles Act 2024:** Establishes statutory safety thresholds, licensing requirements, and dual approval mechanisms for self-driving technologies.
- **Online Safety Act 2023:** Regulates harmful online content, including AI-generated misinformation and abuse.

This distributed, outcomes-oriented model enables the UK to respond rapidly to emerging AI challenges while maintaining regulatory coherence and international interoperability.

**International Alignment:**
The UK actively develops and follows international AI and data frameworks, providing an alternative to EU-style standards with US-style flexibility to avoid the drawbacks of both. Digital trade agreements include AI considerations, and the UK encourages global collaboration on frontier AI risks.

**Public Sentiment and Adoption:**
The UK public displays cautious optimism towards AI, supporting its use in areas like healthcare if properly regulated, but raising concerns about employment and bias. Government initiatives focus on transparency and building trust, shaped by past incidents like the 2020 exam algorithm controversy. The UK is widely recognised as a leader in AI regulation and innovation, aiming to balance progress with safety and public confidence.

**Conclusion**
The UK's approach in 2025 balances promoting innovative AI development with maintaining strong governance, adaptable regulation, and international leadership. By investing in talent, infrastructure, and global partnerships while upholding a robust ethical and legal framework, the UK seeks to keep its AI ecosystem dynamic, trustworthy, and in line with the public interest.

## **5.2** DIMENSION-BY-DIMENSION FINDINGS

**Governance & Institutions:** The UK employs principle-based coordination for AI governance, with regulators utilising domain expertise and central oversight through dedicated offices. Although principles are currently non-statutory, there is a willingness to formalise them if necessary. Devolved regions generally follow the national guidance.

**Ethical & Legal Framework:** AI regulation in the UK upholds core values such as democracy and human rights. Existing legislation addresses privacy and discrimination, with new rules (e.g., AV Act) bridging gaps. The UK leads in AI ethics through institutions and soft law. However, rapidly developing areas like generative AI pose unresolved legal challenges.

**Technical Standards & Infrastructure:** Investment in AI technical standards and strong digital and data infrastructure supports safe innovation. Initiatives like the AI Standards Hub and regulatory sandboxes help shape standards and enable practical adoption. Ongoing efforts are necessary to involve SMEs and expand standards across all sectors.

**Operational & Organisational Capacity:** UK regulators and industry demonstrate strong AI expertise, backed by academic talent and specialised skills development programmes. Local authorities require ongoing support to adopt AI responsibly at the frontline. While national capacity remains high, continuous skill enhancement and diverse recruitment are needed.

**Data Governance & Privacy:** The UK maintains a mature data governance framework, balancing innovation with privacy through the ICO and high standards. Adjustments to data regulations are underway to promote growth while preserving protections, though some remain concerned about potential GDPR weakening.

**International Cooperation & Alignment:** The UK actively shapes global AI policy, hosts major summits, and aligns with international standards. It manages relationships with the EU, US, and China and advocates for worldwide transparency and capacity-building.

**Liability & Accountability:** Accountability is clear through sector-specific guidelines and adaptable legal frameworks. Principles such as contestability and explainability are incorporated into law and guidance. Ongoing efforts address liability for diffuse or emerging AI harms.

**Overall Assessment**
The UK's approach to AI safety integrates strong governance, ethics, technical standards, and international collaboration. Its adaptable and inclusive framework positions it well to handle emerging risks while maintaining trust and interoperability.

## 5.3 SECTOR CASE STUDIES HIGHLIGHTS

This section summarises practices and challenges in key UK sectors:

**Autonomous Vehicles:** The UK's legislative process for AVs, including early adoption of BSI standards and extensive stakeholder consultation, has set a strong example in adaptive governance. The AV Act's new liability and safety system is seen as innovative and could influence other countries. Implementation remains challenging, particularly around licensing, enforcement, and public trust.

**Education (AI in Schools):** The 2020 exam algorithm issue raised awareness and prompted an updated DfE policy emphasising human oversight, risk management, and teacher involvement. The UK prioritises AI as a tool to support, not replace, educators, a stance now adopted internationally. Pilot audits and transparency measures enhance accountability, but school resource gaps remain. Centralised training and potential accreditation schemes are being developed to ensure the safe use of AI.

**Cross-Border Data Flows:** The UK balances data sharing with privacy by securing agreements and regularly reviewing transfer regimes. The UK-US Data Bridge and work on the OECD Declaration demonstrate leadership in protecting rights while facilitating data flows. Ongoing challenges include adapting to shifting geopolitics and EU requirements. Maintaining strong, interoperable frameworks is essential for supporting AI development.

## 5.4 GOOD PRACTICES AND POLICY PRIORITIES

Good Practices Identified:

**Stakeholder Engagement and Expert Input:** In all sectors, the UK's practice of involving experts, industry, and the public in policymaking, such as Law Commission consultations for AV, calls for evidence for AI in education, and expert councils for data, has resulted in more robust and widely accepted outcomes. This inclusive approach acts as a model of good governance.

Principles-Based, Adaptive Regulation: The five AI principles adopted by the UK provide a transparent yet flexible ethical compass across sectors. Regulators tailoring these to context (with central coordination ensuring consistency) are proving effective. The willingness to iterate, start with guidance, and move to more complex regulation if needed is a pragmatic practice that other countries note.

**Standards and Assurance-Led Approach:** The UK's broad utilisation of technical standards (through BSI and international bodies) and assurance mechanisms (such as sandboxes, audits, safety cases) supports practical enforcement of AI safety without impeding innovation. For instance, requiring a safety case for AV trials (PAS 1881) or algorithmic transparency reports in local government are non-legislative measures but guarantee accountability through processes.

**International Thought Leadership:** The UK often pilots ideas that eventually gain wider acceptance. The Children's Code is shaping age-appropriate design standards across the globe; the concept of a regulatory sandbox for AI is now discussed in OECD forums, partly inspired by the UK's example; and the Bletchley Park Summit was the first of its kind, focused on frontier AI safety, which has contributed to creating a model for international AI risk governance. Sharing these initiatives openly is a positive way to spread norms.

**Cross-Sector Collaboration:** UK agencies are increasingly collaborating on AI issues. The Digital Regulation Cooperation Forum (which includes ICO, Ofcom, CMA, and FCA) exemplifies joint efforts to tackle AI challenges related to data, content, and competition. It has already resulted in joint statements on online algorithms and experiments in co-regulation.

**Public Transparency:** The UK government and regulators publish quite a lot. For instance, the ICO publishes its reasoning on adequacy decisions, the DfT provides detailed scoping notes on the AV Act, and the government issues an algorithm transparency standard for the public sector. This openness fosters trust and enables external scrutiny to improve systems, a hallmark of good governance.

### Policy Priorities Going Forward:

Building on current progress, the following priorities emerge for the UK to address in the near future:

**Consolidating the AI Governance Framework:** As the AI White Paper principles advance to the implementation phase, ensure that each sector regulator has the necessary resources and guidance to apply them consistently. By late 2025, the UK will likely assess whether a more formal overarching AI legislation is required (not necessarily an AI Act like the EU, but possibly a statutory basis for the principles or an oversight body). The goal is to prevent any gaps or inconsistencies in regulation, especially as new AI applications emerge.

**Frontier AI Risk Management:** The UK should continue investing in research and monitoring of advanced AI (AGI-like systems). A key policy priority is establishing evaluation infrastructure for frontier models, testing them for dangerous capabilities, robustness, and other factors, possibly in secure facilities. The AISI and international partners can lead this effort. Ensuring the UK maintains oversight of cutting-edge AI safety will protect its society and enhance its credibility in shaping global rules. This might include scenario planning for AI-related catastrophic risks and improving forecasts of AI development.

**Education and Workforce Upskilling:** To ensure the safe use of AI, the general workforce and the public need a better understanding of AI. The UK should incorporate AI literacy into curricula, extending beyond computer science classes to include general digital citizenship lessons. It should also broaden programmes to reskill workers likely impacted by AI to prevent social harm caused by job displacement. AI safety involves technical issues and socio-economic impacts that require management. The key priorities are twofold: first, to reduce any adverse effects of AI on employment through training and transition support; second, to enhance expertise in AI safety via educational incentives, such as more MSc programmes on AI ethics and scholarships for underrepresented groups to diversify the field.

**AI in Public Services Exemplars:** The UK can set examples by applying AI in public services safely and ethically. For instance, deploying AI assistants in the NHS for preliminary diagnoses or administrative tasks, but under strict evaluation and with patient consent, could improve services and demonstrate good practice. Another area is smart infrastructure, such as traffic management AI integrated with autonomous vehicles. The policy priority is implementing high-profile public-sector AI projects demonstrating transparency, efficacy, and citizen benefits, while building public trust and setting industry benchmarks.

**Strengthening International Coalitions:** The UK should strengthen its leadership on AI safety by developing a comprehensive interoperability framework that minimises compliance fragmentation while promoting globally aligned safety outcomes. Building on the momentum of the Bletchley Declaration and utilising the OECD AI Principles framework, the strategy focuses on establishing sustainable institutional mechanisms for technical standards harmonisation and mutual recognition across borders.

The UK should institutionalise Bletchley's legacy through multilateral engagement by transforming the Bletchley process into a permanent "Frontier AI Interoperability Forum" with shared technical workstreams on model evaluations and safety benchmarks (UK Government, 2025). This provides an institutional basis for ongoing cooperation beyond temporary summits. At the same time, the UK must utilise Japan's GPAI 2025 presidency to co-lead practical implementation initiatives, building on the G7's June 2025 commitment to "human-centred development and use of safe, secure and trustworthy AI" (G7, 2025). The UK should lead in developing common AI assurance tools and cross-jurisdictional evaluation frameworks, positioning itself as the bridge between technical standards development and global governance frameworks.

The UK needs to promote convergence in technical standards and cross-border infrastructure among the NIST AI Risk Management Framework, EU AI Act harmonised standards, and the OECD's updated AI Principles, which offer "practical and flexible guidance for policymakers and AI actors" (OECD, 2024). The OECD framework, adopted by over 70 jurisdictions with more than 1000 policy initiatives aligned with its principles, provides a foundation for interoperability (OECD, 2024). The UK AI Authority should produce clear cross-walks showing how domestic guidance relates to these international frameworks, supporting firms' export strategies while maintaining the UK's principles-based regulatory approach (UK Government, 2025).

**Monitoring and Evaluation of Policies:** Finally, the UK must continuously assess whether its AI safety measures are effective. Establishing measurement frameworks, such as KPIs like reducing AI-related incidents, public trust levels, and innovation indices, will aid in adjusting policies. The new central monitoring function promised in the white paper needs to be implemented, and annual reports on AI governance should be submitted to Parliament. Such accountability will sustain momentum and political focus on AI safety (preventing it from being overshadowed by solely pro-innovation narratives). By prioritising these areas, the UK can preserve and strengthen its comprehensive approach, ensuring that AI develops safely, ethically, and beneficially across all sectors of society.

## 6. Conclusion – Consolidated Recommendations

The United Kingdom plays a significant role in developing and regulating artificial intelligence. Its national strategies and regulatory frameworks carry significant influence and could shape emerging international standards. This report offers a comprehensive analysis of the UK's approach to AI safety and governance, along with recommendations from an international viewpoint. The emphasis is on creating an AI ecosystem that is safe, innovative, and compatible with global partners. Achieving this is vital to avoid regulatory fragmentation, defend international human rights standards, and ensure that AI development aligns with the Sustainable Development Goals (SDGs).

**Advancing Harmonised Global Standards and Frameworks**
Establishing coherent, international standards ensures that AI technologies are developed and deployed safely and ethically across borders. The United Kingdom should leverage its position to advance this goal.

**Proactive Leadership in Multilateral Forums: T**he UK should strengthen its leadership role within key international standard-setting bodies, including the UNECE on Autonomous Vehicle (AV) safety and the OECD and GPAI on data governance and AI risk management. By sharing its domestic regulatory models, such as its comprehensive AV liability and safety framework, the UK can help shape global best practices based on public safety and accountability.

**Promotion of Interoperable Technical and Ethical Norms:** The UK should persist in aligning its national standards with established and emerging international benchmarks, such as the ISO 42001 AI Management System standard. This ensures that UK-based industries develop AI systems compliant by design, facilitating seamless integration into the global marketplace and supply chains.

**Establishment of a Coherent National Entity for International Engagement:** To effectively engage with UN bodies and international partners, the UK should formalise its proposed "AI Coordination Council." Such a central body can provide a unified national perspective, ensuring that the UK's contributions to global AI governance are consistent, strategic, and impactful.

**Constructing Trusted Mechanisms for Data and Technology Exchange**
Secure and unrestricted data flow is essential for the global AI ecosystem, but it must be balanced with the right to privacy. Therefore, the UK's strategy should focus on establishing trusted, rights-respecting pathways for international data and technology transfer.

**Expansion of High-Standard Data Transfer Agreements:** The UK should continue building a network of international data transfer mechanisms, such as its "data bridges" and adequacy arrangements. Active engagement in multilateral frameworks like the Global Cross-Border Privacy Rules (CBPR) Forum is essential for establishing a scalable and interoperable system that fosters innovation while safeguarding personal data.

**Upholding Internationally Recognised Data Protection Principles:** The credibility of the UK's data bridges depends on the strength of its domestic data protection regime. Any legislative reforms must uphold apparent equivalence with strong international standards, particularly the principles outlined in the EU's GDPR. This is vital for maintaining the fundamental right to privacy (as expressed in Article 12 of the UDHR) for all individuals whose data is processed.

**Development of Interoperable Digital Public Infrastructure:** Investing in digital infrastructure, such as standardised digital mapping of road regulations for AVs, is vital for increasing global interoperability. These initiatives support SDG 9 (Industry, Innovation and Infrastructure) and help break down barriers to international technology deployment, fostering a competitive and cooperative global market.

**Strengthening Domestic Foundations for Global Leadership**
A country's ability to lead internationally in technology governance depends on the strength and integrity of its domestic ecosystem. A strong national framework based on research, education, and public trust provides the basis for credible global engagement.

**Investment in AI Safety and Alignment Research:** By significantly increasing investment in AI safety research, the UK can provide essential public knowledge to the international community on managing and reducing risks linked to advanced AI. This supports a global effort to ensure AI development remains beneficial to humanity.

**Operationalising Universal Ethical Principles in Sectoral Governance:** Creating clear domestic Codes of Practice, such as those in the education sector, provides a practical model for turning universal ethical principles into action. A framework that ensures algorithmic transparency and fair access to AI tools in education directly supports SDG 4 (Quality Education) and SDG 10 (Reduced Inequalities), offering a helpful blueprint for other member states.

**Cultivating a Whole-of-Society Consensus on Responsible AI:** The long-term success of AI governance relies on public trust. By encouraging a national dialogue on AI's societal impacts and fostering a culture of safety and ethics within its industry, the UK can establish the whole-of-society consensus required for the responsible management of this transformative technology. This domestic legitimacy is crucial for effective international leadership.

**Conclusion**

By systematically applying this three-pronged approach, advancing global standards, building trusted data links, and strengthening internal foundations, the United Kingdom is well-positioned to navigate the complexities of AI governance. In doing so, the UK would safeguard its own economic and societal interests while significantly contributing to the United Nations' collective goal: guiding artificial intelligence towards a future that promotes international peace, sustainable development, and universal human rights.

**1.1 UK Definition of AI Safety:** AI Safety in the United Kingdom refers to the set of principles, processes, and technical measures designed to ensure that artificial intelligence systems, especially advanced and general-purpose AI, are developed, deployed, and operated in ways that are safe, reliable, and do not cause harm to individuals or society.

This includes:

**Preventing unintended or harmful outcomes from AI systems**, whether through technical failure, misuse, or adversarial attack.

**Ensuring systems are robust, secure, and resilient** against manipulation, cyber threats, and operational failures.

**Embedding human-centric values** such as fairness, transparency, accountability, and respect for human rights into AI design and governance.

**Rigorous testing and evaluation** of AI models before and after deployment, including state-led and independent safety assessments, as highlighted by the AI Safety Summit and the International AI Safety Report 2025

**Compliance with legal and ethical standards**, including the UK's Secure by Design principles ('Secure by Design Principles', 2025), the Cyber Security Standard ('The Cyber Security Standard', 2025), and sector-specific regulations

**International cooperation** to align safety standards and risk mitigation approaches, as reflected in the Bletchley Declaration (AI Safety Summit 2023, 2025) and ongoing work with the OECD, EU, and United Nations

The International AI Safety Report 2025 further clarifies that AI safety is about "identifying risks and evaluating methods for mitigating them," focusing on both technical and societal risks from advanced AI systems. The report aims to create a shared international understanding of what constitutes safe AI and how to achieve it. (DSIT/AISI, 2025)

The AI Playbook for the UK Government states that all government AI technologies and services must be "secure and resilient," and that "the AI solution is safe, lawful and compatible with other ethical and legal requirements" (GDS, 2025)

Key Elements in the UK's Approach

**Safety = Prevention of Harm:** Avoiding risks to individuals, society, and critical infrastructure.

**Security = Protection from Threats:** Defending AI systems from cyber-attacks, misuse, and manipulation.

**Assurance = Evidence-Based Evaluation:** Using scientific, technical, and ethical assessments to demonstrate trustworthiness.

**Governance = Oversight and Accountability:** Ensuring clear lines of responsibility and compliance with evolving standards.

COUNTRY REPORTS

# Republic Of Korea Country-Level AI Safety Interoperability Report

HAG-MIN KIM, WENSHUAI SU, KYUNGWON KIM AND MINYU JIANG
DEPARTMENT OF INTERNATIONAL BUSINESS AND TRADE, KYUNG HEE UNIVERSITY,
SEOUL, SOUTH KOREA, EDOCTOR@KHU.AC.KR

# Republic of Korea Country-Level AI Safety Interoperability Report

15 October 2025

## 1. Introduction

### 1.1 PURPOSE AND AUDIENCE

Artificial intelligence increasingly operates across organizational and jurisdictional boundaries. For the Republic of Korea, realizing the benefits of AI while managing its risks requires treating governance, safety, and interoperability as a single, coherent assurance stack (ISO & IEC, 2023). Adopting an international-interoperability lens, this report evaluates Korea's current position and outlines practical steps for aligned, trustworthy deployment in priority sectors.

In this report, three interdependent pillars frame the development and deployment of AI systems:

- **AI safety governance:** the frameworks, policies, and operating practices that ensure AI is developed, deployed, and maintained in a safe, reliable, and ethical manner, minimizing risks and preventing harm to individuals and society (Jobin et al., 2019; Lee et al., 2021; Tabassi, 2023; OECD, 2019/2024).
- **Security:** protection of AI systems and data across the lifecycle (confidentiality, integrity, availability), preventing unauthorized access, attacks, misuse, or model/algorithm tampering (National Cyber Security Centre [NCSC], 2022; Tabassi, 2023).
- **Interoperability:** the ability of heterogeneous systems, tools, and components to work together seamlessly through shared protocols, interfaces, and data models, enabling communication, data exchange, and cohesive operation (Zeng, 2019; Berg, 2024).

AI safety refers to the practices, standards, and governance mechanisms ensuring that AI systems operate reliably, securely, and in alignment with human values—minimizing unintended harm, misuse, and systemic risks. These pillars reinforce one another: credible governance embeds robust security controls and relies on interoperability to scale consistently across platforms and jurisdictions; secure and interoperable implementations, in turn, make governance auditable and enforceable.

**Purpose**

This report assesses the Republic of Korea's AI safety regulatory framework through the lens of international interoperability. It benchmarks current laws, standards, and practices; identifies practical frictions at legal/organizational, technical/interface, and data/semantic layers; and proposes actionable pathways to strengthen assurance and cross-border consistency. The analysis focuses on three high-impact sectors: autonomous vehicles, education, and cross-border data flows.

| Criteria | AI Safety Governance | Security | Interoperability |
|---|---|---|---|
| **Focus** | Ethical and trustworthy AI use | Protection from threats | Seamless cooperation across systems |
| **Objective** | Prevent harm and ensure accountability | Safeguard data and integrity | Enable shared operation via standards |
| **Mechanisms** | Laws, guidelines, ethics codes | Encryption, access control, monitoring | APIs, data formats, standardization |
| **Stakeholders** | Policymakers, developers, ethicists | Cybersecurity experts, IT staff | Regulators, engineers, standards bodies |
| **Challenges** | Balancing innovation and oversight | Keeping pace with evolving risks | Aligning technical and legal norms |
| **Policy Role** | Build public trust | Maintains resilience | Enables cross-sector alignment |
| **Examples** | AI Basic Act (2025) | Cybersecurity Act | APEC CBPR, ISO/IEC 42001 (2023) |

**Table 1:** Comparison of AI Safety, Security and Interoperability - *Source: Author's compilation based on OECD (2019/2024) and ISO/IEC (2023).*

**Audience**

The report is intended for policymakers and regulators, industry leaders and operators, standards, testing, and certification bodies, academia and ethics experts, and international organizations and partners—providing a common vocabulary, comparable reference points, and cooperation opportunities to inform policymaking, implementation, and international alignment.

## 1.2 SCOPE

This analysis examines South Korea's AI safety and interoperability landscape across three priority sectors 'autonomous driving, education, and cross-border data flows' through a comparative assessment of legal and ethical frameworks, technical standards and infrastructures, operational readiness, data governance practices, data quality and safety measures, implementation progress, and oversight mechanisms.

The evaluation draws on Korea's key legal and policy instruments, including:

- Act on the Promotion and Support of Commercialization of Autonomous Vehicles (Act No. 20391).
- Basic Act on AI
- Road Traffic Act (Act No. 20677).
- The 7th Master Plan for Educational Informatization (2024–2028)
- General Data Protection Regulation (GDPR), Regulation (EU) 2016/679, art. 45, 2016 O.J. (L 119) 1 (EU).

### 1.3 AI Safety Taxonomy across Three Priority Sectors

To operationalize the concept of AI safety, this report proposes a taxonomy that distinguishes safety dimensions across three priority sectors: autonomous vehicles, education, and cross-border data flows. While the underlying principles of reliability, fairness, accountability, and privacy are shared, sector-specific manifestations of safety require differentiated attention.

#### 1.3.1 Autonomous Vehicles

Functional safety focuses on dependable sensors, perception, planning, and control so that vehicles avoid hazardous behaviour under normal and degraded conditions. Cybersecurity protection concentrates on resilient vehicle, edge, and cloud components with secure communications and update pathways that withstand adversarial interference. Human machine interaction safety emphasizes interfaces that keep users situationally aware and provide understandable fallback and handover procedures. Testing and certification require rigorous pre-deployment and post-deployment validation aligned with international practice and transparent incident reporting and corrective action.

#### 1.3.2 Education

Content integrity and accuracy require safeguards that prevent biased, discriminatory, or harmful outputs in tutoring, assessment, and feedback. Student data protection entails strict minimization, pseudonymization, secure storage, controlled access, and auditability across the full data lifecycle. Teacher oversight and pedagogical safety ensure that AI augments rather than replaces professional judgment, preserving accountability and learning quality. Equitable access and inclusivity promote designs that serve diverse learners and accessibility needs, preventing technology from widening achievement gaps.

#### 1.3.3 Cross-Border Data Flows

Data security and confidentiality prioritize strong encryption, transfer monitoring, and verified controls when information moves across jurisdictions. Regulatory alignment and compliance safety reduce uncertainty through compatibility with leading frameworks and certification mechanisms that support trusted exchange. Transparency and accountability require traceable processing, meaningful explanation of automated decisions, and clear allocation of responsibilities for harms. Interoperability safeguards integrate technical standards and legal instruments so data exchange remains seamless while upholding ethical and rights-preserving requirements.

This taxonomy supplies a common structure for prioritizing controls, allocating responsibilities, and measuring progress. It supports consistent governance within each sector while enabling cross-sector learning and international interoperability.

## 1.4 ANALYTICAL LENS

**Tripartite analysis**

The report examines domestic laws and policies alongside international norms (e.g. GDC principles, UN guidelines) and bilateral/multilateral linkages such as the Global Cross-Border Privacy Rules (CBPR) Forum and the EU–Korea Digital Trade Agreement. This combined lens highlights how Korea's framework operates both internally and in global context. The analysis emphasizes Korea's unique institutions and practices and compares them with international standards to identify where they align and where gaps or divergences exist.

### **1.5** METHODOLOGY

This study adopts a mixed qualitative approach combining a scoping review of legislation, standards, and policy documents with expert interviews. The document review covered all relevant instruments in force as of 7 August 2025, supplemented by recent news reports to capture the latest policy developments. Policies were systematically coded according to five analytical criteria: (i) policy objectives, (ii) guiding principles, (iii) binding nature, (iv) key components, and (v) linkages to international frameworks.

A sector-specific analysis was conducted for autonomous vehicles, education, and cross-border data flows, tracing major policy changes and implementation milestones within each domain. To complement the documentary analysis, semi-structured interviews were conducted with key stakeholders. Insights from these interviews were triangulated with documentary evidence to ensure analytical robustness and contextual accuracy.

### **1.6** RESEARCH QUESTIONS

The report addresses three core research questions:

**RQ1.** How effective are Korea's AI safety governance, security, and interoperability frameworks in both design and operation across the three priority sectors—autonomous vehicles, education, and cross-border data flows?

- 1a (design): What laws, institutions, standards, and assurance processes are in place?
- 1b (operation): How are they implemented through testing and certification processes?
- 1c (effectiveness): What evidence exists of risk reduction and trustworthy deployment?

**RQ2.** Where do strengths and vulnerabilities appear across the legal/organizational, technical/interface, and data/semantic layers in each sector, and what are the most material risk scenarios including security threats?

**RQ3.** To what extent are Korea's policies aligned and interoperable with UN's frameworks GDC and two UN resolutions (ISO/IEC/IEEE standards, Global CBPR/PRP), and are there pathways to mutual recognition?

- 3a: What is the degree of conformance/maturity by pillar (governance, security, interoperability)?
- 3b: Where do incompatibilities or missing mappings block cross-border operations?

## 2. Comparative Assessment Framework

Korea's approach to AI safety governance can be characterized by the following key features.

**Objective:** Accelerate innovation and build public trust in AI while safeguarding public safety and sustaining economic competitiveness.

**Principles & Values:** Emphasize people-centric design, transparency, accountability, fairness, human dignity, and privacy protection in AI systems and policies Jobin et al., 2019; OECD, 2019/2024.

**Approach:** A predominantly top-down regulatory model driven by national legislation (e.g. the new AI Basic Act, autonomous vehicle laws, the Personal Information Protection Act), complemented by hybrid mechanisms that solicit stakeholder input for guidelines and standards.

**Binding Nature:** A coexistence of strict, legally binding statutes and more flexible, non-binding ethical guidelines. For example, statutory requirements for data protection run in parallel with advisory ethical codes for AI developers.

**Components:** A comprehensive framework combining legal and ethical norms, technical standards, operational capacity building, certification and oversight systems, and data governance frameworks.

**International Linkages:** Korea's international positioning is framed in terms of United Nations instruments. The Global Digital Compact, adopted as an annex to the Pact for the Future in September 2024, commits Member States to responsible and interoperable data governance and advances consultations to facilitate trusted cross-border data flows; these commitments provide a UN anchor for aligning domestic arrangements such as GDPR adequacy and participation in CBPR with multilateral norms on data and AI governance (United Nations General

Assembly, 2024, A/RES/79/1). At the General Assembly level, Resolution 78/265 on seizing the opportunities of safe, secure and trustworthy AI systems sets out principles that emphasize human rights, risk management, capacity building and internationally interoperable safeguards, which are directly relevant to Korea's sectoral interoperability agenda (United Nations General Assembly, 2024, A/RES/78/265). Building on the GDC, Resolution 79/325 in August 2025 establishes the modalities for an Independent International Scientific Panel on AI and a Global Dialogue on AI Governance, creating UN processes through which sector-specific testing, safety and recognition initiatives can be compared and harmonized across domains including autonomous vehicles, education and cross-border data flows.

## 3. Benefits and limitations of interoperability in different areas of digital regulation

The following sections review three key sectors – autonomous vehicles, education, and cross-border data flows – to identify how interoperable Korea's digital regulations are with international norms, highlighting benefits and limitations in each domain. An interoperability matrix summarizing the assessment across dimensions is provided at the end of this section.

### **3.1** AUTONOMOUS VEHICLES

**Legal and ethical framework**

Korea has established an evolving legal foundation for autonomous vehicles. The 2019 Act on the Promotion and Support for the Commercialization of Autonomous Vehicles (National Assembly of the Republic of Korea, 2019) and the 2020 Autonomous Vehicle Management Act created a temporary permit regime and a legal basis for paid autonomous transport services (MOLIT, 2024). A March 2024 amendment to the Road Traffic Act (effective March 2025) further requires drivers of partially autonomous cars to undergo safety training (National Assembly of the Republic of Korea, 2025). In January 2025, Korea enacted an AI Basic Act (to be enforced from January 2026) which defines certain "high-impact AI" systems (including autonomous driving systems) and mandates that foreign AI service providers appoint a local representative in Korea (Ministry of Science and ICT, 2024). The AI Basic Act will also establish a dedicated AI Safety Research Institute. On the ethical side, the Ministry of Land and Transport issued non-binding Ethical Guidelines for Autonomous Vehicles in December 2020, emphasizing principles of safety, accountability, and transparency (MOLIT, 2020). These laws and guidelines collectively aim to ensure that autonomous driving technologies develop with public safety in mind, though the ethical guidelines are voluntary and the new AI Basic Act provisions (such as high-risk AI requirements) are not yet in force.

**Safety and ethics considerations**

While Korea's legal and policy frameworks address functional and regulatory requirements, safety and ethical concerns remain central to the autonomous driving agenda. Safety risks such as sensor failures, algorithmic misjudgments, or malicious cyberattacks highlight the importance of redundant safety mechanisms, secure vehicle-to-cloud communication, and real-time monitoring systems. Ethically, the guidelines stress that autonomous driving systems should prioritize human life, fairness in accident scenarios, and transparency in decision-making processes. For instance, ethical discussions in Korea have emphasized how vehicles should be programmed to respond in unavoidable accident situations, ensuring decisions are made in ways that align with human dignity and fairness principles. Furthermore, as these vehicles become "moving data hubs," ethical considerations around privacy, data ownership, and the potential misuse of biometric or location data require stronger safeguards. Linking domestic initiatives with global frameworks such as the OECD AI Principles (2019) and the UNESCO Recommendation on the Ethics of AI (2021) helps reinforce Korea's commitment to embedding safety and ethics into the governance of autonomous vehicles (OECD, 2019; UNESCO, 2021).

**Technical standards**

Korea is actively pursuing technical standardization and infrastructure to support autonomous driving. The Third Automotive Policy Master Plan (2022–2026), along with a national mobility innovation roadmap, targets commercial rollout of Level 4 autonomous buses and shuttles by 2025 and Level 4 passenger vehicles by 2027. These plans set ambitious goals (e.g. aiming for half of all new vehicles sold by 2030 to be Level 4 or above) and call for nationwide deployment of enabling infrastructure such as high-resolution mapping, 5G/V2X communication networks, and dedicated autonomous vehicle lanes on roads (Ministry of Science and ICT, 2023). Korea also aligns its vehicle standards with international benchmarks; for instance, it works with the UN Economic Commission for Europe's WP.29 on vehicle regulations, although full harmonization with UNECE safety standards for autonomous vehicles remains underway. As for testing protocols, by 2023 over 1,400 autonomous test vehicles had been issued temporary permits. In November 2023, MOLIT(Ministry of Land, Infrastructure and Transport) expanded designated pilot testing zones, reaching a total of 34 pilot zones covering all 17 provinces and major cities. This extensive pilot zone network demonstrates Korea's commitment to providing a controlled environment for technical validation and standard convergence of autonomous driving technologies (UNECE, 2020; MOLIT, 2024).

**Operational capacity**
Korean authorities and industry have made significant investments in autonomous driving pilots and infrastructure. Demonstrations: Autonomous shuttles and robo-taxis are already operating in selected districts of Seoul (e.g. Sangam and Gangnam) and in smart-city testbeds like Pangyo. Other cities such as Sejong, Jeju, and various localities are running pilot programs for self-driving public transport. On highways, freight platooning trials allow convoys of autonomous trucks to coordinate movements and braking, showcasing advanced vehicle-to-vehicle interoperability. Regulatory sandboxes and the K-City test site (a dedicated autonomous driving proving ground) enable start-ups and corporations to experiment with new mobility services under relaxed regulatory conditions. Investment: The government plans to invest roughly KRW 4 trillion (around US$3 billion) in supporting infrastructure, including high-precision digital maps and cooperative intelligent transportation systems. This strong public investment in R&D and infrastructure boosts operational readiness and signals long-term commitment to autonomous mobility. Overall, Korea's extensive network of pilot zones and test vehicles illustrates a high operational capacity, though scaling up nationwide deployment will require ongoing training, public awareness, and maintenance of infrastructure (Seoul Metropolitan Government, 2021).

**Data governance and liability**
Interoperability in data and safety governance is addressed through specific requirements. Event recorders and cybersecurity: All autonomous vehicles in Korea must be equipped with event data recorders ("black boxes") to log driving data (National Assembly of the Republic of Korea, 2024a). Manufacturers and service operators are also required to obtain cybersecurity certification and conduct real-time monitoring of vehicle systems to guard against cyberattacks – a practice aligning with emerging international standards (such as ISO 21434 for automotive cybersecurity). Liability allocation: Korea has begun to clarify liability in the event of accidents involving self-driving cars. An amendment to the Guarantee of Automobile Accident Compensation Act (2020) assigns primary liability for accidents to the vehicle's owner(The term "motor vehicle owner" means a person who owns a motor vehicle or a person entitled to use a motor vehicle, and who operates the motor vehicle for personal use), even if the vehicle is operating autonomously, while also allowing recourse against manufacturers or software providers if a defect in the vehicle or its algorithms contributed to the incident (ATIC Global Vehicle Regulation Research Department, 2025; National Assembly of the Republic of Korea, 2024b; National Assembly of the Republic of Korea, 2025). This shared liability model is one of the early attempts globally to legislate accident responsibility in mixed human–AI driving contexts. Remaining challenges include refining these rules for higher levels of autonomy (Levels 4–5 where a human driver may not be present) and ensuring insurance products and legal frameworks adapt accordingly.

**AI risks and challenges**
In autonomous vehicles, material risks involve system and perception failures under edge or degraded conditions that can produce unsafe behavior, adversarial and cyber-physical threats against connected vehicles and over-the-air updates, human–machine interaction hazards including automation complacency and unclear handover, and complex liability and data-governance exposure as continuous location and biometric collection crosses jurisdictions; responsibility tracing becomes harder when software updates or map data originate abroad, which calls for clearer cross-border legal mappings and evidence chains.

**Complex liability**
Determining responsibility for crashes or violations is complex when multiple parties (drivers, vehicle owners, automakers, and software developers) are involved in an autonomous driving ecosystem. Clear interoperability of legal frameworks (domestic and foreign) is needed when software updates or map data come from international sources.

**Data privacy**
Autonomous vehicles generate continuous data (location, sensor recordings, even biometric data of occupants). Rules for cross-border data transfer and use of this sensitive data need clarification, especially if data is processed on cloud servers outside Korea. Privacy protections must keep pace as vehicles become "moving data hubs" across jurisdictions.

## **3.2** EDUCATION

**Legal and ethical framework**
Korea has embarked on an ambitious but contentious initiative to introduce AI-enabled digital textbooks in education. Starting in March 2025 (The Straits Times, 2025; World Bank, 2024), the Ministry of Education began piloting AI-powered digital textbooks in select grades and subjects (English and mathematics for 3rd–4th grade in primary school; English, math, and computer science in secondary school). These textbooks incorporate generative AI to personalize learning content and feedback for each student, and they include features like real-time captions and translations to improve accessibility for students with different needs. Initially, these AI textbooks were slated to become part of the official national curriculum.

However, on August 4, 2025, the National Assembly unexpectedly passed a law downgrading the status of AI digital textbooks from “official textbooks” to optional supplementary materials (Good e-Reader, 2024). This policy reversal means each school can now decide whether or not to use the AI textbooks, and centralized government funding for the program is no longer guaranteed. The sudden change – largely driven by public concern over untested AI in classrooms – threatens the viability of the hundreds of billions of won invested in developing the AI textbooks and has prompted legal challenges from educational publishers. It also highlights a governance issue, inconsistency in policy can undermine long-term interoperability and trust. On the ethical front, Korea has issued guidelines to protect student privacy and well-being in digital learning (MSIT & KISDI, 2023; Ministry of Education Implementation Plan, 2024). For example, the education authorities require that any student data used by AI systems be anonymized or pseudonymized, and that access be limited to authorized educational personnel. Child protection rules mandate that AI education tools avoid harmful content or bias (World Bank, 2024). These guidelines, aligned with the national Personal Information Protection Act (PIPA), emphasize that even as AI personalizes education, it must respect students’ rights and safety.

| Principle | Domestic Legal Basis | International Alignment |
|---|---|---|
| **Learner agency & diversity** | Framework Act on Education (Arts. 3, 12, 14) | UNESCO Recommendation on the Ethics of AI (2021) |
| **Transparency & explainability** | Personal Information Protection Act (PIPA) (Art. 3); Framework Act on Education (Art. 23) | OECD AI Principles (2019 / 2024) |
| **Privacy protection** | PIPA (Arts. 3, 5); Framework Act on Education (Art. 23-3) | UNESCO & OECD alignment |
| **Teacher professionalism & oversight** | Framework Act on Education (Art. 14) | UNESCO Recommendation on the Ethics of AI (2021) |
| **Challenges** | Balancing innovation and oversight | Keeping pace with evolving risks |
| **Policy Role** | Build public trust | Maintains resilience |
| **Examples** | AI Basic Act (2025) | Cybersecurity Act |

**Table 2:** Summary of Interoperable Principles and Legal Alignment

**Safety and ethics considerations**
Beyond regulatory frameworks, the introduction of AI-powered education tools raises significant safety and ethical concerns. From a safety perspective, risks include exposing students to biased, inaccurate, or potentially harmful content generated by AI, as well as over-reliance on automated feedback that could undermine pedagogical integrity. Ethical concerns focus on protecting children's rights, ensuring equitable access to high-quality education, and preventing algorithmic bias from disadvantaging specific groups of learners.

The question of how much decision-making should be delegated to AI versus teachers remains contentious, with many educators arguing that human oversight must remain central to maintain accountability and safeguard student development. In addition, the increasing amount of sensitive data collected by AI educational platforms, ranging from academic performance to behavioral patterns, raises ethical issues about privacy, data ownership, and the possibility of commercial misuse. To mitigate these concerns, Korea has introduced anonymization requirements, strict access control, and child protection rules, but gaps remain in terms of algorithmic transparency and accountability.

**Technical standards**
The use of AI in education is at an early stage, and Korea is laying groundwork for technical interoperability and standards. Personalization algorithms: The AI textbook platforms use algorithms to analyze each learner's performance, preferences, and learning style, thereby generating customized exercises and interactive feedback. While effective personalization can benefit students, ensuring these algorithms meet fairness and transparency standards is a challenge. Korea is exploring international standards (e.g., IEEE/ISO) for AI in education and has begun taking concrete steps: it is leading work in ISO/IEC JTC 1/SC 42, most recently securing approval of the ISO/IEC 5259-1:2024 data-quality standard for AI; KERIS is establishing technical standards, publishing development guidelines, and operating an accreditation/review system for AI digital textbooks; KATS is deepening cross-border standards coordination via the 2024 U.S.–Korea Standards Forum; and public bodies have started adopting ISO/IEC 42001 AI management systems (e.g., the National Tax Service and Samsung SDS), with applicability to education now being assessed.

**Data protection technologies**
On the backend, the cloud platforms hosting AI textbooks employ encryption and strict access controls. Student records are encrypted and stored with measures to prevent re-identification of personal data, and all access and activity are logged for auditability. By adopting such technical and administrative measures, Korea aims to align its data-protection framework with international norms (e.g., the EU's GDPR) to ensure interoperability, while prioritizing PIPA-based requirements in the education sector. However, technical standards for efficacy (ensuring the AI's recommendations are pedagogically sound and unbiased) are still in development, indicating a limitation in current interoperability, Rather than the absence of a benchmark, UNESCO (2023) articulates concrete, testable criteria for "safe and effective" AI in education—ethical/pedagogical validation across the lifecycle, bias audits and consent safeguards, and policy requirements for data privacy and age-appropriate use—providing a de facto international reference point for adoption decisions.

**Operational capacity**
The mixed reception of Korea's new AI textbooks illustrates both institutional progress and persisting operational gaps. During the first semester of 2025, about 30 percent of schools offered the AI textbooks chose to integrate them into their curricula (The Straits Times, 2025), signaling strong initial interest and willingness to experiment. However, a national survey revealed that 98.5 percent of teachers felt inadequately trained to use AI-based tools effectively (The Straits Times, 2025). This widespread lack of teacher readiness represents a critical implementation challenge, constraining the practical interoperability of AI technologies with classroom routines and pedagogical methods.

To address these gaps, the government allocated substantial resources to strengthen digital education capacity. Approximately USD 700 million (around KRW 760 billion) was earmarked over three years for upgrading smart classroom infrastructure and providing teacher training in digital and AI pedagogy (World Bank, 2024). These investments signify the government's commitment and are a positive factor in building capacity. These investments demonstrate clear policy commitment and lay an essential foundation for building long-term operational resilience and interoperability within the education sector.

**Data governance**
Managing student data and privacy in AI-enhanced education is a critical concern in Korea's regulatory framework. All learning and assessment records generated by AI textbooks must be stored in pseudonymized form, with Only the Ministry of Education and individual schools authorized to access identifiable information. Publishers or technology companies that have created AI systems are not allowed to access raw student data unless they receive prior authorization (World Bank, 2024). This arrangement aligns with national privacy law (PIPA) and mirrors best practices in data governance, aiming to prevent misuse of minors' personal information.

In practice, multiple stakeholders share responsibility for ensuring AI tools are used ethically – the education authorities (ministries and school boards), the textbook publishers providing content, and the platform operators providing the AI technology.

Together they oversee data protection compliance and conduct bias and fairness reviews of the algorithms. Accountability between private publishers and public authorities is still unclear, especially for monitoring algorithmic bias or errors. This uncertainty suggests that governance interoperability is still evolving, as clear and enforceable standards for liability and accountability in cases of harmful AI errors or data breaches have yet to be fully established.

**AI risks and challenges**

The integration of AI into education brings several risks that Korea is grappling with: - Bias and inequality: AI-generated practice questions or explanatory content could inadvertently reflect biases (for example, cultural or gender biases present in training data), potentially disadvantaerging certain groups of students. If not carefully managed, AI tutors might widen achievement gaps rather than close them. - Child data protection: There is a risk of sensitive educational or personal data about children being leaked or misused. Strict controls on external access (Ministry of Education, AI Ethics Principles for Education — Detailed Guidance, 2022; Ministry of Education, 7th Master Plan for Educational Informatization (2024–2028), 2024; Ministry of Education, 2025 Intelligent Information Society Implementation Plan, 2024; Government of the Republic of Korea, Digital Bill of Rights — Commentary, 2023) need to be rigorously enforced and regularly audited. International interoperability here means ensuring any third-party AI service (possibly cloud-based or foreign-developed) adheres to Korea's child data protection standards.

Over-reliance on AI tools could undermine the role of teachers and the development of students' self-directed learning skills. Korean educators have voiced concerns that students might become passive recipients of AI-curated content. Striking the right balance between AI assistance and human instruction is a challenge that requires continuous policy attention and probably further guidelines or training standards.

## **3.3** CROSS BORDER DATA FLOWS

AI safety, AI security and interoperability are interlinked concepts that shape cross border data governance. AI safety aims to ensure that AI systems do not cause unintended harm and adhere to ethical principles such as reliability, fairness, transparency, accountability and privacy (Huang, 2024). AI security focuses on protecting AI systems and data pipelines from unauthorized access, data breaches and adversarial attacks, emphasizing confidentiality, integrity and availability (Huang, 2024). Interoperability refers to the ability of different systems and regulatory regimes to work together, both technically—by enabling data sharing—and normatively—through aligning laws and standards (Onikepe, 2024). All three elements must work in concert: securing AI systems and ensuring they behave safely requires interoperable frameworks across borders.

**Legal and ethical framework**

South Korea has developed a robust legal framework for AI and cross border data flows. The AI Framework Act (2024) designates high impact AI sectors (including energy, drinking water, healthcare, nuclear facilities, biometrics, law enforcement and public services) and imposes obligations for risk assessments, impact evaluations and labelling of generative AI content. Foreign AI providers serving Korean users must appoint a domestic representative with an address in Korea and report this to the Ministry of Science and ICT; failure to do so may result in administrative fines. The Ministry may conduct inspections and issue corrective orders if violations are suspected.

The Personal Information Protection Act (PIPA) was amended in 2023 to introduce data portability and the right to explanation for automated decisions, enabling individuals to request that their personal data be transferred to other services and to challenge algorithmic decisions (PIPC, 2025a). The 2025 enforcement rules require large data controllers to offer encrypted downloads and API based transfers for international data flows and expand the MyData program from finance to healthcare, telecoms and energy sectors (Delara Derakhshani, 2025). South Korea also aligns with global norms: it supports the OECD AI Principles and UNESCO's AI Ethics Recommendation, emphasizing human rights, transparency and accountability. The Seoul Declaration (2024) commits Korea and other signatories to harmonizing AI safety standards across jurisdictions.

**Safety and ethics considerations**

AI safety encompasses ethical and social impacts, while AI security focuses on technical protections; both are essential for public trust (Huang, 2024). Cross border data flows raise privacy and fairness concerns, as personal data may be transferred to jurisdictions with differing protection levels. OECD's concept of "data free flow with trust" advocates strong safeguards and oversight for international transfers (OECD, 2022). Although Korea has EU adequacy recognition, differences in the definitions of sensitive data, children's privacy and lawful bases for processing create compliance challenges (Kuner & Zanfir Fortuna, 2025). When AI models are developed or trained abroad, it becomes difficult to identify the data used and to explain the decision making logic, raising questions about accountability for harm.

**Technical standards**
South Korea supports secure data transfer mechanisms. The MyData program uses standard APIs and end to end encryption; before exporting data abroad, controllers must conduct risk assessments and monitor how the data are used after transfer. The 2025 expansion requires controllers to provide encrypted file downloads and API connections to other services, emphasizing robust technical safeguards (Delara Derakhshani, 2025). South Korea participates in the Global Cross Border Privacy Rules (CBPR) / Privacy Recognition for Processors (PRP) certification system launched in June 2025; certified companies can transfer data across member economies (Korea, the United States, Japan and others) with fewer regulatory barriers, and the program plans to add requirements for sensitive and children's data (PIPC, 2025b). The AI Framework Act requires clear labelling of AI generated content and transparency reporting for high impact systems.

**Operational capacity**
The Personal Information Protection Commission (PIPC) and the Korea Internet & Security Agency (KISA) oversee implementation, including risk assessment guidance, certification and enforcement. Foreign providers must designate domestic representatives responsible for liaising with regulators; if a representative violates the law, the AI provider remains liable. The AI Framework Act allows a one year transition period, during which the Ministry of Science and ICT will draft detailed regulations and guidelines (Shivhare & Park, 2025). PIPA amendments mandate pre transfer risk assessments and post transfer monitoring to ensure safe data flows.

**Data governance**
Onward transfer restrictions stipulate that personal data exported from Korea may not be transferred to third countries unless equivalent protection is ensured, and all transfers must be documented. Despite Korea's compliance with EU adequacy requirements, geopolitical tensions and competition in AI have led to divergent data transfer rules; researchers warn that this fragmentation weakens EU influence and encourages alternative frameworks like CBPR (Kuner & Zanfir Fortuna, 2025). PIPC's MyData expansion guidelines propose encrypted file downloads, API based data delivery and automated data access through certified agencies to maintain accountability.

**International cooperation**
Korea engages in bilateral and multilateral efforts to harmonize AI governance. The EU–Korea Digital Trade Agreement (negotiated in 2024) prohibits data localization, promotes free flow of data and requires algorithmic transparency, further aligning Korea's standards with those of the EU (Ministry of Trade, 2025). Korea participates in OECD, UNESCO, the Global Privacy Assembly, G20, and the Global CBPR Forum to promote interoperability of AI safety standards. As a middle power nation, Korea seeks to bridge regulatory approaches and advocate for trusted AI norms globally.

**AI risks and challenges**
Regulatory fragmentation can limit access to diverse datasets and hinder AI model performance (Kuner & Zanfir Fortuna, 2025). Growing data localization measures in some countries create compliance burdens for international businesses. When cross border AI services cause harm, liability may be unclear and multiple legal systems might apply. Finally, generative AI misuse—including deepfakes—continues to spread across borders despite labelling requirements, highlighting the need for international cooperation on detection technologies and common standards.

**Interoperability Matrix: Dimension × Sector**
To compare Korea's interoperability performance across the three sectors, the table below summarizes key assessment dimensions and an approximate implementation score (on a 1–5 scale, with 5 being most favorable) for each sector:

| Dimension | Autonomous Vehicles | Education | Cross-Border Data |
|---|---|---|---|
| **Clarity of norms & values** | 3 – Core laws exist; ethics non-binding. | 4 – Clear learner-centered ethics; some policy shifts. | 4 – High clarity via PIPA reform; global norm alignment. |
| **Alignment with standards** | 3 – Partial UNECE alignment; ongoing. | 3 – Early IEEE/ISO adoption. | 4 – Strong GDPR/CBPR/DFFT alignment. |
| **Operational capacity** | 4 – Pilot zones, R&D investment strong. | 3 – Uneven teacher training, limited uptake. | 3 – DPO networks active; SME gaps. |
| **Data governance** | 3 – Basic logging, cybersecurity; sharing rules pending. | – N/A (PIPA-based privacy). | 4 – Advanced controls; pre/post-transfer checks. |
| **Legal enforceability** | 3 – Permits, insurance in place; top-level rules pending. | 2 – Voluntary AI use; weak obligations. | 3 – Strong PIPC enforcement; certification evolving. |

**Table 3:** Comparative Assessment of Interoperability across Key Sectors

*Note: Education data governance follows general PIPA provisions; no separate rating assigned.*
*Source: Author's assessment based on legislation, standards, and policy documents (as of Aug 2025).*

This matrix highlights that South Korea's cross-border data frameworks are the most internationally interoperable (due to alignment with global standards and robust enforcement), while the education sector currently lags in enforceability and consistent implementation. The autonomous vehicle sector shows strong operational readiness but still faces regulatory and standards alignment gaps as international rules evolve.

## 4. Jurisdiction Profile – South Korea

### **4.1** SNAPSHOT

South Korea's digital regulatory ecosystem is characterized by proactive government initiatives, strong legal protections, and advanced technological infrastructure: - Governance approach: Korea employs a mix of stringent laws and adaptive guidelines. For example, the new AI Basic Act (2025) establishes broad governance for high-risk AI, while sector-specific ministries issue ethical codes to guide industry. This top-down yet consultative approach reflects Korea's commitment to a people-centered, "data sovereignty" paradigm for AI governance – ensuring technology serves citizens under clear rules. - Digital infrastructure and innovation: The country boasts one of the world's highest connectivity rates (nationwide 5G coverage) and has designated multiple innovation zones (like smart cities and autonomous vehicle pilot zones). Government-backed projects in AI R&D, from autonomous driving testbeds to AI education pilots, demonstrate capacity for rapid deployment of new technologies, albeit with some public debate. - Global integration: Korea actively aligns its regulations with international frameworks to enable interoperability. It has obtained EU adequacy for data protection, joined the Global CBPR Forum for privacy, and engages in bilateral agreements on digital trade and AI cooperation. Korean standards and certifications increasingly reference or incorporate global norms, making the country a bridge between East Asian regulatory practices and Western frameworks.

### **4.2** DIMENSION BY DIMENSION FINDINGS

Examining Korea's AI and digital regulation across key dimensions reveals the following strengths and gaps:

- **Legal and Ethical Frameworks:** Korea has a comprehensive set of laws addressing AI and data. PIPA provides a strict privacy regime, and the forthcoming AI Basic Act will introduce dedicated AI risk management obligations. Sectoral laws (in transport, etc.) and guidelines embed ethics like transparency and human accountability. These legal instruments align well with global principles (e.g. OECD AI Principles, UN AI ethics guidelines). A challenge remains in keeping legislation agile – for instance, education policy shifts showed a need for more consultative, stable rule-making to maintain trust. Overall, the legal foundation is strong, but some ethical norms (like AI ethics in private sector use) still rely on voluntary uptake, which may require bolstering to ensure consistent adherence.

- **Technical Standards and Infrastructure:** South Korea generally seeks to adopt or influence international technical standards rather than go it alone. In autonomous vehicles, it is involved in UNECE vehicle regulations and ISO technical committees; in AI, Korean researchers contribute to global standard-setting for AI risk management. Domestically, the government funds the development of infrastructure (from high-speed networks to AI computing clusters and data trusts) that not only serve local needs but can also interoperate internationally (e.g. common data formats or API standards in MyData align with global norms). One limitation is that in emerging domains like AI in education, clear technical standards have not yet matured – Korea is essentially part of the global learning process to develop them. The country's strength is in quickly building test environments and sandboxes which can generate data to inform standardization.
- **Operational Readiness and Institutions:** Korea has established a robust institutional framework to enforce AI-related regulations. The PIPC stands out as a capable regulator with increasing expertise in algorithmic issues. Other bodies like KISA provide technical support and certification. Ministries have dedicated teams or sub-organizations (e.g. an AI policy bureau in MSIT) focusing on AI strategy and safety. The presence of trained DPOs across companies and the offering of government-led AI ethics training programs for officials and developers further indicate growing operational readiness. However, the breadth of adoption is uneven: large companies and government agencies have embraced these roles, whereas many SMEs and local governments still have limited capacity to implement AI governance measures. Strengthening capacity at those levels (through funding, training, international cooperation) is a priority to ensure interoperability isn't just on paper but in practice countrywide.
- **Data Governance and Privacy:** Personal data protection is a particularly strong suit for Korea. It has a single, unified privacy law (PIPA) that is broadly compatible with GDPR, and a dedicated enforcement commission (PIPC) ensuring compliance. Cross-border mechanisms, such as requiring contractual clauses or certifications for international data transfers, are in place and mirror global practices. Innovations like MyData highlight a data empowerment approach – giving individuals rights and tools to control their data – which resonates with emerging global digital rights discussions. The main gaps here relate to new forms of data like non-personal big data or AI model transparency. Korea is beginning to address data governance for AI (for instance, considering how to audit AI models' training data for biases or privacy issues), but these efforts are in early stages. Aligning such AI-specific data governance with global discussions (like at OECD or GPAI) will be important.
- **Oversight and Enforcement Mechanisms:** Korea's ability to enforce rules is evident in its use of heavy penalties and novel sanctions (like deleting AI models) when regulations are breached. The interoperability advantage is that Korea sets precedents that can inspire or pressure regulators elsewhere to consider similar actions, thereby leading to more consistent global enforcement norms over time. Korea also actively participates in international forums – for example, it will host the Global Privacy Assembly in 2025 – using those platforms to push for coordinated oversight of AI and data practices. One area for improvement is increasing transparency of enforcement outcomes domestically (so that companies can learn from each case) and enhancing cooperative oversight with other jurisdictions (through MOUs or joint investigations). Aligning definitions – what constitutes "high-risk AI," for example – with trading partners and international guidelines will also help ensure that oversight in Korea is understood and respected globally, and vice versa.

## 4.3 SECTOR CASE STUDIES TABLE

To illustrate Korea's practical approach to AI safety and interoperability, the table below summarizes representative initiatives and corresponding interoperability strengths across key sectors.

Each sector showcases South Korea's willingness to pilot innovative solutions while embedding international interoperability considerations (be it through global standards, protective measures, or cooperative frameworks).

| Sector | Notable Initiative / Example | Interoperability Strength |
|---|---|---|
| **Autonomous Vehicles** | Nationwide Pilot Zones & Freight Platooning: 34 autonomous driving pilot zones established across all provinces; Seoul testing robo-taxis, and highways hosting platoons of self-driving trucks (MOLIT,2023). | Strong government facilitation of testing and data-sharing has created a rich environment aligned with evolving international vehicle safety standards. Korea's engagement in UNECE mutual recognition talks for autonomous vehicle certifications further boosts cross-border operability. |
| **Education** | AI Digital Textbook Pilot (2025): AI-driven textbooks introduced in primary and secondary schools (personalizing lessons in English, math, etc.), with nearly one-third of schools initially participating. | Significant public investment in digital infrastructure and teacher training reflects a commitment to modernize education. Privacy-by-design in these tools (strict student data protection) demonstrates alignment with global child data protection norms, even as Korea refines the program based on public feedback. |
| **Cross-Border Data** | MyData Expansion & Global CBPR Membership: The MyData program enables individuals to port their financial and soon health/telco data between service providers. Simultaneously, Korea joined the Global CBPR Forum to streamline international data exchange. | A comprehensive legal toolkit for data interoperability: Korea has high standards (PIPA) but also mechanisms (CBPR, trade agreements) to recognize equivalent protections abroad. Enforcement actions (e.g. ordering deletion of an AI model trained on improperly exported data) signal to international partners that Korea upholds "data free flow with trust," reinforcing global norms through action. |

**Table 4:** *Sectoral Highlights and Interoperability Strengths*

## **4.4** GOOD PRACTICES AND PRIORITY

### Good practices

South Korea has implemented several noteworthy practices that could serve as models for other jurisdictions:

- **Holistic privacy enforcement:** The PIPC's aggressive enforcement (including fines and requiring algorithm deletion for violations) is a good practice in holding companies accountable for AI and data misuse. This approach enforces interoperability by ensuring foreign companies meet Korean standards, thus indirectly raising their compliance globally.

- **Regulatory sandboxes and pilot programs:** By allowing autonomous vehicles and other AI innovations to operate in controlled environments, Korea gathers data and experience to inform standards. This pragmatic approach balances innovation with safety and can be shared internationally (e.g., mutual recognition of test results with other countries).

- **Stakeholder engagement in policy shifts:** The decision to adjust the AI textbooks rollout based on teacher and parent feedback, while contentious, shows a responsiveness to public concerns. Engaging educators, industry, and civil society early and often (through forums, task forces, etc.) is becoming a standard practice in Korean AI policymaking, aligning with democratic governance values emphasized in UN and OECD guidelines.

- **International cooperation networks:** Korea's active participation in global bodies (Global CBPR Forum, OECD AI groups, GPAI, etc.) and pursuit of bilateral agreements is a good practice in interoperability. By being present at the table, Korea helps shape rules that it can adopt domestically with less friction and contributes its on-the-ground insights to global policy development.

### Priorities

To further enhance AI safety and interoperability, the following areas should be prioritized in Korea's policy agenda:

- **Consistency and clarity in education AI policy:** The abrupt change in the AI textbook program highlighted a need for more stable, well-communicated policy frameworks for edu-tech. Korea should clarify the legal status of AI in classrooms, possibly through dedicated legislation or standards, to give schools confidence. Expanding teacher training and establishing evaluation tools for AI educational content (to monitor bias or efficacy) will be critical moving forward.

- **Liability and safety for advanced autonomous vehicles:** As technology approaches full self-driving capabilities, Korea should refine its liability rules (especially for Level 4 and 5 vehicles) and update insurance and compensation systems accordingly. Codifying ethical guidelines for AI decision-making in vehicles (for instance, how to program choices in unavoidable accident scenarios) would also strengthen trust. Internationally, pursuing mutual recognition of safety tests and certifications with partners (through UNECE or bilateral agreements) will ensure Korean autonomous vehicles can operate abroad and vice versa without redundant regulatory hurdles.

- **Effective SME inclusion in data governance:** While large Korean firms are adapting to global data rules, many SMEs may have difficulty. Simplifying certification processes (e.g., for CBPR) and providing financial or technical assistance can help smaller companies become interoperable in the global digital economy. This might involve government-subsidized tools for compliance (like template privacy impact assessments or open-source AI audit frameworks) so that high standards do not become a barrier to entry.

- **Integrating AI risk management into business practice:** Korean regulators should encourage (or require) companies to embed AI risk assessment and mitigation processes into their operations. For example, developing sector-specific checklists or "AI audit" requirements (aligned with frameworks like the EU's upcoming AI Act or NIST's AI Risk Management Framework) will prepare Korean companies to demonstrate safety and fairness of their AI systems globally. Establishing procedures for model review and data provenance – such as the ability to retrace and delete training data that was collected improperly – should become standard practice.

- **Enhanced transparency and public accountability:** As AI systems become more widespread in Korea, building public trust is essential. Authorities could implement regular AI policy evaluations (perhaps annual white papers or audits of key programs like autonomous vehicles and MyData) with results made public. Broadening stakeholder consultations – including citizen juries or public forums on AI ethics – will also improve transparency and ensure that Korea's AI governance remains aligned with societal values, which is a priority echoed in international discussions.

By focusing on these priorities, South Korea can strengthen its leadership role in safe and interoperable AI development, ensuring that its domestic innovations and rules remain in harmony with global norms and expectations.

## 5. Conclusion

This report uses South Korea as a case to evaluate how domestic regulatory refinement, international rule alignment, and scenario-based pilots can jointly advance AI safety and interoperability. Overall, Korea shows that an innovation strategy grounded in privacy protection and accountability can be converted into a working governance architecture when high-level principles are translated into auditable practices and measurable outcomes. First, cross-sector evidence indicates that ethical principles have been embedded as operational and testable requirements. In cross-border data governance, statutory rights such as data portability and visible enforcement have strengthened individual protection while improving regulatory predictability for firms. In autonomous vehicles, a sequence of piloting, evaluation, and legalization has reduced real-world deployment risk and created a traceable safety loop. In education, the cautious expansion and subsequent recalibration of generative AI textbooks demonstrate a policy feedback mechanism that is bounded by public trust. This layered progression from principle to standard to implementation turns interoperability from a value statement into verifiable processes and documentation.

Second, at the international level, Korea's EU GDPR adequacy status, participation in the Global and APEC CBPR systems, and forward-looking digital trade arrangements have created a high-alignment compliance corridor with major economies. The two-way circulation between domestic rules and multilateral frameworks lowers cross-border frictions, increases mutual recognition and the portability of compliance, and supports data free flow with trust. In practice, data and AI-enabled services can move to and from Korea with greater legal certainty and shared expectations around risk management.

Third, important gaps remain. Implementation is not yet consistent across regions and sectors, which limits the diffusion of best practices from advanced testbeds to schools and small and medium enterprises. SME participation in high-standard frameworks is still too limited to make interoperability inclusive across the whole economy. Most critically, soft ethical guidance for high-risk AI must be hardened into enforceable and auditable obligations that cover data governance, model evaluation, pre-deployment and post-deployment testing, incident reporting, and corrective action, so that ethical commitments map to concrete engineering, process, and evidence requirements.

Looking ahead, Korea's trajectory is moving from principle led governance to an evidence and outcomes orientation that resonates with global initiatives and converges with emerging international AI safety frameworks. Continued progress will depend on deepening normative specificity, expanding standards adoption, widening mutual recognition, and diffusing capabilities. This includes codifying minimum compliance packages for priority use cases, scaling conformity assessment and certification capacity, and mainstreaming support for SMEs so that interoperability benefits the full economy.

In sum, the Korean case shows that achieving AI safety and interoperability is a dynamic and iterative process in which social trust sets the boundary conditions, standards and enforcement provide the engine, and international cooperation acts as the multiplier. For other jurisdictions, transferable lessons include grounding rights and accountability in auditable controls, building interoperable compliance corridors with key partners, and using pilots with rigorous feedback to update rules. For Korea itself, the next step is to close execution gaps, broaden SME participation, and fully institutionalize enforceable governance for high-risk AI, thereby achieving a higher order balance between local effectiveness and global consistency.

## Acknowledgement

We extend our sincere gratitude to Young Sik Kim (UNU) and Hogun Lee (Society for EdTech) for their valuable insights and constructive feedback that informed the development of this report. We also thank experts from academia, public agencies, and industry in Korea whose anonymous contributions enriched the analysis. Finally, we acknowledge the UNU project team for their coordination and support under the *Interoperability of AI Safety Governance Project (China, Korea, Singapore, and the UK).*

## Interview Responses Table

[Understand how the interviewee defines and connects AI safety and interoperability; assess alignment and gaps between domestic rules and international norms; describe organizational AI-risk governance and responsiveness; outline international collaboration and its feedback into policy; and, in Korea's push for technological sovereignty, propose ways to balance sovereignty–interoperability–trust while navigating trade-offs among safety, security, and interoperability.]

| Respondent | Key Insights |
|---|---|
| **A** | Define AI safety as "the social risks arising from insufficient validation of black-box systems," and interoperability as "the ability of systems to connect and exchange information." The recommendations include establishing a dedicated task force to close risk-governance gaps, prioritizing the consolidation of technological advantages, leveraging Korea's social dynamism and agility, and setting a long-term roadmap while promptly reviewing current R&D. |
| **B** | The core tension is that anonymized releases for safety and security reduce inter-agency interoperability, while directly identifiable data creates severe disclosure risk. With resident registration numbers legally restricted, tokenization or pseudonymization can balance safety and interoperability. Over the next 2–3 years, Korea should align domestic rules and technical profiles with the EU AI Act, Global CBPR, DFFT, and NIST AI RMF to enable compliance while closely tracking U.S. policy moves. |
| **C** | The respondent treats AI safety and interoperability as one system, favors trust-based openness with global alignment and domestic safeguards, prioritizes AV logging/cybersecurity/real-time monitoring, education anonymization and child protection, and cross-border ex-ante risk with ex-post tracking, notes weak SME compliance capacity, and cites OECD/UNESCO/UNECE/Global CBPR engagement plus firm PIPC enforcement including model deletion orders as trust anchors. |
| **D** | Governance is still at early-stage without clear roles, processes, metrics, or audits. Before easing controls, policymakers should roll out trust mechanisms that citizens can tangibly experience, then institutionalize them with education as the starting platform. Embed AI ethics and data governance as ESG defaults, backed by public indicators such as privacy protection rate, bias mitigation, and appeal-handling speed. |
| **E** | Align with the UNESCO Recommendation on AI ethics; note Korea's AI Basic Act; require from 2025; require from 2025 that public bodies deploying AI conduct personal-information impact assessments with AI-specific fields; apply the Ministry of Education's AI ethics principles; and ensure interoperability and auditability by broadly adopting 1EdTech standards with digital-textbook competencies mapped via CASE. |
| **F** | There is inherent tension between safety and interoperability since security safeguards often reduce interoperability; current governance leans on data-protection regulation and DPO mechanisms and uses de-identification and pseudonymization to make data both usable and safe; the OECD AI Principles are cited as most influential, and the respondent favors networks of cities and nations to jointly address the expanding power of large platforms over data and AI. |

[Education: Assess outcomes and risks of Korea's AI digital textbooks and literacy programs; identify hurdles in adopting foreign AI tools and how PIPA/education policies help or hinder; explain educator capacity-building and standards collaboration with TTA/KATS/UNESCO/OECD; and examine ethics around IP, student over-reliance, and the human–AI balance.]

| Respondent | Key Insights |
|---|---|
| A | AI digital textbooks and AI literacy initiatives mainly create an enabling atmosphere but may widen inequality over time, and the primary barrier to introducing overseas tools is language localization. |
| B | Although AI digital textbooks were slated for March 2025, a late-2024 legislative amendment reclassified them as educational materials, and in August 2025 the National Assembly removed their legal status as official textbooks, creating frontline confusion over use, evaluation, and scale-up. Key ethics risks include student over-reliance that leads to automation bias and weaker critical thinking, as well as opaque training data, copyright disputes, and security concerns when sensitive personal content is used for training. |
| C | Programs expand personalized learning and international collaboration but expose privacy risks, algorithmic bias, and teacher capacity gaps; weakening of "official textbook" status creates policy consistency issues, calling for expanded teacher training and adoption of UNESCO and OECD standards. |
| D | Programs are still at a pilot stage, the main risk is students shifting from assistance to replacement, and major hurdles include language localization, strict privacy rules and platform specification mismatches that create integration and security burdens, so ethics and safety should be institutionalized alongside technology adoption with clearer operating guidance and better teacher readiness. |
| E | AI digital textbooks are positioned as teaching aids and are governed by privacy controls and, where applicable, ISMS-P requirements. Implementation is being rolled out through KERIS-led teacher cohorts and systematic training. Broad adoption of open standards is planned, including 1EdTech's LTI, OneRoster, CASE, and QTI, as well as Caliper Analytics or xAPI and Open Badges. Textbook competencies are mapped using CASE. |
| F | Pilots reveal more problems than gains, including opaque decision criteria, potential misuse, and negative effects on student capability; importing foreign tools faces corpus and value-gap issues that yield irrelevant answers, plus blind spots in the education system and in PIPA; teachers and administrators are under-prepared, and student over-reliance is rising without clear countermeasures. |

[Cross-border data flows: Under the Data Bridge, EU adequacy, APEC CBPR, and DFFT frameworks, determine a proportionality-based balance between AI-enabling cross-border data sharing and privacy/security; pinpoint binding frictions (consent, localization, export controls, liability); recommend updates to PIPA and the forthcoming AI Basic Act; and shape coordinated government–business strategies that stay competitive, compliant, and trust-building.]

| Respondent | Key Insights |
|---|---|
| A | Data localization is identified as the largest obstacle, and the advised response is active participation in international standardization with close tracking of global trends. |
| B | NO Answer |
| C | Korea attains GDPR adequacy, joins Global CBPR, advances digital trade with the EU, and PIPC has ordered model deletion for unlawful cross-border transfers, illustrating a path that balances technological sovereignty with international trust. |
| D | The respondent calls for institutional regulation together with technical reinforcement, identifies unclear responsibility across countries as a key obstacle, and recommends common principles such as data minimization, purpose limitation, the right to deletion and transparency with risk tiering by sensitivity and clear responsibility assignment. |
| E | NO Answer |
| F | Back openness with transmission/monitoring tech plus transparent outcome-sharing; tackle data-localization as the chief barrier and GDPR constraints on global scaling; engage early in rulemaking, balance rights and national interests, and apply proportionality to allow limited flows of public-benefit data. |

[Autonomous vehicles: Evaluate the effectiveness of Korea's alignment on autonomous-driving interoperability and safety (e.g., K-City); identify the technical-legal barriers in cross-border V2X, cybersecurity, and map-data APIs and the approaches for real-time data and cross-API compatibility; clarify the governance framework for responsibility allocation in Level-4-and-above systems and the role of ethical impact assessments; and weigh the benefits and risks of mutual recognition with foreign certification regimes.]

| Respondent | Key Insights |
|---|---|
| A | K-City, together with UN R155 and UN R156, is viewed as effective, and the respondent supports creating international standards and incorporating Ethical Impact Assessment, with rapid expert development needed. |
| B | NO Answer |
| C | K-City and thirty-four pilot sites provide the validation and standards base, yet V2X and map data API interoperability remain challenges; Korea is expanding UN R155 and R156 adoption and needs mutual recognition for testing and certification. |
| D | The respondent stresses API harmonization and regular technical inspections for efficient and safe operations, supports mutual recognition of K City testing with foreign certifications, and urges supplementary verification and responsibility sharing rules to account for differences in road environments and legal systems. |
| E | NO Answer |
| F | Korea adheres to UN R155 and UN R156 yet lacks robust empirical data sharing and research-facility utilization. Policy on data management and interoperability should advance in step with technical R&D; V2X collaboration with leading institutions should continue; Ethical Impact Assessment should guide decisions; K-City test results should seek mutual recognition with foreign certifications; and responsibility allocation and data-leak response procedures should be defined in advance. |

[Conclusion: Identify Korea's standout strengths and long-term challenges in AI safety/governance, the structural barriers to fix and the practices worth exporting; specify the next 2–3 years' priority reforms, collaborations; and research to build an interoperable, trustworthy ecosystem aligned with evolving international standards.]

| Respondent | Key Insights |
|---|---|
| A | Prioritize a long-term research-and-education roadmap with continuous review—avoid quick fixes and reassess current R&D; the main obstacles are limited long-range vision and policy short-termism; therefore, Korea should link domestic agility and openness to a sustained roadmap that also delivers meaningful contributions to international governance. |
| B | The immediate priority is to align domestic standards internationally to minimize conflicts with the EU AI Act, NIST, CBPR, and DFFT; collaboration and learning are hindered by ministerial silos (MOTIE and MSIT) and a failure-avoidance culture; over the next two to three years the focus should be on refining standards and guidelines and closely tracking international developments, including shifts in U.S. policy. |
| C | Priority actions are restoring education policy consistency, clarifying responsibility for Level 4 autonomy, helping SMEs achieve international certification and compliance, and embedding AI risk management for the next two to three years; despite strong test beds, coordination and SME capability gaps persist, so Korea should establish a national model that combines firm PIPA enforcement, regulatory sandboxes, and a twin strategy of sovereignty and openness. |
| D | Institutionalize ESG-style, socially acceptable governance by disclosing indicators, tying them to procurement and certification, and running a permanent program that trains teachers and administrators; address weak institutionalization for openness and innovation and low social acceptance; balance Korea's strong legal framework and public-sector execution so openness and innovation advance with rights protection. |
| E | Partner with industry, academia, and UNESCO/1EdTech to operate education-data standards and joint governance, achieving openness and interoperability while reinforcing protection and sovereignty; due to weak legal embedding and inter-agency conflicts, apply "openness via standards, protection via governance" through consistent standards, APIs, robust data governance, and anonymization/pseudonymization. |
| F | Korea's strength is a gradual, oversight-driven openness, but regulatory and implementation capacity must grow. Over the next two to three years, the respondent urges creating a multilateral platform to tackle AI challenges jointly and continuing sandbox pilots to enable early validation and interoperability testing. |

COUNTRY REPORTS

# People's Republic of China Country-Level AI Safety Interoperability Report

CHUNLI BI, LEILEI ZHANG, TIANYU WANG AND NA FU
CHINA ACADEMY OF INFORMATION AND COMMUNICATIONS TECHNOLOGY
WANGTIANYU@CAICT.AC.CN

# People's Republic of China Country-Level AI Safety Interoperability Report

15 October 2025

## 1. Introduction

Artificial Intelligence (AI) is advancing rapidly across key industrial sectors and increasingly influencing public governance and international cooperation. For People's Republic of China - hereafter "China" - the aim is to maximize the benefits of AI while managing its risks. This requires an approach that coordinates governance, security, and international cooperation to reach a unified safeguard system. This report evaluates China's current AI safety and interoperability framework from the perspective of international alignment, identifies frictions across legal and organizational practices, technical and interface standards, and data and semantic layers, and proposes actionable pathways to enhance assurance and promote consistency with global frameworks.

### **1.1** CHINA'S DEFINITION OF AI SAFETY

In China, AI security encompasses aspects such as legal norms, regulatory mechanisms, technical standards, and ethical requirements. Its purpose is to ensure that the research and development (R&D), deployment, and operation processes of AI systems—especially high-risk systems like those for autonomous driving, generative AI, and AI in education—are **safe**, **controllable**, and **responsible**, without compromising individual rights, public interests, or national security. Its core objective is to **develop responsible artificial intelligence**, which specifically includes the following contents:

- **Preventing unintended harms caused by technical failures** (e.g., algorithmic misjudgments), misuse (e.g., biased data application), or malicious attacks (e.g., cyberattacks on connected vehicles).
- **Safeguarding data security and privacy throughout the AI lifecycle**, including compliance with "minimization of collection," "local storage of important data," and cross-border data security assessment requirements.
- **Embedding human-centric ethical values** (fairness, transparency, accountability, respect for human dignity) into AI design and governance.
- **Implementing full-lifecycle risk management**, including pre-deployment safety testing (e.g., autonomous vehicle road tests), in-operation monitoring (e.g., real-time supervision of AI educational tools), and post-incident accountability (e.g., data breach remediation).
- **Aligning with national security and public interest objectives,** ensuring AI applications do not undermine critical infrastructure, social stability, or public health (e.g., regulating facial recognition in public spaces).

Key policy documents defining China's AI safety framework include:

- *New Generation Artificial Intelligence Governance Principles—Developing Responsible Artificial Intelligence* (2019)
- *New Generation Artificial Intelligence Ethical Guidelines* (2021)
- *Measures for the Review of Science and Technology Ethics* (Trial) (2023)
- *Cybersecurity Law, Data Security Law, and Personal Information Protection Law* (PIPL, 2021)

### **1.2** PURPOSE AND AUDIENCE

**Purpose**
This report evaluates China's AI safety regulatory framework through the lens of international interoperability. It benchmarks existing laws, standards, and practices; analyses frictions across legal, organizational , technical, interface, data and semantic layers; and outlines actionable pathways to enhance assurance and cross-border data flows. The analysis focuses on three high-impact sectors: **Autonomous Vehicles (AVs), Education,** and **Cross-Border Data Flows**.

**Audience**
The report seeks to reach different sets of stake-holders. Including private individuals such as policymakers and regulators - to refine sector-specific governance and international alignment strategies -, industry leaders and operators - to clarify compliance pathways for AI deployment -, and academics and ethics experts - to inform evidence-based research on AI safety. Not only that, legal entities such as standards, testing, and certification bodies and international organizations are also of interest to identify gaps in technical standardization and facilitate mutual recognition of AI safety standards.

## **1.3** SCOPE

This analysis explores China's AI safety and interoperability landscape across three priority sectors: autonomous vehicles, education, and cross-border data flows. The assessment is drawn on five key dimensions: legal and ethical frameworks, technical standards and infrastructure, data governance practices, indicators of implementation progress, and oversight mechanisms.

The evaluation is based on China's key legal and policy instruments for AI, including:

- *Personal Information Protection Law* (PIPL, 2021)
- *Data Security Law* (2021)
- *Cybersecurity Law* (2017)
- *Measures for the Management of Automotive Data Security* (Trial) (2021)
- *Guidelines for the Safe Transportation Services of Autonomous Vehicles* (Trial) (2023)
- 7*th* *Master Plan for Educational Informatization* (2024–2028)
- *Provisions on Promoting and Regulating Cross-Border Data Flows* (2024)
- *Measures for the Safety Management of Facial Recognition Technology Application* (2025)

Meanwhile, the comparative framework is anchored in international norms, including the UN *Global Digital Compact* (GDC), OECD *Principles on Artificial Intelligence*, UNESCO *Recommendation on the Ethics of Artificial Intelligence*, UNECE WP.29 regulations on automated driving.

## **1.4** ANALYTICAL LENS

**Tripartite Analysis**
The report examines domestic laws and policies alongside international norms (e.g., GDC, OECD AI Principles) and bilateral or multilateral engagements (e.g., participation in ISO/IEC standardization, regional digital cooperation). This lens highlights how China's "scenario-driven, pilot-first" governance model aligns with or diverges from global practices. Not only that, it showcases both the frictions in cross-border AI operations (e.g., data export compliance, standard mutual recognition) and the synergies between China's sector-specific rules (e.g., automotive data) and international frameworks (e.g., UNECE vehicle regulations).

## **1.5** METHODOLOGY

This study employs a qualitative document analysis approach, complemented by policy implementation tracking. The analysis systematically reviewed laws, regulations, standards, and policy documents in force as of September 2025, covering national legislation, sector-specific regulation, local pilot rules, and international cooperation agreements. Policies are categorized according to five criteria: objectives, guiding principles, binding nature, key components, and linkages to international frameworks. Sectoral implementation is tracked in three areas—autonomous vehicles, education, and cross-border data flows—through milestones such as road test coverage, AI textbook pilots, and security assessment cases. Finally, representative case studies, are used to triangulate policy effectiveness and interoperability challenges.

## **1.6** RESEARCH QUESTIONS

The report addresses three core research questions:

1. How effective are China's AI safety governance, security, and interoperability frameworks in design and operation across Autonomous Vehicles, Education, and Cross-Border Data Flows?
   - 1a (Design): What laws, institutions, standards, and assurance processes are in place?
   - 1b (Operation): How are they implemented (e.g., testing, certification, enforcement)?
   - 1c (Effectiveness): What evidence exists of risk reduction (e.g., pilot safety data, compliance rates)?

2. What strengths and vulnerabilities exist across the legal/organizational, technical/interface, and data/semantic layers in each sector, and what are the key risk scenarios?

3. To what extent are China's policies aligned with UN frameworks (GDC), ISO/IEC standards, and global data privacy mechanisms (e.g., CBPR), and what pathways exist for mutual recognition?
   - 3a: What is the maturity of alignment by pillar (governance, security, interoperability)?
   - 3b: What incompatibilities block cross-border AI operations (e.g., data localization, standard differences)?

## 2. Comparative Assessment Framework (China AI Safety Governance Overview)

China's approach to AI safety governance is characterized by a **multi-layered**, **scenario-driven model** that balances innovation with risk mitigation. The framework is defined by the following core elements:

| Element | Details |
|---|---|
| **Objective** | Accelerate AI innovation while safeguarding national security, public interests, and individual rights; achieve "responsible AI development" through lifecycle governance. |
| **Principles & Values** | People-centric design, safety controllability, fairness transparency, accountability, privacy protection, and alignment with national development goals (e.g., "Digital China"). |
| **Governance Approach** | Top-down legal/regulatory framework supplemented by bottom-up pilot programs:<br>- **Laws**: Cybersecurity Law, Data Security Law, PIPL (foundational)<br>- **Regulations**: Sector-specific rules (e.g., automotive data, algorithm recommendation)<br>- **Pilots**: Local demonstration zones (e.g., Beijing AV pilot zone) and "sandboxes" (e.g., 20-city V2X pilots)<br>- **Ethics**: Mandatory ethical reviews for AI research (per Science and Technology Ethics Review Measures). |
| **Binding Nature** | Coexistence of **legally binding statutes** (e.g., data localization for important data) and **non-binding guidelines** (e.g., AI ethical norms for education). Critical sectors (AVs, cross-border data) have stricter mandatory requirements. |
| **Key Components** | 1. Legal/ethical norms (e.g., PIPL's minor protection provisions)<br>2. Technical standards (e.g., GB/T 40429-2021 for AV grading)<br>3. Operational capacity (e.g., local regulatory teams for AV testing)<br>4. Data governance (e.g., cross-border assessment, Classification and categorization of data)<br>5. Oversight mechanisms (e.g., Algorithm Filing annual compliance audits). |
| **International Linkages** | Active participation in global standardization (ISO/IEC JTC1, UNECE WP.29) and regional digital cooperation. Aligns with OECD AI Principles and UNESCO AI Ethics Recommendation. Maintains bilateral digital agreements (e.g., China-Singapore Free Trade Agreement) and participates in multilateral forums (GPAI). |

**Table 1**

## 3. Sectoral Analysis: Autonomous Vehicles, Education, and Cross-Border Data Flows

The analysis that follows reviews three priority sectors—autonomous vehicles, education, and cross-border data flows— aiming to evaluate the interoperability of China's digital regulations with international standards.

### 3.1 AUTONOMOUS VEHICLES (AVS)

#### 3.1.1 AI Safety Taxonomy for AVs

| Safety Dimension | Definition & Scope |
|---|---|
| **Functional Safety** | Ensuring AV systems operate as intended (e.g., avoiding collisions) per ISO 26262 (GB/T 34590 series) and SOTIF (Safety of the Intended Functionality) standards. |
| **Data Safety** | Protecting vehicle-generated data (sensor, location, biometric) via "in-vehicle processing," "minimum collection," and local storage of important data (per Automotive Data Security Measures). |
| **Cybersecurity** | Defending AVs from cyberattacks (e.g., OTA tampering, V2X communication breaches) via network security grading and compliance with Cybersecurity Law. |
| **Ethical Safety** | Ensuring AI decision-making complies with ethical requirements (e.g., prioritizing pedestrians homo sapiens in accident scenarios), and conduct ethical reviews in research and development activities. |

**Table 2**

#### 3.1.2 AI Risk Taxonomy for AVs

| Risk Category | Specific Scenarios |
|---|---|
| **Technical Risk** | Sensor failure in extreme weather, algorithmic misjudgment in complex traffic, cyberattacks on vehicle control systems. |
| **Data Risk** | Leakage of sensitive data (e.g., driver biometrics), non-compliant cross-border data transfer, re-identification of anonymized driving data. |
| **Liability Risk** | Unclear responsibility for accidents (manufacturer/operator/user), difficulty in proving software defects, inconsistent local accident handling rules. |
| **Compliance Risk** | Failure to obtain road test permits, non-compliance with AV grading standards (GB/T 40429-2021), inadequate cybersecurity certification. |

**Table 3**

#### 3.1.3 Legal & Ethical Framework

China's governance of autonomous vehicles (AVs) follows a four-tiered structure. Foundational laws being the first. It includes the Cybersecurity Law, Data Security Law, and Personal Information Protection Law (PIPL), setting baseline requirements for data and network safety. Sector-specific regulations follows, providing oversights based on the Measures for the Management of Automotive Data Security (Trial, 2021), it mandates "in-vehicle data processing", "minimal collection", and local storage of important data, by this regulation cross-border data security assessments are required. The second sector specific regulation in that of the Guidelines for the Safe Transportation Services of Autonomous Vehicles (Trial, 2023), which allows AVs to engage in public transport, such as buses and taxis, after they pass designated safety assessments.

Technical standards also play a central role, with GB/T 40429-2021 defining six levels of driving automation – from L0 to L5-, serving as a reference for policy and testing. The last of the four tiers has to do with local level pilot rules. The Beijing's Autonomous Vehicle Regulations (2025) formalizes a three-stage process of "road testing, demonstration application, and road piloting," establishes a unified operation data platform, and expands application scenarios (e.g., private cars, urban logistics). Meanwhile the Shenzhen Special Economic Zone Intelligent Connected Vehicle Management Regulations (2022), became the first local legislation to codify commercial operation requirements and insurance obligations for AVs.

In parallel, ethical frameworks are embedded in China's AV governance. This is supported by the New Generation Artificial Intelligence Ethical Guidelines (2021) which requires AVs to prioritize human life, ensure decision transparency, and avoid algorithmic bias.

**3.1.4 AI Safety Considerations**

China's regulatory approach to AV safety considers four key dimensions. Functional safety is ensured through requirements of closed-field testing and open-road validation - with Beijing reporting over 120 million kilometers of cumulative test mileage by 2024. Product access evaluations are embedded in international standards such as ISO 26262 (GB/T 34590) and SOTIF principles. Data safety considerations require all AVs to be equipped with event data recorders ("black boxes") to log operational data, while sensitive information (eg. facial recognition, high-precision maps) is subject to "minimum collection" and anonymization requirements. Cybersecurity obligations specify that manufacturers must obtain certification, conduct real-time system monitoring, and report vulnerabilities to regulators. V2X communication systems are required to use encrypted transmission protocols. Finally, ethical risks considerations focus on "unavoidable accident scenarios" (e.g., prioritizing pedestrians vs. passengers), with policies mandating that AV decision-making aligns with public moral consensus and is publicly explainable.

**3.1.5 Technical Standards**

China's automated driving technology standards not only align with international benchmarks but also fully meet domestic regulatory and industry needs. Classification standards such as GB/T 40429-2021 adopt the automated classification system of the United Nations Economic Commission for Europe (UNECE) World Forum for Harmonization of Vehicle Regulations (WP.29), ensuring compatibility with global automated driving terminology.Functional safety standards are established through the GB/T 34590 series (equivalent to ISO 26262) and GB/T 43267-2023 Road Vehicles - Safety of the Intended Functionality (equivalent to ISO 21448), which jointly set out requirements for system design, testing, and verification.In terms of cybersecurity, GB 44495-2024 specifies technical requirements for on-board information security, including engineering practices and risk assessment processes. Additionally, V2X (Vehicle-to-Everything) communication standards and national standards for high-precision maps coordinated and developed by the Ministry of Natural Resources provide support for "vehicle-road-cloud collaboration" pilot projects.

**3.1.6 Operational Capacity**

China's operational capacity for AVs has four main pillars: pilot zones, infrastructure testing, funding support, and personnel training. As of 2024, the country has 17 AV pilot zones established across all provinces, along with 20 "vehicle-road-cloud integration" pilot cities. Beijing's high-level AV demonstration zone has more than 30 enterprises and nearly 1,000 test vehicles. Supporting this expansion, testing facilities for infrastructure such as Shanghai International Automobile City and urban testbeds in districts like Shenzhen's Gangnam provide controlled environments for AV validation. In parallel, regulators in the form of the Ministry of Industry and Information Technology (MIIT) and the Ministry of Transport (MOT) offer certification programs for AV test engineers and safety supervisors, while some of the country's leading universities (e.g., Tsinghua, Shanghai Jiao Tong) have launched academic majors related to Autonomous Driving.

**3.1.7 Sustainability**

China's approach to AV governance factors environmental, social, and long-term viability into its considerations. On the environmental front, AV pilots prioritize electric vehicles (EVs) to align with "dual carbon" goals, with cities such as Beijing and Shenzhen requiring their public transport fleet to consist fully of electric vehicles. Social sustainability is advanced through accessibility initiatives. In these AVs are tested in accessibility scenarios (e.g., autonomous shuttles for the elderly and disabled) to reduce mobility gaps. The "Vehicle-Road-Cloud Collaboration" model is committed to enhancing road safety, and has been proven in pilot areas to effectively reduce the accident rate caused by human error. To ensure long-term viability, policymakers have sought to avoid "pilot fatigue" by linking test results to commercialization.

**3.1.8 Data Governance**

Chinese data governance framework for AVs is structured around classification, privacy protection, and data sharing requirements. Vehicle-related data is divided into three categories: general data, sensitive personal information (e.g. driver facial data), and important data (e.g. high-precision maps and fleet operation data). Important data must be stored domestically, and any cross-border transfer is subject to a security assessment. Privacy protection is reinforced through PIPL, which requires explicit consent for the collection of sensitive personal information (e.g. biometrics). In addition, AV operators must anonymize data used for research and development and are prohibited from secondary use without user consent. To support it, Beijing has introduced a unified AV operation data platform that requires enterprises to share test safety data, such as accident records with regulators, while protecting commercial secrets via data desensitization.

#### 3.1.9 International Cooperation

China’s approach to international cooperation in AV governance include standardization, bilateral cooperation, and participation in multilateral forums. In the area of standardization, China actively participates in UNECE WP.29 on global AV regulations and ISO/IEC JTC1 on AI safety standards, with standards such as GB/T 40429-2021 being referenced in international AV standardization discussions. China and Germany have signed the Joint Statement of Intent on Cooperation in the Field of Autonomous and Connected Driving, and will jointly develop vehicle-to-everything (V2X) technology.

#### 3.1.10 Liability & Risk Management

China’s regulatory framework AVs encompasses liability allocation, insurance, and incident response considerations. At present, China continues to adopt the presumption of liability framework under the Road Traffic Safety Law. Specifically, when a vehicle is operating in automated driving mode (as defined by GB/T 40429-2021), the vehicle operator is deemed liable for accidents by default, unless proven otherwise to be caused by human error or force majeure events. Finally, considering incident management, regulators require AV operators to establish 24/7 emergency response teams, and that accidents must be reported to authorities within 1 hour, with black box data submitted for investigation.

## 3.2 EDUCATION

#### 3.2.1 AI Safety Taxonomy for Education

| Safety Dimension | Definition & Scope |
|---|---|
| **Data Safety** | Protecting student data (learning behavior, grades, biometrics) via "minimum collection," anonymization, and compliance with PIPL’s minor protection provisions. |
| **Content Safety** | Ensuring AI-generated educational content (e.g., textbooks, homework feedback) is accurate, non-biased, and free of harmful information (e.g., violence, misinformation). |
| **Algorithmic Safety** | Preventing algorithmic bias (e.g., discriminatory resource recommendations) and ensuring transparency in AI-driven decisions (e.g., grading, learning path planning). |
| **Ethical Safety** | Safeguarding student well-being (e.g., avoiding over-reliance on AI) and upholding educational equity (e.g., preventing "digital divides" in AI tool access). |

**Table 4**

#### 3.2.2 AI Risk Taxonomy for Education

| Risk Category | Specific Scenarios |
|---|---|
| **Data Risk** | Leakage of student biometrics (e.g., facial recognition for attendance), unauthorized cross-border transfer of learning data, over-collection of psychological assessment data. |
| **Content Risk** | AI-generated textbooks containing factual errors, biased teaching materials (e.g., cultural stereotypes), or inappropriate content (e.g., extreme ideologies). |
| **Algorithmic Risk** | Personalized recommendations reinforcing learning gaps (e.g., low-level resources for disadvantaged students), "labeling" students via algorithmic grading (e.g., "low-potential" tags). |
| **Ethical Risk** | Students becoming overly dependent on AI (e.g., AI writing essays), teachers losing oversight of AI-driven instruction, widening equity gaps (e.g., wealthy schools accessing better AI tools). |

**Table 5**

**3.2.3 Legal & Ethical Framework**

China’s governance of AI in education is structured around a combination of foundational laws, sector-specific regulations, and ethical guidelines. At the foundational level, PIPL (2021) provides special protection for minors under the age of 14, requiring guardian consent for data collection and limiting data processing to the minimum necessary. The Minors Internet Protection Regulations (2024) further mandate age-appropriate design for educational AI tools and prohibit algorithmic practices that encourage addiction (e.g., unlimited AI homework help). Sector-specific rules complement these measures. The Education Mobile Internet Application Recordation System (2019) encourages a “no recordation, no entry into schools” rule for educational apps, a significant measure considering over 2,000 of new apps have been recorded as of 2024. Meanwhile the Notice on Strengthening Primary and Secondary School Mobile Phone Management (2021) restricts classroom mobile phone use to reduce AI tool abuse, such cheating through AI chatbots. The 7th Master Plan for Educational Informatization (2024–2028) emphasizes “safe and controllable” AI application, with a focus on nurturing AI literacy education and teacher training. Ethical are taken into consideration through the New Generation Artificial Intelligence Ethical Guidelines (2021), which require educational AI to prioritize student development, avoid replacing teachers, and ensure fairness.

**3.2.4 AI Safety Considerations**

AI safety in education in China is led by a multi-dimensional framework that holds data, content, algorithmic, and ethical risks as pillars. Student data (e.g., grades, attendance) is classified as “sensitive personal information” requiring educational platforms to use pseudonymization and encryption in data in transit and storage. The Minors Internet Protection Regulations specifies the prohibition of third-party sharing of student data without guardian consent. Consent safety is maintained through an obligatory “double review” - AI plus human oversight - to ensure accuracy for AI-generated educational content (e.g., textbooks, virtual experiments). Algorithmic safety considerations apply to platforms using personalized recommendation algorithms, specifying algorithm record filing with the Cyberspace Administration of China (CAC), the provision of “one-click adjustment” tools for teachers to modify AI recommendations, and the disclosure of algorithm logic to parents (e.g., explaining why a student was recommended a specific learning resource). Ethical safeguards complement these technical requirements policies prohibit AI from replacing core teaching tasks such as essay grading or moral education, while the Ministry of Education “AI Learning” column (on the National Smart Education Platform) includes modules on “ethical AI use” to prevent student over-reliance.

**3.2.5 Technical Standards**

Educational AI tools must comply with GB/T 35273-2020 Information Security Technology—Personal Information Security Specification and adopt standardized APIs to ensure integration with school management systems, including compatibility with the National Smart Education Platform. The Guidelines for the Use of Generative Artificial Intelligence (Generative AI) in Primary and Secondary Schools (2025 Edition), released by the Ministry of Education, centers closely on the application scenarios of generative AI in primary and secondary education. These guidelines clarify usage norms for all school stages to ensure that technology assists teaching, promotes students’ personalized learning, and drives the intelligentization of educational management in a safe, appropriate, and effective manner, while strictly upholding the bottom lines of data security and ethical principles.

In terms of cybersecurity, educational platforms must meet the requirements of GB/T 22239-2019 Information Security Technology—Grade Protection of Information Systems (Level 2 or above) and conduct annual penetration testing. Cloud-based AI tools are required to use domestic cloud services in accordance with data localization guidelines. Additionally, technical accessibility standards mandate that AI educational tools support text-to-speech functionalities, adjustable font sizes, and compatibility with technologies such as screen readers to accommodate students with disabilities.

**3.2.6 Operational Capacity**

As of 2024, the National Smart Education Platform has recorded 40.54 billion page views, with 12.57 billion from students and 11.208 billion from teachers. Furthermore, 95% of primary and secondary schools across China have access to high-speed internet, supporting AI tool development. To support this, the Ministry of Education offers online courses on “AI in Education”, which were completed by 1.2 million teachers in 2024. These courses cover AI safety, algorithm interpretation, and ethical application, while local education bureaus have appointed designated “AI education leads” in 80% of schools to oversee implementation.

In addition, the Ministry of Education launched 100 “AI Education Pilot Schools” in 2024 to test AI tutors, intelligent homework systems, and virtual teachers. The outcomes of these pilot programs are reported and later used to shape national policy, including the refinement of AI textbook standards. Provincial education departments conduct quarterly inspections of AI tools used in schools, as a form of operational oversight, focusing on data compliance and content safety. As a result of these efforts, more than 30 non-compliant AI apps were removed from school environments in 2024.

**3.2.7 Sustainability**

To promote equity, the government provides free access to the National Smart Education Platform's AI tools for rural and underdeveloped regions, helping to reduce the "digital divide". In 2024, 60% of rural students made use of the platform's AI homework help function, demonstrating its growing role in supporting learners. Chinese policies also emphasize a vision that positions "AI as a tool, not a replacement". While AI is increasingly used in administrative tasks such as grading in order to reduce teacher workload, the core aspects of teaching – such as critical thinking - remain human-led.

In addition, environmental sustainability is a core principle. Educational AI platforms are hosted on green data centers powered by renewable energy sources, which help in reducing carbon footprints. The Ministry of Education has further mandated that cloud service providers must meet "net-zero" targets by 2030.

**3.2.8 Data Governance**

According to the Chinese framework, Educational AI tools must meet the requirements of data minimization, meaning they may only collect the data necessary for their function. For instance, an AI tutor may track a student's learning progress data but is not allowed to gather details such as family income. Furthermore, any unused data must be deleted within six months after the end of the school year. With regard to cross-border data transfers, student data may not be transferred abroad unless approved by a data export security assessment. Educational institutions using foreign AI tools, such as international tutoring platforms, are required to store data domestically via localized servers.

Moreover, the framework establishes guardian rights in data governance. Guardians have the right to access, correct, and delete their child's data. To support this, educational platforms must provide a "parent portal" that allows guardians to view AI-generated learning reports and opt out of non-essential data collection.

**3.2.9 International Cooperation**

China actively engages in international cooperation on AI in education through multiple channels. It participates in UNESCO's Global Education Coalition on AI, aligning domestic AI-in-education policies with the UNESCO's Recommendation on the Ethics of AI, with emphasises equity and the preservation of teachers' roles. In the field of standardization, China contributes to ISO/IEC JTC1 SC36 (Learning Technologies), helping to develop international standards for AI educational tools as well as sharing experiences from the National Smart Education Platform.
At the academic level, Sino-foreign university partnerships (e.g., Tsinghua-MIT AI Education Lab) are conducting joint research on AI safety in education. These collaborations focus on key issues such as mitigating algorithmic bias and strengthening the protection of minors' data.

**3.2.10 Liability & Risk Management**

In China schools for AI tool-related harms (e.g., data leaks), bearing primary responsibility if they fail to conduct proper due diligence, for example by using unrecorded apps. To enforce accountability, headmasters are required to sign annual "AI safety responsibility letters." On the provider side, educational AI companies face significant fines in case of non-compliance, including penalties of up to RMB 50 million for incidents such as data breaches or the spread of harmful content.

In addition, policies mandate "human-in-the-loop" oversight for high-stakes AI decisions (e.g., exam grading, student placement). Teachers are required to review on AI-generated grades or recommendations.

## 3.3 CROSS-BORDER DATA FLOWS

**3.3.1 AI Safety Taxonomy for Cross-Border Data Flows**

| Safety Dimension | Definition & Scope |
|---|---|
| **Data Export Safety** | Ensuring cross-border data transfers comply with legal pathways (assessment, standard contracts, certification) and do not compromise national security or individual privacy. |
| **Data Integrity Safety** | Protecting data from tampering (e.g., during cross-border transmission) via encryption, secure APIs, and real-time monitoring. |
| **Jurisdictional Safety** | Mitigating risks from divergent foreign data regulations (e.g., conflicting privacy laws) via cross-border cooperation and mutual recognition mechanisms. |
| **Supply Chain Safety** | Ensuring third-party service providers (e.g., foreign cloud vendors) adhere to China's data security standards and do not misuse transferred data. |

### 3.3.2 AI Risk Taxonomy for Cross-Border Data Flows

| Risk Category | Specific Scenarios |
|---|---|
| **Compliance Risk** | Transferring important data without a security assessment, using expired standard contracts, or failing to meet foreign regulatory requirements (e.g., GDPR). |
| **Security Risk** | Data breaches during cross-border transmission (e.g., unencrypted channels), unauthorized onward transfer by foreign recipients, or cyberattacks on overseas servers. |
| **Jurisdictional Risk** | Conflicting legal requirements (e.g., China's data localization vs. foreign data access demands), leading to "double compliance" burdens or penalties. |
| **AI Model Risk** | Training AI models on cross-border data that is biased or non-compliant (e.g., using foreign personal data without consent), leading to model unfairness or legal liability. |

### 3.3.3 Legal & Ethical Framework

China's framework for cross-border data governance is anchored on a foundation of laws, specialized rules, and ethical principles. The Cybersecurity Law (2017) requires critical information infrastructure operators (CIIOs) to store personal information and important data domestically, while any cross-border transfer must be approved by the CAC. Complementing this, the Data Security Law (2021) establishes a data classification and categorization system and requires security assessments for the transfer of "important data" abroad. The PIPL (2021) further refines the framework by defining three cross-border pathways for personal information transfers: security assessment, standard contracts, and personal information protection certification.
In addition to these foundational laws, specialized rules provide more guidance. The Measures for Data Export Safety Assessment (2022) mandate assessments for transfers involving the personal information of 100,000 or more individuals or any "important data," with more than 250 assessments conducted in 2024. The Provisions on Promoting and Regulating Cross-Border Data Flows (2024) expand exemptions to cover activities such as academic cooperation and emergency rescue, while also clarifying that "unidentified important data" does not require an assessment. The Measures for the Administration of Personal Information Export Standard Contracts (2023), meanwhile, provide templates for transfer contracts involving fewer than 100,000 individuals' personal information.

Underlying these rules are key ethical principles. The concept of "data sovereignty with trust" emphasizes that cross-border data flows must uphold national security and individual rights, for example by preventing foreign surveillance via unauthorized data access. At the same time, the principle of "fair and equitable" governance discourages data localization requirements that hinder legitimate cross-border cooperation as per the China-Singapore Free Trade Agreement.

### 3.3.4 AI Safety Considerations

AI safety in the context of cross-border data flows requires careful consideration of multiple dimensions. One major concern is data classification risk, as AI training data may include "important data" (e.g., industrial AI uses manufacturing data) that require assessment before cross-border transfer. Another key aspect is model training safety, since AI models trained on cross-border data must comply with both Chinese and foreign laws (e.g., avoiding GDPR-protected data without consent).

The security of transmission is equally important: cross-border data transfers must use encrypted channels (e.g., TLS 1.3) and secure APIs. Finally, enterprises are required to conduct due diligence on foreign recipients (e.g., verifying their data security capabilities) and include "onward transfer" restrictions in contracts, which prohibit recipients from transferring data to third parties without explicit approval.

### 3.3.5 Technical Standards

In terms of data security standards, the GB/T 35273-2020 Information Security Technology—Personal Information Security Specification and the GB/T 43697-2024 Data Security Technology—Data Classification and Grading Rules provide technical criteria for identifying sensitive data and assessing cross-border risks. For certification standards, the Personal Information Protection Certification Implementation Rules (2022) sets technical requirements for certification, such as data anonymization and access control. These standards have been recognized in more than ten countries via bilateral agreements. With respect to AI model standards, the TC260 Artificial Intelligence Safety Governance Framework (2024) includes guidelines for auditing AI models trained on cross-border data. These guidelines ensure compliance with data origin laws.

### 3.3.6 Operational Capacity

China's operational capacity for managing cross-border data flows is supported by coordinated efforts at the regulatory, enterprise, and technical levels. At the regulatory level, the Cyberspace Administration of China (CAC) takes the lead in establishing the "Expert Database for Cross-Border Data Flow Supervision," bringing together university scholars and industry specialists. In 2023, the CAC also launched the Data Export Service Platform, which simplifies processes of assessments and contract filings.

At the enterprise level, large companies have developed dedicated data governance teams to manage cross-border data flows, while small and medium-sized enterprises (SMEs) can access government-subsidized consulting services through local IT associations. The system is furthered by technical tools provided by vendors who offer cross-border data security tools, including data classification software, encryption modules, and real-time monitoring systems – tools that have been adopted of large enterprises in 2024.

**3.3.7 Sustainability**
China's approach to sustainable cross-border data governance is supported by multilateral cooperation, innovation facilitation, and green data flows. In terms of multilateral cooperation, China participates in the Global Cross-Border Privacy Rules (CBPR) Forum as an observer, promoting the mutual recognition of data protection standards and helping to reduce "double compliance" costs for enterprises. On the innovation side, the Provisions on Promoting and Regulating Cross-Border Data Flows (2024) exempt certain low-risk scenarios from assessment. This flexibility aids in fostering international collaboration in AI research and development.

At the same time, Chinese policies encourage the use of "edge computing" and "federated learning" to reduce cross-border data volume (e.g., training AI models locally and sharing only model parameters). Such initiatives contribute to the lowering of the global carbon footprint.

**3.3.8 Data Governance**
China's cross-border data governance framework requires enterprises to implement safeguards throughout the whole process of transferring data abroad. Prior to any transfer, enterprises must conduct a cross-border data transfer risk assessment evaluating the recipient's data security capabilities, the risks of data misuse - such as unauthorized access by foreign governments - and the impact on national security and individual rights. Once data has been transferred, enterprises must monitor foreign recipients' data use – including the use of audit logs. Any breaches must be reported to regulators within 72 hours.

The governance framework also incorporates measures for individual rights. PIPL's (2021), "right to data portability" allows individuals to request transfer of their personal data to domestic or foreign service providers, provided the recipient meets safety standards. As for overseas data storage, critical data - including national defense and public health information - must be stored domestically, while non-critical data may be stored abroad if it is subject to a security assessment or standard contract.

**3.3.9 International Cooperation**
China actively advances international cooperation in cross-border data governance through bilateral, multilateral, and mutual recognition initiatives. China has established a comprehensive free trade relationship with Singapore, committing to the principle of "data free flow with trust" and establishing mutual recognition of data security standards. In multilateral forums, China contributes to the OECD's Working Party on Data Governance and Privacy as well as the GPAI's Data Governance Working Group, where it advocates for balanced cross-border data rules.

Beyond the mentioned before, China provides technical assistance to developing countries – via initiatives such as the Digital Silk Road -, to help partner nations in establishing data security frameworks, expanding the global network of trusted data partners.

**3.3.10 Liability & Risk Management**
China's cross-border data flow framework imposes liability and risk management measures on enterprises and foreign recipients. Enterprises face penalties, including fines for non-compliant cross-border data transfers.

To minimize risks, the government encourages companies to purchase "cross-border data liability insurance", which provides compensation in the event of breaches. By 2024, 30% of large enterprises had adopted such insurance. In addition, China has established cross-border dispute resolution mechanisms with the EU and ASEAN, using mediation to resolve conflicts, such as conflicting data access requests.

## 4. Interoperability

### **4.1** OVERVIEW OF INTEROPERABILITY IN CHINA'S AI SAFETY GOVERNANCE

Interoperability in China's AI safety framework refers to the ability of laws, standards, and systems to work consistently across sectors, regions, and international borders. This framework is organized into three core layers. The governance layer focuses on the alignment of principles - such as safety and fairness - across sectoral regulations and international norms. The security layer emphasizes consistency in cybersecurity, data protection, and risk management practices across sectors. Finally, the technical and data layer ensures compatibility of technical standards - including data formats and APIs - and data governance mechanisms.

The table below summarizes interoperability performance across the three priority sectors:

| Sector | Legal/Organizational | Technical/Interface | Data/Semantic | Overall Assessment |
|---|---|---|---|---|
| **Autonomous Vehicles** | Comprehensive laws (e.g., Beijing Regulations) with clear liability; aligned with UNECE WP.29. | Aligned with ISO/UNECE standards but domestic V2X standards need global harmonization. | Strict data localization; limited cross-border data sharing frameworks. | Strong foundation with international standard alignment gaps. |
| **Education** | Clear minor protection rules; aligned with UNESCO ethics. | Dedicated AI-in-education standards still being formulated; relies on general data standards. | Child-specific data protections; incompatible with some foreign education platforms. | Developing framework with technical standardization needs. |
| **Cross-Border Data Flows** | Clear pathways (assessment/contracts); aligned with OECD principles. | Adopts global encryption/ API standards; CBPR alignment in progress. | Data classification aligns with GDPR but minor differences in "important data" definition. | Effective operational framework with minor jurisdictional frictions. |

### **4.2** INTEROPERABILITY MATRIX (GOVERNANCE, SECURITY, TECHNICAL/DATA LAYERS)

| Sector | Governance Layer | Security Layer | Technical/Data Layer |
|---|---|---|---|
| **Autonomous Vehicles** | - Automotive Data Security Measures (2021)<br>- Beijing Autonomous Vehicle Regulations (2025)<br>- Alignment with UNECE WP.29 principles. | -GB 44495—2024 Technical requirements for vehicle cybersecurity<br>- Cybersecurity certification for manufacturers. | -GB/T 40429-2021 Taxonomy of driving automation for vehicles<br>- Digital event data recorders (black boxes). |
| **Education** | - Minors Internet Protection Regulations (2024)<br>- Education APP Recordation System<br>- Alignment with UNESCO AI Ethics Recommendation. | -GB/T 22239-2019 Information security technology — Baseline for classified protection of cybersecurity<br>- Content double review (AI + human)<br>- Annual penetration testing for platforms. | - GB/T 35273-2020 Information security technology—Personal information security specification<br>- National Smart Education Platform APIs<br>- WCAG 2.1 (accessibility standards). |
| **Cross-Border Data Flows** | - Data Export Safety Assessment Measures (2022)<br>- Provisions on Promoting Cross-Border Data Flows (2024)<br>- Alignment with OECD Data Governance Principles. | - TLS 1.3 encryption - Pre-transfer risk assessment<br>- Post-transfer monitoring (audit logs). | - GB T 43697-2024 Data Security Technology - Rules for Data Classification and Grading<br>- Standard contract templates (2023)<br>- ISO/IEC 27701 (privacy management). |

## **4.3** INTEROPERABILITY STRENGTHS AND GAPS

**4.3.1 Strengths**

China's AI safety governance framework shows several distinguished strengths. First, governance coherence is achieved by consistently applying core principles such as safety, privacy, and fairness across sectors via laws including the Cybersecurity Law, the Data Security Law, and PIPL. This ensures major regulatory contradictions will not happen. Second, the framework accounts for scenario-driven interoperability, where pilot zones - for example, AVs in Beijing or AI education programs in 100 schools - are used to test interoperable practices, such as cross-city data sharing for AVs, before national rollout.

A third strength is the rigor of data security, as strict data classification and localization rules create a "trust baseline" for cross-sector data sharing (e.g., AV and smart city data integration). Finally, international standard alignment is of benefit to the frameworks; key technical standards, such as GB/T 40429-2021 for AVs, reference ISO and UNECE benchmarks, facilitating global compatibility.

**4.3.2 Gaps**

Despite its progress, China's AI safety governance framework still faces several gaps. One challenge is the lack of international standard mutual recognition. While China has developed domestic standards (e.g., V2X), these are not yet fully recognized by major economies like the European Union and the United States, creating barriers for Chinese AV exports. Another issue is cross-border data compliance frictions. Differences in how "important data" is defined - for example, between China and the EU – lead to "double compliance" burdens for AI enterprises operating globally.

A further gap lies in the sector-specific standard maturity. While fields like AVs already benefit from stablished frameworks such as ISO 34503, dedicated AI safety standards for Education are still being developed. As a result, relying on general data standards that are less precise. Finally, there are regional implementation inconsistencies. For instance, local variations in AV accident handling - such as the different practices in Beijing versus Guangzhou - hinder cross-region interoperability for national AV fleets.

# **5. Jurisdiction Profile – China**

## **5.1** SNAPSHOT

China, with a population of approximately 1.4 billion, has a digital economy that accounted for 41.5% of GDP in 2024. The nation has set a strategic goal of becoming a global AI leader by 2030. Its AI innovation capacity is significant as the country is home to leading technology enterprises such as Baidu, Alibaba, and Tencent, and ranks second globally in AI patent applications as of 2024, while also hosting the World Artificial Intelligence Conference (WAIC).

Key policies and legislation provide the foundation for this governance model. These include the New Generation Artificial Intelligence Development Plan (2017), which sets the national AI strategy, as well as the Cybersecurity Law (2017), the Data Security Law (2021), PIPL (2021), which establish foundational and security legal framework. Sector-specific rules complement these laws; some examples are the Automotive Data Security Measures (2021) and the Minors Internet Protection Regulations (2024).

China also engages in international alignment, being a signatory to the OECD AI Principles, the UNESCO AI Ethics Recommendation, and the UN Global Digital Compact. The country also participates in bodies such as ISO/IEC JTC1 and UNECE WP.29 and has signed digital economy agreements with more than ten countries.

Public sentiment toward AI in China reflects a stance of cautious optimism. A 2024 survey showed that 68% of citizens support the use of AI in healthcare and transport, but 75% express concern about data privacy. The 2022 Didi incident increased public demand for stricter cross-border data regulation.

## **5.2** DIMENSION-BY-DIMENSION FINDINGS

**5.2.1 Governance & Institutions**

China's governance and institutional framework for AI safety shows notable strengths but also important gaps. The adoption of a "pilot-first" model (e.g., AV zones) allows iterative rule refinement before national implementation. In addition, dedicated regulators play key roles through initiatives who have established specialized AI safety teams.

Despite these strengths, gaps remain. Local regulatory capacity varies, with rural areas lacking expertise in AI data governance. Moreover, voluntary ethical guidelines – such as the one for AI in education - rely on industry compliance, with limited enforcement.

**5.2.2 Ethical & Legal Framework**

China's ethical and legal framework for AI governance combines strong legal protections, but even so challenges remain. On the strengths side, the PIPL includes minor protection provisions and Data Security Law's classification system embed ethical values such as privacy and fairness into law. The "Measures for the Ethical Review of Science and Technology (2023)" was passed, requiring ethical review of artificial intelligence activities in the list of documents, further strengthening supervision and ensuring compliance with human dignity standards. In addition, a range of soft-law instruments, such as the AI Ethical Guidelines, provide flexible guidance for emerging AI.

At the same time, the framework faces gaps. Legal grey areas for generative AI (e.g., IP rights for AI-generated content) require further clarification. Furthermore, cross-border ethical coordination (e.g., AI model training on global data) lacks formal mechanisms.

**5.2.3 Technical Standards & Infrastructure**
China has made significant progress in developing technical standards and infrastructure to support AI safety. Among its strengths, the country has established robust technical standards for critical sectors, including GB/T 40429-2021 for AVs and GB/T 35273-2020 for data security. Its advanced digital infrastructure accounts for developments such as the 5G coverage extending to 98% of the population, the establishment of national AI compute clusters such as the Shanghai AI Lab, and the National Smart Education Platform. Beyond that, standards hubs such as TC260 (information security) and the China Electronics Standardization Association (CESA) play a central role in advancing AI safety standardization.

Nonetheless, gaps can still be found. Education and smaller-scale AI applications still lack dedicated safety standards. Furthermore, international mutual recognition of domestic standards (e.g. V2X) is limited.

**5.2.4 Operational & Organizational Capacity**
China demonstrates strong operational and organizational capacity in implementing AI safety measures. Among the strengths, large enterprises have established dedicated AI safety teams, such as Baidu's AV Safety Office, and comply to strict audit requirements. Regulators also contribute by offering training programs, most notably the CAC's "Cross-Border Data Governance Training", which trained over 10,000 participants in 2024. In addition, the government has created an array of testbeds, including 17 AV zones, 20 "vehicle-road-cloud integration" cities, and 100 AI education pilot schools.

However, there are gaps in capacity. SMEs lack the resources for AI safety compliance, in areas like data classification tools. Moreover, local governments in underdeveloped regions require more support to strengthen their AI safety oversight.

**5.2.5 Data Governance & Privacy**
China boosts from a solid data governance and privacy framework that combines legal clarity, innovative mechanisms, and strict enforcement. Among the strengths, the country has a unified data governance structure framework through the Data Security Law, which sets classification requirements, the PIPL, which focuses on privacy rights, and cross-border rules that provide clarity. Innovative mechanisms further support this system. The "Data Export Service Platform," which streamlines compliance, and the "federated learning" enables cross-border AI training without raw data transfer. Enforcement is also strict, with high-profile cases (e.g. Didi) serving as deterrents to non-compliance and reinforcing public trust.

At the same time, there are gaps. The definition of "important data" remains sector-specific and lacks global harmonization, creating cross-border friction. In addition, anonymization techniques for AI training data require further standardization.

**5.2.6 International Cooperation & Alignment**
China plays an increasingly active role in international cooperation and alignment on AI governance. Among the strengths, the country actively participates in global standardization efforts through organizations such as ISO/IEC and UNECE, and also in multilateral forums like GPAI and the OECD. Bilateral initiatives also play a key role, with digital agreements such as the China-Singapore one promote cross-border data flows with trust. Furthermore, China invests in capacity building by supporting developing countries in the establishing data security frameworks, through initiatives like the "Digital Silk Road".

Nonetheless, gaps can still be found. Mutual recognition of data protection regimes with major economies is still limited - for example, equivalence with the EU GDPR is still pending. In addition, geopolitical tensions hinder collaboration on frontier AI safety (e.g., AGI research).

**5.2.7 Liability & Accountability**
China has made notable progress in clarifying liability and accountability within its AI governance framework. There is clear liability allocation in critical areas, such as operator liability in AVs autonomous mode and enterprise liability for cross-border data breaches. The country also enforces "human-in-the-loop" requirements to ensure accountability for AI-driven decision-making - for example in education grading. In addition, China has promoted insurance mechanisms such as AV liability insurance and cross-border data liability insurance.

Despite these advances, gaps remain to be addressed. National-level accident liability rules for AVs are pending, which has resulted in local inconsistencies. Furthermore, liability for diffuse AI harms (e.g., algorithmic discrimination) is not clearly defined.

## 5.3 SECTOR CASE STUDIES HIGHLIGHTS

**5.3.1 Autonomous Vehicles: Beijing Autonomous Vehicle Regulations (2025)**
Beijing's 2025 Autonomous Vehicle Regulations formalizes the staged process of road testing, demonstration, and commercial operation process for AVs, expanding scenarios to include private cars, taxis, and logistics The regulations introduce several key innovations. A unified operation data platform requires all AV enterprises to share safety-related data, such as accident records, with regulators, enabling cross-enterprise safety analysis. They also emphasize public participation, allowing residents to apply to test AV services with their feedback incorporated into policy refinement. In addition, strict insurance requirements mandate that operators purchase RMB 5 million in liability coverage, ensuring accident compensation.

From an interoperability perspective, the regulations align with UNECE WP.29 principles and use GB/T 40429-2021 for grading, facilitating future mutual recognition of test results with foreign countries.

**5.3.2 Education: National Smart Education Platform (2024)**
The National Smart Education Platform, is China's primary hub for AI-in-education hub, with 40.54 billion page views by 2024. It offers AI tools for personalized learning, homework help, and teacher training to students all over the country. Several key innovations underpin its success. The platform boosts from standardized APIs, ensuring that AI tools are fully compatible with school management systems nationwide. It also establishes a framework for content review, where AI-generated materials undergo "AI + human" review, with a 95% accuracy requirement. The platform also emphasizes equity, offering free access to rural schools to help bridge the digital divide.

From an interoperability standpoint, the platform's data governance aligns with PIPL and UNESCO ethics, serving as a model for safe AI-in-education interoperability.

**5.3.3 Cross-Border Data Flows: Didi Data Security Incident (2022)**
In 2022, the CAC fined Didi RMB 8.026 billion for illegally collecting and transferring user data overseas to train AI models, marking China's largest cross-border data enforcement action to date. The case produced several key outcomes. First, it led to strengthened cross-border data assessment requirements, where enterprises must conduct "data mapping" to identify sensitive fields before transfer. Second, it established model training compliance rules, specifying that AI models trained on Chinese data may not be deployed overseas without approval. Third, it had a strong deterrent effect across the industry, with more than 300 enterprises voluntarily conducting cross-border data audits following the case.

The 2021 Didi case highlighted the need for aligning AI model training with cross-border data rules, prompting the 2024 Provisions on Promoting Cross-Border Data Flows to clarify exemptions for low-risk research.

### **5.4** GOOD PRACTICES AND POLICY PRIORITIES

**5.4.1 Good Practices**
Chinese AI safety governance showcases several good practices, one of the most significant being the multi-layered governance framework, which progresses from laws to regulations to standards to pilots, balancing flexibility and rigor, allowing for iterative improvement (e.g., AV rules evolved from 2018 test guidelines to 2025 regulations). Another is the emphasis on scenario-driven pilots, such as AV test zones and AI education programs, which test interoperable practices in controlled environments, reducing risk before national rollout.

Data governance also reflects a "security by design" approach, with principles such as minimum data collection, local storage, and pre-transfer assessments embedded in rules, ensuring data safety throughout the AI lifecycle. Finally, China has prioritized public trust building through both high-profile enforcement actions (e.g. Didi) and transparency mechanisms (e.g. AV safety data platforms) which together reinforce public trust in AI safety.

**5.4.2 Policy Priorities**
Looking ahead, several policy priorities are critical for strengthening China's AI safety governance and interoperability. First, there is a need to **Unify National AV Liability Rules** by developing national legislation to clarify accident liability for L4/L5 AVs, resolving local inconsistencies and facilitating cross-region operations. Second, China should work to **Enhance Cross-Border Data Facilitation** by expanding exemptions for low-risk AI research, as well as clarify "important data" definitions.

A third priority is to **Develop Education AI Standards** by launching dedicated technical standards for AI-in-education in areas such as content quality and algorithm fairness to fill current gaps. Additionally, it is important to **Strengthen SME Capacity** by providing subsidies for AI safety tools, such as data classification software, and offering training on cross-border compliance.

Finally, China should **Deepen International Standard Cooperation**, leading or co-leading ISO/IEC working groups on AV V2X and AI education standards to promote mutual recognition.

## 6. Conclusion – Consolidated Recommendations

China has established a robust AI safety governance framework characterized by **multi-layered laws, scenario-driven pilots, and strict data security**. To enhance interoperability and align with global best practices, the following recommendations build on existing policies and address identified gaps:

### **6.1** FOR AUTONOMOUS VEHICLES

For autonomous vehicles, first, China should **implement a Implement National AV Liability Rules** meaning that the country should develop a national-level “Autonomous Vehicle Accident Liability Law” to unify local practices, clarify rules for presenting evidence of software defects, and align with UNECE WP.29 liability principles.

Second, the country should aim to **Harmonize V2X Standards Globally**. To further this objective, China should lead ISO and IEC working groups to align Chinese V2X standards with EU and US standards, enabling cross-border AV communication.

Third, the nation should be encouraged to **Expand “Vehicle-Road-Cloud Integration” Pilots**, the number of participant cities should be increased, to test cross-city data sharing and infrastructure interoperability, with the objective of developing national standards. Finally, China could **Facilitate Test Result Mutual Recognition,** reducing export barriers for Chinese AV enterprises.

### **6.2** FOR EDUCATION

In the education sector, China should prioritize the **Development of Dedicated AI-in-Education Standards**. This could be achieved by the launch of a Technical Specifications for AI Educational Tools - covering content quality, algorithm fairness, and data security - referencing on global AI-in-education benchmarks such as IEEE 3527.1. **Teacher capacity must also be strengthened by AI training**. Safety and ethics training should be integrated into both initial teacher training and continuous education, with the target of full teacher coverage. To reduce inequality, China must **Expand Equity-Focused AI Access** by providing additional funding for rural schools to adopt AI tools and train teachers, reducing the “digital divide”. Furthermore, the **Establishment of an AI Content Review Platform** for centralized national-level review of AI-generated educational content, ensuring consistency and accuracy across regions is highly recommended.

### **6.3** FOR CROSS-BORDER DATA FLOWS

For cross-border data flows, there is an urgent need to **Clarify “Important Data” Definitions**. Sector-specific lists, including categories such as AI training datasets, should be issued to reduce enterprise uncertainty and align with global practices, such as those of the OECD.

**Expanding Mutual Recognition Agreements** - would further facilitate international AI research and development. At the technical level, China should **Promote Privacy-Enhancing Technologies**, by funding research and development on federated learning and synthetic data, to enable secure cross-border AI training without raw data transfers. Finally, the country should **Simplify SME Compliance** through the launch of a a national “Cross-Border Data Compliance Toolkit” for SMEs, providing template assessments and free consulting services.

### **6.4** CROSS-CUTTING RECOMMENDATIONS

Several cross-cutting measures are proposed to strengthen China’s AI safety governance. First, the **Establish a National AI Safety Coordination Council** would help to resolve regulatory inconsistencies and represent China in global AI governance forums by centralizing the coordination of cross-sector AI safety policies. Second, sustained **Invest in Frontier AI Safety Research** is essential, to advance research on AGI safety, algorithmic bias mitigation, and cross-border AI risk assessment.

Third, the **Enhancement of Public Engagement** through the launch of a national “AI Safety Dialogue” series is highly recommended. This initiative would compile museums, and schools to collect public feedback and build trust in AI systems through media. Finally, a **Monitor Policy Effectiveness** should put into place to Develop KPIs for AI safety (e.g., AV accident rates, data breach incidents) and publish an annual China AI Safety Report to track progress and refine policies.

United Nations University
Institute in Macau

COUNTRY REPORTS

# Singapore Country-Level AI Safety Interoperability Report

JAMES ONG, SAMEER GAHLOT
ARTIFICIAL INTELLIGENCE INTERNATIONAL INSTITUTE (AIII)
JAMES.ONG@ORIGAMI-FRONTIERS.COM
GAHLOT.LEGAL@GMAIL.COM

UNU Macau

www.unu.edu

# Singapore Country-Level AI Safety Interoperability Report

15 October 2025

## Background

With the launch of Smart Nation initiative to National AI Strategy 1.0 in 2014 and 2018, respectively, the focus of the Singapore Government has always been on digitising, ensuring convenience and safety of the netizens [1, 2, 3, 4, 5, 6,]. This AI Safety Interoperability Report aims to provide a comprehensive review of the AI related policies and trace certain development(s) reflected through its rules, regulations and policies. The report further delves deeper into the challenges and opportunities presented with the development and deployment of Artificial Intelligence (AI).

Lastly, it provides a concrete set of recommendations, aligning with the Global Digital Compact and Smart Nation Initiative 2.0. It is intended for a professional audience, including policymakers, industry leaders, technologists, and legal and academic researchers, who require a detailed understanding of the frameworks, policies, and practical applications of AI governance within Singapore's key strategic sectors. The purpose of this analysis is to provide a reference resource that can inform decision-making and foster a deeper appreciation for Singapore's nuanced, multi-stakeholder approach to responsible AI.

## Jurisdiction Profile – Singapore

**Snapshot**

As a city-state, Singapore is one of the world's first in introducing AI governance framework in 2019 with the Model AI Governance Framework [9, 10, 11, 12, 13]. It takes a very pragmatic, hybrid top-down and bottom-up approach with collaboration between government and industry to provide AI safety guidelines while empowering industry with a community to provide industry-ready tools to support multilingual safety evaluation that translate high-level principles into day-to-day practice and use for the companies.

Singapore does have a national AI strategy that has evolved from its initial 2019 plan (NAIS 1.0) to a more comprehensive 2023 update (NAIS 2.0). NAIS 1.0 (2019) focused on a project-based approach, identifying five key national AI projects in sectors such as healthcare, education, logistics, and border clearance. The primary goal was to deepen the use of AI to transform the economy. NAIS 2.0 (2023) represents a shift from "opportunity to necessity" and from a "local to global" perspective. It expands upon the first strategy by taking a systems-based approach. The strategy is built on three key systems:

- Activity Drivers: This system focuses on integrating AI across industry, government, and research.
- People & Communities: This system aims to grow the AI talent pool and empower individuals and businesses with AI skills.
- Infrastructure & Environment: This system is dedicated to building the necessary computational and data infrastructure while fostering a trusted environment for AI innovation.

The overall goals of NAIS 2.0 are to achieve Excellence by developing specific peaks of AI excellence and Empowerment by enabling people and businesses to use AI confidently.

Further, Singapore's position is to actively contribute to and build upon international efforts. The Singapore Consensus on Global AI Safety Research Priorities, a document that builds on the International AI Safety Report, organizes AI safety research into a "defence-in-depth model" [14, 15]. This framework is divided into three areas:

1. Risk Assessment: Understanding the severity and likelihood of potential harms.
2. Developing Trustworthy, Secure and Reliable Systems: Focusing on building safe systems from the ground up.
3. Control: Monitoring & Intervention: Implementing tools for monitoring and intervening after a system has been deployed.

The Singapore Consensus serves as a roadmap for future collaborations and aims to facilitate conversations between global AI scientists and policymakers. It highlights areas of mutual interest for cooperation, where sharing safety advances offers minimal competitive advantage, much like collaboration on aviation safety.

## 1. Introduction

Singapore's approach to AI safety and governance is a strategic and pragmatic blend of national vision, flexible frameworks, and targeted legislation. Rather than relying on a single, omnibus AI law, the nation has established a cohesive ecosystem designed to foster innovation while building public trust. This is a core component of its overarching **Smart Nation** initiative, with a central focus on **Trust, Growth, and Community**.

The National AI Strategy 2.0 (**NAIS 2.0**) serves as the operational blueprint, laying out a 15-point plan to position Singapore as a global leader in responsible AI. [1,2,3,4]. Singapore's national AI safety policies are an integrated component of its broader **Smart Nation** initiative. Launched with the goal of harnessing information technology to boost economic growth and improve citizens' lives, the initiative has evolved into the **Smart Nation 2.0** blueprint, unveiled in 2024 by Prime Minister Lawrence Wong. [5] This refreshed strategy is built on three core pillars: **Trust**, Growth, and Community. The emphasis on Trust—ensuring citizens can confidently use digital tools with protected safety and privacy—is a direct recognition of the societal importance of AI governance. This strategic imperative is also being executed through the government's own use of AI to deliver public services, such as leveraging the Scam Analytics and Tactical Intervention System (SATIS) [6] and deploying AI-enabled productivity tools like Pair [7] and SmartCompose [8].

To realize this vision, Singapore introduced its **National AI Strategy 2.0 (NAIS 2.0)**, a comprehensive "whole-of-government, whole-of-economy" approach. The strategy is not merely a statement of intent; it is a meticulously detailed plan with 15 courses of action to be executed over the next three to five years. These actions range from anchoring new AI "centers of excellence" and strengthening the AI startup ecosystem to attracting top global talent and upskilling the workforce. A significant investment has already been committed, with more than S$500 million allocated through AI Singapore (AISG) under the Research, Innovation and Enterprise (RIE) 2020 and 2025 plans.

This substantial public investment in AI research and development creates a direct need for a robust governance framework. The government's proactive role in fostering AI as a transformative force for the public good means that policies for responsible use must be developed concurrently with technological advancement. The strategic philosophy is that AI regulation should not be a reactive measure implemented after technology is deployed and risks have materialized. Instead, governance is an inseparable part of the development lifecycle, designed to build a trusted environment where AI can flourish responsibly from its inception. A key pillar of NAIS 2.0 is, therefore, to "ensure fit-for-purpose regulatory environment" and "improve security and resilience baseline".

**The Governance Architecture**

Singapore's approach to AI governance is characterized by its reliance on **voluntary frameworks and targeted legislation**, a deliberate "quasi-regulation" strategy that provides agility in a fast-evolving technological landscape. The cornerstone of this approach is the **Model AI Governance Framework**, initially released in 2019 for traditional AI and updated in 2024 to address the unique characteristics and risks of Generative AI (GenAI). This framework provides "readily-implementable guidance" for private sector organizations, detailing how they can responsibly develop and deploy AI solutions by addressing ethical and governance issues [9, 10, 11, 12, 13] and AI Safety issues [14, 15]

The GenAI Framework introduces several "dimensions of trust" that organizations are encouraged to implement [9, 36, 37]. These include **accountability**, which involves allocating responsibility across the full AI ecosystem and considering indemnities for end-users ; **data quality** and **transparency**, which advises the use of privacy-enhancing technologies (PETs) like anonymization and the disclosure of information about training data through a "food label" approach; and **trusted and safe development**, which recommends implementing best practices like fine-tuning models with human feedback to reduce harmful output and developing a comprehensive safety evaluation framework.

A critical tool for implementing these principles is **AI Verify**, an AI governance testing framework and software toolkit. Developed by the **Infocomm Media Development Authority (IMDA**) and its not-for-profit subsidiary, the **AI Verify Foundation**, this toolkit helps organizations validate their AI systems against 11 ethical principles, including transparency, fairness, security, and robustness [13]. This focus on promulgating standards and audit-like frameworks is a key characteristic of Singapore's digital governance to share the future together.

The Model AI Governance Framework, while voluntary, includes provisions that suggest a potential shift toward a more specific, mandatory legal regime for certain high-stakes applications. The framework contains "signposts" that "may signal future changes in Singapore law" and recommends that governments need "greater transparency for higher-risk models, especially if they have national security or societal implications". This indicates a flexible, risk-based approach where the government is prepared to move from soft law to hard law as the technology and its societal implications mature. This pragmatic adaptability ensures that regulations do not stifle innovation prematurely but can be tightened when necessary to mitigate emerging risks.

**Key Entities and Their Roles**

Singapore's AI governance is managed by a network of specialized government agencies and advisory bodies. The **Infocomm Media Development Authority (IMDA)** and the **Personal Data Protection Commission (PDPC)** are the principal regulatory bodies, working in close collaboration. The PDPC, established in 2013, administers and enforces the **Personal Data Protection Act (PDPA)**, which is the foundational law for governing the collection, use, disclosure, and care of personal data. The PDPC's role extends to providing guidance to private sector organizations on addressing AI governance issues when using AI solutions.

Moreover, an additional layer of governance is provided by the **Advisory Council on the Ethical Use of AI and Data** [11, 38]. Established in 2018, the council comprises representatives from technology companies, AI users, and other key stakeholders.[19] Its primary role is to advise the government and engage with stakeholders to provide guidance on the responsible development and deployment of AI. Singapore's national AI strategy is dynamic and responsive, prepared to evolve from voluntary guidelines to mandatory law as technology matures and risks emerge. The multi-stakeholder and globally-minded approach positions the nation not only as a hub for AI innovation but also as a key international partner in AI governance. The focus on consensus-building between government, industry, and citizens is a deeply embedded characteristic of Singapore's broader digital governance ecosystem.

| Authority | Framework/Source | Function and Scope |
|---|---|---|
| **IMDA** | Model AI Governance Framework (GenAI & Traditional AI), AI Verify [9, 10, 11, 13] | Provides voluntary, principles-based guidance and a technical testing framework for organizations. |
| **PDPC** | Personal Data Protection Act (PDPA), Proposed Advisory Guidelines on AI Recommendation and Decision Systems [12] | Enforces data protection laws and provides specific guidance for the use of personal data in AI systems. |
| **LTA** | Road Traffic Act and Rules for AVs, CETRAN Assessment Framework [17, 18, 19] | Regulates and certifies Autonomous Vehicles to ensure physical safety and mitigate public liability risks. |
| **MOE** | EdTech Masterplan 2030, Student Learning Space (SLS) [20, 21, 22, 23] | Integrates AI into education to enhance learning outcomes while providing ethical guidelines for use within the school system. |

The following table provides a clear overview of the key sources and authorities governing AI safety across the identified sectors:

### **1.2** SCOPE

The scope of this report is a focused examination of Singapore's national AI safety strategy as it applies to three specific sectors: Autonomous Driving, Education, and Cross-Border Data Flow. The analysis is limited to the risks posed by advanced AI systems with broad capabilities, often referred to as "generative AI" or "general-purpose AI (GPAI)," which can perform or be adapted to perform a wide range of tasks. We distinguish between "societal" risks (e.g., job displacement, misinformation) and "catastrophic" risks (e.g., loss of human control). The report does not delve into sector-specific rules for narrower applications unless they directly inform the broader governance framework.

### **1.3** ANALYTICAL LENS

The analysis is conducted through a lens of pragmatic governance, which is a hallmark of Singapore's governance framework. [1, 2, 3, 4, 5] The report explores how the nation balances a commitment to fostering technological innovation and economic growth with the proactive development of safety standards and frameworks to build public trust. It highlights the strategic use of voluntary guidelines alongside targeted, risk-based legislation to maintain agility in a rapidly evolving technological landscape.

### **1.4** METHODOLOGY

The research is based on a comprehensive review of a wide range of public and professional sources, primarily relying on secondary research. This includes official government publications and policy documents from agencies such as the Infocomm Media Development Authority (IMDA), the Personal Data Protection Commission (PDPC), the Land Transport Authority (LTA), and the Ministry of Education (MOE). The analysis also incorporates insights from legal research papers and industry reports to provide a holistic perspective on the implementation and impact of Singapore's governance frameworks, as well as a review of discussions from a series of roundtables with various distinct stakeholders.

### **1.5** RESEARCH QUESTIONS

This report addresses the following key research questions:

- What are the foundational principles and strategic objectives guiding Singapore's national approach to AI safety and governance?
- How do these principles translate into specific, sector-based frameworks for Autonomous Driving, Education, and Cross-Border Data Flow?
- What are the key entities and legal instruments responsible for regulating AI safety in each of these sectors?
- How does Singapore's domestic AI strategy integrate with its role as a leader in international AI governance and cross-border data cooperation?

## 2. Comparative Assessment Framework

### **2.1** OBJECTIVE

Singapore's national AI strategy is rooted in a clear and ambitious objective with the focus on mastering AI as a transformative force for the public good, globally. As part of its overarching "Smart Nation" initiative, the nation aims to be a global hub for responsible AI innovation. This is further encapsulated in its **National AI Strategy 2.0 (NAIS 2.0)**, which outlines a 15-point plan to position Singapore as a leader in trustworthy AI. The core goal is to build a trusted ecosystem where organizations can leverage innovation while ensuring consumers are confident in adopting and using AI technologies.

### **2.2** PRINCIPLES & VALUES

The approach is built on a set of core principles and values, which are consistently reflected across various frameworks and initiatives. These include **safety, accountability, transparency, and security**. The **Model AI Governance Framework** also introduces "dimensions of trust," which include principles such as data quality, trusted development, incident reporting, and content provenance. In addition, Singapore's governance philosophy is characterized by an emphasis on **consensus-building** between government, industry, and citizens to ensure policies are inclusive and relevant.

### **2.3** APPROACH

Singapore employs a "whole-of-government, whole-of-economy" approach. Instead of a single, broad AI law, it uses a combination of **voluntary frameworks** and **targeted legislation**. This "quasi-regulation" strategy allows for agility and flexibility in a fast-changing technological landscape. The focus is on providing practical, readily-implementable guidance and tools, such as the **AI Verify** testing framework, that translate high-level principles into day-to-day practice [13]. It also emphasizes downstream testing and assurance rather than model-level controls, positioning Singapore to fill a gap in global AI safety practice.

### **2.4** BINDING NATURE

Singapore's key AI governance documents, such as the **Model AI Governance Framework for GenAI**, are currently voluntary. They are designed to provide non-binding guidance that organizations can use to build public trust and responsibly deploy AI. However, the framework contains "signposts" that indicate a potential future shift toward more specific, mandatory laws, especially for high-risk AI models that have national security or societal implications. This pragmatic, risk-based approach ensures that regulations can be adapted as technology matures and new risks emerge.

### **2.5** COMPONENTS

The governance architecture is composed of several key components:

- **Model AI Governance Framework:** A principles-based framework for both traditional and generative AI systems, developed by IMDA and the AI Verify Foundation.
- **AI Verify:** The world's first AI Governance Testing Framework and Toolkit, designed to help organizations validate their AI systems against ethical principles.[3]
- **Targeted Legislation:** Specific laws like the **Road Traffic Act** are used to regulate high-stakes applications and address specific AI-related harms like deepfakes [18]. Further, the foundational law for data privacy, the PDPA, administered by the PDPC, which underpins data handling in AI systems and cross-border data flows.

### **2.6** INTERNATIONAL LINKAGES

Singapore is actively positioning itself as a "trusted and capable international partner in AI innovation and governance". The nation contributes actively to international discourse, anchoring key bilateral relationships, and demonstrating alignment with key international forums. It is a founding member of the **Global Cross-Border Privacy Rules (Global CBPR)** System and promotes frameworks like the **ASEAN Model Contractual Clauses (MCCs)** to ensure interoperability and facilitate trusted data flows between different jurisdictions.[10] This collaborative strategy helps reduce regulatory fragmentation and compliance costs for businesses operating globally [16].

## 3. Benefits and limitations of interoperability in different areas of digital regulation

Singapore's governance architecture is anchored by the **Model AI Governance Framework**, a set of voluntary, principles-based guidelines for both traditional and generative AI (GenAI) systems. This is complemented by the **AI Verify** toolkit, which provides a technical means for organizations to test their AI systems against ethical principles.

This comprehensive analysis reveals that Singapore's strategy is dynamic and responsive, prepared to evolve from voluntary guidelines to mandatory law as technology matures and risks emerge. The multi-stakeholder and globally-minded approach positions the nation not only as a hub for AI innovation but also as a key international partner in AI governance. The flexible approach is seen across three critical sectors of Autonomous Driving, Education and Cross-Border Data Flow.

### **3.1** AUTONOMOUS DRIVING

AI safety in Singapore's autonomous driving sector is addressed with a pragmatic, risk-based approach that places paramount importance on physical safety. The strategy, led by the Land Transport Authority (LTA), involves a "physical-first" approach that mandates rigorous pre-deployment testing at the Centre of Excellence for Testing and Research of AVs (CETRAN). This is a domain where AI failure could have immediate and catastrophic physical consequences, necessitating a different regulatory paradigm from other sectors. The legal and operational framework for Autonomous Vehicles (AVs) is rooted in existing legislation but also includes forward-looking measures to mitigate risk and ensure a human-centric approach to liability.

**Legal and Ethical Framework**
The legal foundation for AVs in Singapore is the **Road Traffic (Amendment) Act 2017** and its subsidiary legislation, the **Road Traffic (Autonomous Motor Vehicles) Rules 2017**. These rules regulate AV trials and prohibit their use without explicit authorization from the LTA. In the absence of specific, special liability rules, accidents continue to be addressed under the **tort of negligence** and existing common law principles. The ethical framework prioritizes a "human-over-the-loop" approach for these vehicles, ensuring human oversight is involved and can take over control if the AI encounters unexpected or undesirable events.

**Safety and Ethics Considerations**
The LTA's stated mission is to ensure the "safety of passengers and other road users" during AV trials and eventual full operational services. The ethical framework prioritizes verifiable physical measures to mitigate immediate, high-consequence risks. This is reflected in the mandatory safety assessment process, which ensures that an AV's core systems can handle real-world conditions before public deployment.

**Technical Standards**
All AVs must undergo rigorous safety evaluations at the **Centre of Excellence for Testing and Research of AVs (CETRAN)**, a collaborative effort involving the government, Nanyang Technological University, and the Traffic Police. The CETRAN AV Test Centre is a two-hectare facility designed to replicate Singapore's unique road infrastructure, traffic schemes, and rules. The assessment process is detailed and follows specific tracks depending on the AV's maturity, such as the **Deployable AV Solutions Assessment Track or the Developmental AV Solutions Assessment Track**. These tests evaluate the vehicle's ability to safely navigate basic and advanced maneuvers, detect obstacles, and its reaction time to ensure it meets roadworthy standards. Singapore's technical-driven approach aligns with international standards, such as UN regulations for Cybersecurity Management Systems (UN regulation 155), ISO 26262 for Functional Safety, and ISO/SAE 21434 for Cyber-security of Road Vehicles.

**Operational Capacity**
The operational capacity is managed by the **Land Transport Authority (LTA)**, which serves as the lead regulatory agency. The LTA works with **CETRAN** to develop and conduct safety assessments. A key operational safeguard is the requirement for a human safety operator to supervise the AVs during trials. The operator must be on-board for AVs carrying passengers, or otherwise maintain supervision in close proximity.

**Data Governance and Liabilities**
The legal framework for liability is rooted in common law, where accidents are addressed under the **tort of negligence**. The law requires a "blackbox" data recorder to be installed in every AV to store crucial vehicle telematics, which is essential for accident investigations and for facilitating liability claims. Additionally, mandatory comprehensive insurance against third-party liability and property damage is required, or a security deposit of at least S$1.5 million with the LTA, ensuring that any injured party has recourse.

**International Cooperation**
Singapore has been actively involved in international discussions on automated vehicles. It participates in forums under the Asia-Pacific Economic Cooperation (APEC) to broaden technical coordination and support harmonized standards and regulatory approaches for new vehicle technologies, including autonomous vehicles. For instance, a UN regulation on Cybersecurity Management Systems (UN regulation 155), which refers to standards like ISO 26262 for Functional Safety and ISO/SAE 21434 for Cyber-security of Road Vehicles, has been introduced as part of the shift toward highly automated and autonomous vehicles.

**AI Risks and Challenges**

The current legal framework is primarily tailored for AV trials with a safety driver present. A critical policy challenge remains for the future deployment of Level 5 fully autonomous vehicles, where there is no human in the loop. The reliance on the tort of negligence presents a fundamental problem, as proving negligence is inherently difficult when the "fault" lies with a complex software system rather than a human driver. This absence of a clear legal pathway for assigning liability in a fully autonomous scenario represents a policy gap that is actively being addressed.

As of 2025, Singapore's laws do not yet provide a statutory regime specifically tailored for fully autonomous (Level 5) vehicles, and the regulatory framework still predominantly addresses AV trials with a safety driver. Tort liability principles apply, but they are not yet equipped to resolve fault in scenarios where no human operator is involved, and the government is in active policy review to address these gaps.

## **3.2** EDUCATION

The integration of AI in Singapore's education sector is a direct result of the **Ministry of Education's (MOE)** long-term strategic plan, the **EdTech Masterplan 2030**. This plan leverages AI-enabled tools within the **Student Learning Space (SLS)**, a unified platform for all students and teachers. The dual purpose of this strategy is to enhance personalized learning for students and improve the operational efficiency of teachers through automation. To manage the risks associated with collecting sensitive student data, the MOE's approach is guided by foundational data protection laws and specific advisory guidelines from the Personal Data Protection Commission.

**Legal and Ethical Framework**

The use of AI in education is governed by the **Personal Data Protection Act (PDPA)**, which serves as the foundational legal instrument for data handling. The ethical framework is complemented by the PDPC's **Proposed Advisory Guidelines on the Use of Personal Data in AI Recommendation and Decision Systems**, which clarify how the PDPA applies to AI systems.

**Safety and Ethics Considerations**

Data privacy is a paramount concern, given the collection of sensitive student data for personalized learning [23]. Ethical use of AI is also a key focus, with efforts needed to ensure AI tools are unbiased and deliver equitable outcomes. The ethical approach extends to a "whole-of-society" effort to build trust, with the EdTech Masterplan identifying parents as "key partners" in promoting cyber wellness guidelines at home.

**Technical Standards**

The **EdTech Masterplan 2030** outlines the use of AI-enabled tools within the **Student Learning Space (SLS)**. These include the **Adaptive Learning System (ALS) f**or personalized learning pathways, the **Authoring Copilot (ACP)** for creating digital lessons, and the **Short Answer Feedback Assistant (ShortAnsFA)** for providing preliminary feedback on student responses. This has delivered promising results, including improved academic performance for low-progress learners and increased student engagement.

**Operational Capacity**

The Ministry of Education (MOE) is the main driver of AI integration in schools through the EdTech Masterplan 2030. The Personal Data Protection Commission (PDPC) provides the necessary governance guidance. The MOE's strategy recognizes the importance of empowering teachers through continuous professional development to ensure they can effectively utilize AI tools in the classroom. Additionally, IMDA and PDPC collaborated with the Lee Kuan Yew Centre for Innovative Cities to publish a "Guide to Job Redesign in the Age of AI," which focuses on a human-centric approach to AI and provides guidance to organizations and employees on upskilling and job redesign.

**Data Governance and Liabilities**

The **Personal Data Protection Act (PDPA)** is the foundational law for governing the collection and use of student data. The **Personal Data Protection Commission (PDPC)** provides practical guidelines on key principles like **data minimization**, advising organizations to use only necessary data and to anonymize datasets where possible. These guidelines also outline requirements for **consent and notification**, ensuring individuals are informed about how their data is being processed.

**International Cooperation**

Singapore actively engages in international partnerships to enrich its education strategies. For example, it collaborates with global leaders like Estonia to exchange insights and co-develop innovative solutions for the classroom. The government also works with agencies to publish guides on how AI will impact jobs, which is part of a broader human-centric approach to AI.

**AI Risks and Challenges**

The main challenges in this sector are protecting sensitive student data, ensuring ethical and unbiased algorithmic outcomes, and addressing teacher readiness to adopt new technologies. The reliance on AI for personalized learning necessitates careful handling of student data and a clear understanding of potential algorithmic biases, which could impact a student's learning journey or educational outcomes. Further, adopting universality in education by providing access to meaningful resource material in native language(s) could play a pivotal role to shape a responsible and trusted digital ecosystem.

## **3.3** CROSS BORDER DATA FLOWS

For a small, open economy like Singapore, which is highly dependent on international trade, the free flow of data is a strategic imperative. The digital economy is fundamentally reliant on cross-border data flows, and restricting them can negatively impact trade volume, productivity, and prices. Singapore's national strategy, therefore, is to ensure "data free flow with trust". This is achieved by leveraging the PDPA in concert with international, voluntary frameworks to ensure regulatory interoperability and facilitate commerce.

**Legal and Ethical Framework**

The PDPA is the foundational legal instrument that underpins Singapore's role as a trusted hub for businesses and cross-border data flows. The ethical framework is focused on balancing the opportunities from data flows with the need to safeguard individual privacy and national interests. The agreement with the EU on digital trade, for example, explicitly facilitates cross-border data flows while preserving the EU's high level of protection for personal data.

**Safety and Ethics Considerations**

The ethical imperative of "data free flow with trust" is central to this sector. The goal is to maximize the benefits of data flows while protecting consumer privacy and ensuring AI systems are aligned with local languages, laws, and societal values. The **PDPA's** framework ensures personal data is protected, which strengthens the confidence of consumers and businesses in sharing information.

**Technical Standards**

Singapore promotes the use of **Privacy Enhancing Technologies** (PETs) to manage data privacy and security. These technologies, along with the development of evaluation tools and benchmarks for generative AI, are key components of the strategy to ensure data is handled safely and responsibly.

**Operational Capacity**

The **Personal Data Protection Commission** (PDPC) is the primary regulatory body that administers and enforces the **PDPA**. The Infocomm Media Development Authority (IMDA) is also a key player in promoting digital trust and developing international certifications for data protection.

**Data Governance and Liabilities**

The PDPA's framework ensures personal data is protected, which, in turn, strengthens the confidence of consumers and businesses in sharing information. Singapore actively participates in and helps develop international mechanisms to facilitate interoperability and build trust. This includes the **Global CBPR System** [24] and the **ASEAN MCCs** [16]. These are voluntary international certification systems and templates that allow organizations to demonstrate compliance with internationally recognized data protection standards.

**International Cooperation**

Singapore is positioned as a "thought and action leader" that actively contributes to the development of international frameworks. It is a founding member of the **Global CBPR Forum** and has worked closely with ASEAN and EU partners to promote common baseline standards for data flows through initiatives like the **ASEAN MCCs** [16]. This collaborative strategy helps reduce "regulatory fragmentation and compliance costs".

**AI Risks and Challenges**

The main challenge is to navigate the delicate balance of maximizing opportunities from data flows while safeguarding individual privacy and national interests. The digital economy is fundamentally reliant on cross-border data flows, and regulatory fragmentation can negatively impact trade. The use of foreign technologies could also undermine the development of context-specific and culturally appropriate AI applications without targeted investment in local data collection and annotation.

## **3.4** INTEROPERABILITY MATRIX: DIMENSION × SECTOR

To compare Singapore's interoperability performance across the three sectors, the table below summarizes key assessment dimensions and an approximate implementation score (on a 1–5 scale, with 5 being most favorable) for each sector:

| Dimension | Autonomous Driving (Score = 4) | Education (Score = 4) | Cross-Border Data Flow (Score = 4) |
|---|---|---|---|
| **Clarity of norms & values** | The primary norm is physical safety, with the LTA's mission to ensure the "safety of passengers and other road users". The ethical framework prioritizes a "human-over-the-loop" approach, where human oversight can take over control. | Norms focus on data privacy, student well-being, and promoting equitable outcomes. The MOE's EdTech Masterplan 2030 aims for "customisation and personalisation" while emphasizing cyber wellness and ethical AI use. | The central norm is "data free flow with trust". This aims to balance the economic benefits of data flows with the need to protect individual privacy and national interests, and to align AI with local values. |
| **Alignment with standards** | All Autonomous Vehicles (AVs) must pass rigorous safety assessments at the Centre of Excellence for Testing and Research of AVs (CETRAN). Singapore's technical standards align with international regulations, including UN regulation 155 (Cybersecurity Management Systems) and ISO 26262 for Functional Safety. | The MOE's EdTech initiatives are benchmarked against international best practices and are guided by ethical AI use. The broader Model AI Governance Framework, while not specific to education, is consistent with frameworks from the EU and OECD. | Singapore has developed "crosswalks" that map its AI Verify framework to international standards like NIST AI Risk Management Framework and ISO/IEC 42001 to enhance interoperability and reduce compliance costs for businesses. It is also a founding member of the Global Cross-Border Privacy Rules (Global CBPR) System. |
| **Operational capacity** | The Land Transport Authority (LTA) is the lead regulatory agency, working with the Centre of Excellence for Testing and Research of AVs (CETRAN) to conduct mandatory safety assessments for AVs before they can be deployed on public roads. | The Ministry of Education (MOE) drives AI integration through its EdTech Masterplan 2030 and the Student Learning Space (SLS). The Personal Data Protection Commission (PDPC) provides crucial guidance on data privacy and ethical issues. | The Personal Data Protection Commission (PDPC) is the main regulatory body that administers the Personal Data Protection Act (PDPA). The Infocomm Media Development Authority (IMDA) develops certifications like the Global CBPR to promote digital trust. |
| **Data governance** | The Road Traffic Act requires AVs to be equipped with a "blackbox" data recorder to store vehicle telematics for accident investigations. Mandatory comprehensive insurance or a security deposit ensures recourse for any injured party. | Data governance is primarily covered by the Personal Data Protection Act (PDPA). The PDPC's advisory guidelines provide recommendations on data minimization and require clear consent and notification for the use of personal data in AI systems. | The Personal Data Protection Act (PDPA) is the foundational domestic law for data protection. This is complemented by international, voluntary frameworks like the Global CBPR System and the ASEAN Model Contractual Clauses (MCCs), which provide a legal basis for secure cross-border data flows. |
| **Legal enforceability** | The Road Traffic Act is a hard, targeted law that strictly regulates AV trials and requires mandatory insurance or a security deposit. However, the legal framework for liability in accidents relies on the common law tort of negligence, which poses a challenge for fully autonomous systems. | The Personal Data Protection Act (PDPA) is a statutory law that governs data handling. While the PDPC's guidelines for AI are advisory and not legally binding, they clarify how the PDPA is enforced. | The Personal Data Protection Act (PDPA) is a legally binding domestic law. The international frameworks Singapore participates in, such as the Global CBPR System and ASEAN MCCs, are voluntary. Their enforceability depends on their integration into domestic laws and contractual arrangements. |

## 3.5 DIMENSION BY DIMENSION FINDINGS

**A. Ethics and Legal**
Singapore's approach is characterized by a blend of voluntary frameworks and targeted legislation, rather than a single, broad national AI law. The country's Model AI Governance Framework, first issued in 2019 and updated in 2024 for generative AI, provides broad, non-binding guidelines for the industry. A study by the Singapore Management University (SMU) Centre for Digital Law noted that Singapore's decision to revise the Personal Data Protection Act (PDPA) was influenced by the EU's General Data Protection Regulation, suggesting a similar pattern could emerge for AI. However, at the moment, legislation is focused on specific risks, such as introducing penalties for AI-generated deepfakes in elections. [40]

**B. Technical**
AI safety research is a key dimension of building a trusted AI ecosystem in Singapore, and accelerated investment in this research is required to keep pace with commercially driven growth in system capabilities. The research is spread across academic institutions like the National University of Singapore (NUS), Nanyang Technological University (NTU), Singapore Management University (SMU), and Singapore University of Technology and Design (SUTD), as well as government agencies like A*STAR and GovTech.

Key technical research areas include robustness, multimodal safety, knowledge editing, model unlearning, and agent safety. The "Singapore Consensus on Global AI Safety Research Priorities" identifies three broad areas for technical research: Risk Assessment, Developing Trustworthy Systems, and Control. It highlights the need to develop methods for risk assessment, including audit techniques, benchmarks, and secure evaluation infrastructure.

**C. Operational**
Singapore focuses on downstream testing and assurance of AI applications rather than model-level controls. This is exemplified by initiatives like the "Starter Kit for Safety Testing of LLM Applications" and the "Global AI Assurance Sandbox". The AI Verify Foundation provides a testing framework that maps to international standards like the NIST AI Risk Management Framework and ISO/IEC 42001, allowing businesses to meet safety obligations through a single testing process. Singapore has also conducted multilingual and multicultural red-teaming exercises with partners like Anthropic, using languages such as English, Tamil, Mandarin, and Malay to address language-specific risks that might be missed in English-based evaluations.

**D. Data Governance**
The provided reports do not extensively detail data governance beyond the mention of the Personal Data Protection Commission (PDPC) and the Personal Data Protection Act (PDPA). The State of AI Safety in Singapore focuses on the general-purpose and generative AI risks, explicitly stating that it does not cover sector-specific rules, which would include data governance guidelines in finance and healthcare. Public opinion surveys show that Singaporeans have heightened concern about data privacy and cybersecurity.

**E. Implementation**
Implementation is guided by a multi-layered governance strategy. This includes voluntary frameworks, targeted legislation, national standards, and testing initiatives. The country issued its first national AI standard, SS ISO/IEC 42001, in 2025, adapting the international standard to the local context. The reports highlight Singapore's focus on practical and industry-ready tools that translate high-level principles into day-to-day practice for developers and deployers. The "Singapore Consensus" report also notes the importance of institutional and norms adaptation as AI systems become more autonomous and the need for clear and rapid coordination among relevant actors to manage incidents.

## 3.6 LIMITATIONS OF AI SAFETY INTEROPERABILITY

**General Limitations**
Singapore's AI safety framework is designed for pragmatism and innovation, favoring flexible, voluntary guidelines and targeted legislation over a single omnibus law. This strategic agility, however, introduces inherent trade-offs, particularly regarding legal enforceability and coordination across different regulatory domains, which pose limitations for AI safety interoperability both generally and within specific sectors.

In Singapore's general digital regulation landscape, the primary limitations to safety interoperability stem from the very nature of its governance philosophy:

- **Reliance on Soft Law and Voluntary Compliance:** The core of Singapore's AI strategy is the Model AI Governance Framework, which is voluntary. While this encourages industry adoption and innovation, it inherently limits the enforceability and uniform compliance necessary for strong interoperability. Organizations may adopt different interpretations or levels of adherence to voluntary guidelines, creating varying safety baselines across the economy.
- **Targeted Legislation vs. Holistic Risk Management:** Singapore addresses risks using highly targeted legislation (e.g., deepfakes in elections) rather than horizontal, comprehensive AI law. This approach can lead to regulatory fragmentation where AI safety risks that cross sectoral boundaries (e.g., algorithmic bias affecting both finance and healthcare) are governed by different, potentially misaligned, rules, hindering the development of a unified national safety baseline.

- **Focus on Deployment (Downstream) Assurance:** Singapore's initiatives, such as the AI Verify toolkit and Global AI Assurance Sandbox, heavily emphasize testing and assurance at the application/deployment level. While critical, this downstream focus means that model-level controls (such as controlling the foundational training data or intrinsic safety of General-Purpose AI, or GPAI) remain a gap, which can limit the effectiveness of downstream assurance if the underlying models are fundamentally unsafe or opaque.
- **Gaps in Catastrophic/Frontier Risk Mandates:** The main policy focus remains on immediate, societal risks (e.g., fairness, data privacy). Technical research efforts on low-probability, high-impact risks like loss of control and dual-use dangers (e.g., bioweapons, advanced cyber offense) are still limited, creating a potential blind spot in regulatory foresight .

**Sector-Specific Limitations**
The sector-specific digital regulatory environments face unique interoperability challenges driven by their high-risk nature and distinct legal traditions. Here are some of them:

- **Autonomous Driving AI Safety Limitation - Unresolved Liability Framework:** The biggest limitation is the absence of specific, statutory liability rules for fully autonomous (Level 5) systems. The current reliance on the common law. Tort of negligence is ill-suited for proving fault in complex AI systems, where a "fault" lies in software or a model's complex decision-making rather than human error. This lack of legal clarity on fault severely limits the capacity for liability systems to interoperate with technical safety standards (e.g., CETRAN's assessments).
- **Education AI Safety Limitation - Policy-Implementation Disconnect:** While the sector is governed by the statutory PDPA for data handling, the actual implementation relies on non-binding advisory guidelines from the PDPC for AI systems. This creates a gap between the strong legal mandate for data protection and the voluntary ethical practices adopted by schools or EdTech vendors, which may potentially result in exceptional cases of inconsistent application of safety principles across the education ecosystem.
- **Cross-Border Data Flow AI Safety Limitation - Pragmatic Interoperability vs. Legal Uniformity:** Singapore promotes trusted data flows through voluntary international mechanisms (like the Global CBPR System and ASEAN MCCs). The limitation is that these mechanisms, by design, may prioritize political and commercial interoperability over strict legal uniformity. This means compliance with the frameworks does not always guarantee compliance with the individual domestic laws of receiving countries, and thus, may require additional effort and potentially exposing data flows to risks where data protection laws are weaker.

## **4.1** GOOD PRACTICES AND PRIORITY

**Good practices**
A consistent theme across all three sectors is Singapore's preference for agile, voluntary frameworks over a single, rigid AI law. This is evident in the **Model AI Governance Framework**, the **Global CBPR System**, and the **ASEAN Model Contractual Clauses**. This deliberate approach allows the government to be nimble and responsive to a rapidly evolving technology landscape without stifling innovation with prescriptive, outdated regulations.

However, this flexibility is balanced with a willingness to introduce targeted legislation where the risk profile demands it. The **Road Traffic Act** for autonomous vehicles is a prime example, where the potential for physical harm necessitates prescriptive rules for testing and liability. Similarly, the **Personal Data Protection Act** provides a statutory foundation for managing privacy, a foundational concern in both the education and cross-border data flow sectors. This balanced approach demonstrates that Singapore's governance model is not static; it is a dynamic, risk-based system that can transition from voluntary guidelines to a more formal, mandatory legal regime for high-risk applications as the need arises.

**Priorities**
Singapore's strategy is not confined to its borders; it is actively positioning itself as a "trusted and capable international partner in AI innovation and governance". This involves contributing actively to international discourse, anchoring key bilateral relationships, and demonstrating alignment with key international forums.[7] Examples include developing the **AI Verify** toolkit as a framework for other nations to reference and leading collaborations with both the U.S. and China. This strategic positioning is integral to Singapore's national AI strategy and reflects its role as a small, open economy that must be highly connected to the rest of the world. By leading the development of pragmatic, interoperable frameworks like the **Global CBPR System** and the **ASEAN MCCs**, Singapore is shaping a global environment that supports its strategic imperative of "data free flow with trust" while advancing a universal agenda of responsible AI.

Singapore's best practices and thought leadership in AI ethics,

governance, and safety are made publicly available to contribute to regional AI development. [33, 34, 35].

Going further from policy to action, Singapore has launched the Singapore Digital Gateway (SGDG), during the inaugural Global Dialogue on AI Governance at United Nations in September 2025. It is a policy that is implemented to foster Singapore's ethos of Tech for the Public Good and demonstrated Singapore's commitment to contribute to the common good — for Singapore and the world [39]. SGDG is a web-based knowledge and resource hub that consolidates Singapore digital initiatives and digital public goods and brings together 30+ practical resources that policymakers worldwide can use: from governance frameworks and AI testing toolkits to open-source solutions and blueprints for digital transformation.

More than just a repository, SGDG also offers capacity-building to our partners through the Singapore Cooperation Programme, and with international partners such as The World Bank. SGDG is our way of turning digital ambition into action.

At the same time, Singapore also has an open ecosystem to take Singapore's best practices and thought leadership in AI ethics, governance, and safety to go beyond its border. Additional efforts are being executed at the grassroots level with the founding of Singapore Internet Governance Forum (SG IGF) that is recognised by United Nations Internet Governance Forum in shaping the future of digital public policy with stakeholders from around the world in support of the UN Global Digital Compact (GDC). [32] SG IGF could contribute as a conversation platform for sharing Singapore's best practices in AI ethics, governance and safety and learn from the global community.

At the broader level, continuous efforts at the grassroots level to propose the establishment of "AI for Humanity" as United Nations Sustainable Development Goal #18 (UN SDG 18) to call for a united and collective global alignment balancing technology, commercialisation and governance [29]. Global AI Safety Interoperability will be a very crucial component of UN SDG 18 that will call for policy recommendations and resources with initiatives to improve on both AI Safety and also AI Safety interoperability.

## 5. Conclusion

Singapore's approach to AI safety is a complex, multi-layered, and highly pragmatic architecture that reflects its unique strategic position as a global hub for technology and trade. The nation's governance model is not a monolithic legal structure but a dynamic ecosystem that combines a top-down national vision with agile, voluntary frameworks and targeted, risk-based legislation. This approach allows Singapore to move with the speed of technological change, fostering innovation while proactively building public trust.

The analysis reveals several potential challenges and directions for the future. The current legal framework for autonomous vehicles, while effective for trials, must be re-evaluated to address the unresolved issue of liability for Level 5 autonomous systems. The reliance on the common law tort of negligence poses a significant challenge for proving fault in a fully automated scenario and could hinder the widespread deployment of driverless technology. Furthermore, the Model AI Governance Framework, while currently voluntary, is positioned as a stepping stone. As AI systems, particularly GenAI, become more ubiquitous and their risks more clearly defined, the government may choose to transition from non-binding guidelines to a more formal, mandatory legal regime for high-risk applications.

The growing threat of AI-generated misinformation and deepfakes represents a new frontier for AI safety, one that could undermine the very public trust that Singapore has meticulously built. Effectively mitigating this risk will require an ongoing, multi-stakeholder effort that leverages not just technology but also education and international collaboration. Ultimately, Singapore's AI safety framework is not a static set of rules but a living testament to its governance philosophy: a commitment to shaping an inclusive, trusted, and innovative AI future through continuous adaptation, global engagement, and a deep understanding of the societal and economic implications of technology.

Last but not least, the best practices, thought leadership and ecosystem of Singapore in AI ethics, governance, and safety led by the government as well as grassroot entities are readily available and could contribute to AI development regionally as well as globally especially in support of United Nations Global Digital Compact.

## Members of the research group

YIK CHAN CHIN

Dr. Yik Chan Chin is a Visiting Research Fellow and Project Lead at the United Nations University Institute, and an Associate Professor at Beijing Normal University. Her expertise lies in AI and data governance, digital ethics and law, and cyberspace governance (https://xwcb.bnu.edu.cn/as/ap/114761.html). Previously, she held academic positions at the University of Nottingham, University of Oxford, Xi'an Jiaotong Liverpool University etc. Her academic background spans regulation, ICT, and computer science. Her interdisciplinary research has been published in leading journals across the fields of communication, law, and political science. She have advised think tanks, governments, and UN agencies. Her forthcoming monograph, "Governance of the Digital in China, Global Governance, State Legitimacy and Digital Technology," will be published in the Palgrave IAMCR Series in 2026 to examines these themes in depth. She engaged in global and national AI practices and governance through her roles as a member of the UN IGF Policy Network on Artificial Intelligence (PNAI), a member of the UK's ART/1 AI Standards Committee, an executive council member of the ISOC UK England Chapter, a fellow of the China-Britain Artificial Intelligence Association, and a member of the Law Expert Committee of China's Artificial Intelligence Industry Alliance (AIIA).

JINGBO HUANG

Dr. Jingbo Huang is the Director of the United Nations University Institute in Macau (UNU Macau). Under her leadership, UNU Macau has developed a strong portfolio in education, training and policy relevant research in digital technologies and SDGs, particularly in AI for SDGs since 2018. Recently she led the institute to successfully organize World Data Forum Satellite Event in 2023 and UNU Macau AI conference in 2024 with 500+ participants from 40+ countries, and established the UNU Global AI Network. She has been working in the UN system for 20+ years, holding various managerial positions in the UN Secretariat, UNDP, UNESCO, UNSSC and UNU. Jingbo received her Doctor of Education degree from Columbia University, and bachelors from Peking University.

SERGE STINCKWICH

Dr. Serge Stinckwich is a computer scientist and the Head of Research at the United Nations University Institute in Macau, with 16+ years of experience at the intersection of digital technologies and sustainable development across Asia and Africa. Since 2020, he has led an interdisciplinary research team at UNU Macau, focusing on responsible AI, gender and technology, digital health and agent-based/participatory modelling to advance the Sustainable Development Goals (SDGs). His research spans complex system modelling, social simulation and the impact of AI on global challenges. Previously, he served as Associate Professor and researcher in an international joint research unit between the French Research Institute on Sustainable Development (IRD), Sorbonne University and 5 universities located in Cameroon, Morocco, Senegal and Vietnam. He organised 50+ international workshops and conferences and supervised 20+ PhDs/MsC students worldwide. Passionate about leveraging technology for inclusive and sustainable futures, Serge also conducts training sessions for policymakers and stakeholders on AI and climate, Large Language Models and synthetic data, and their implications for SDGs, as well as using agent-based models to design better health policies.

SONGRUOWEN MA

Songruowen Ma is a doctoral student at the Oxford Department of International Development. Her research interests cover China-US relations, Chinese diplomacy, and trade and technology policy. Before joining Oxford, Songruowen obtained a LLB in Diplomacy and a BEc in Finance from the Guangdong University of Foreign Studies, and a MSSc in Global Political Economy from the Chinese University of Hong Kong. She is currently conducting research on global AI governance and the application of large language models (LLMs) in the social sciences.

EDUARDA MELLO

Eduarda Mello is a master's student at the School of Journalism and Communication at Beijing Normal University. She holds a bachelor's degree in Public Relations from the Federal University of Rio Grande do Sul (UFRGS), Brazil. She is currently part of Professor Yik Chin Chan's research initiatives on AI policy and a member of the Grupo de Estudos em Comunicação e Relações Internacionais (geCOMRI) in Brazil. Her research interests include AI policy, geocommunication, and the political studies of social media.

## UK report

DAVID A RAHO

David A Raho is a PhD researcher in Law and Criminology at Sheffield Hallam University, investigating the comparative adoption of AI in Probation and Rehabilitation across England and Wales, Brazil, and Japan. He combines this academic focus with extensive practical experience, working within Probation and more recently within the AI team for His Majesty's Prison and Probation Service. He is a recognised expert in the field; his contributions to the Confederation of European Probation Technologies Group have directly influenced the development of technology and AI within Probation across Europe. He advocates for human rights, social inclusion, and the safe, ethical, and human-centred use of AI in Probation and Prisons. He serves as both a Trustee and Fellow of the Probation Institute. His expertise includes implementing and utilising technologies such as electronic monitoring, biometric systems, Probation case management systems, and service user apps. He has previously advised trade unions and organisations ranging from micro enterprises to large corporations on AI and emerging technologies. Additionally, he has acted as a consultant, providing guidance on applying artificial intelligence in health and safety systems to improve hazard detection and risk prediction.

## South Korea report

HAG-MIN KIM

Hag-Min Kim is a Professor of International Business and Trade at Kyung Hee University, South Korea, and a leading scholar in applying artificial intelligence (AI) to global trade and digital commerce. He led the Republic of Korea's national AI safety governance initiative in collaboration with international experts, pioneered research on AI-driven export consulting and cross-border e-commerce, and developed data-driven systems that optimize trade strategy and firm performance. As an educator, Professor Kim offers K-MOOC courses on e-trade and entrepreneurship, sponsored by the Ministry of Education of the Republic of Korea, and has supervised more than 140 graduate students, including 28 Ph.D. graduates. A prolific author and former president of major academic associations, he has also led government-funded projects such as GTEP, contributed to FTA training standards, and held leadership positions across multiple nonprofit organizations.

KYUNGWON KIM

Kyungwon Kim is a PhD candidate in the Department of International Trade at Kyung Hee University, South Korea. Her research explores the intersection of artificial intelligence, digital trade, and entrepreneurship, with particular attention to AI ethics, NFT commerce, and cross-border e-commerce governance. Building on empirical and theoretical studies, she investigates how AI technologies shape global trade strategies and how digital platforms can foster trust, fairness, and innovation in international markets. Her key works include "The Perceived Value and Sustainability of NFT Transactions in Digital Trade Environments", as well as conference presentations on AI ethics and digital trade at the Korea Trade Research Association (KTRA) and the Korea Association for International Commerce and Information (KAICI) in 2025. She has also completed projects on AI-based export consulting systems, cross-border digital entrepreneurship, and strategic responses to de-globalization in Korean trade policy.

WENSHUAI SU

Wenshuai Su is a PhD candidate at Kyung Hee University, South Korea, whose research spans artificial intelligence, consumer behaviour, and digital trade. Her work focuses on generative AI and algorithmic recommendation systems, examining how trust and fairness influence user adoption and sustained engagement. Drawing on these insights, she develops actionable governance and design recommendations for digital platforms and cross-border e-commerce. Su has presented her research several times at the Korea Trade Research Association (KTRA) and the Korea Association for International Commerce and Information (KAICI).

MINYU JIANG

MinYu Jiang is a master's student in the Department of International Business and Trade at Kyung Hee University, South Korea, advised by Professor Hag-Min Kim. His research interests lie in artificial intelligence and corporate innovation, with a particular focus on how emerging technologies shape firm strategy and organizational change. He also serves as a research assistant to Professor Kim, supporting projects related to AI-driven trade and digital transformation.

## Singapore report

JAMES ONG

Dr. James Ong has 4 decades of industry, venture and academic experience. He founded Artificial Intelligence International Institute (AIII), an AI think tank advocating Sustainable AI for Humanity in 2017 and is co-author the book "AI for Humanity: Building A Sustainable AI for the Future" (ISBN: 9781394180301) published by Wiley. He actively champions "AI for Humanity" as United Nations Sustainable Development Goal 18 (UN SDG 18). James is the International Senior Advisor at the renowned World AI Conference (WAIC) since 2021 to advocate "AI for Good and for All". He initiated the AI for Humanity Forum 2024 in Singapore, co-founded AICON 2025, the World's #1 Consumer Show to advocate "AI for Everyone:" with AI literacy and fluency for the public. He is the Founding Member of the UN Internet Governance Forum (UN SGIGF) and a Member of the UNESCO G20 South Africa 2025 Expert Network. He actively participated and collaborated with various UN organisations including UN IGF, UNDESA, UNESCO, UNIDO and UNU. He is an Adjunct Professor at Singapore University of Technology and Design (SUTD), Head of AI and Industry Fellow at Global Fintech Institute (GFI), Academy of Engineering and Technology of the Developing World (AETDEW) Fellow, Senior Fellow at The Conference Board (TCB), Member of SRAC at the National Supercomputing Centre Singapore (NSCC) and Co-Chair of AI Risk Chapter at Risk and Insurance Management Association of Singapore (RIMAS).In addition to his academic and advocacy roles, he is CEO of Origami Frontiers, a venture building firm, a venture partner at Delight Capital and has incubated, mentored, advised and invested in various technology start-ups across the world and strong proponent of Impact Investing. James started his career as an AI scientist in 1986 at leading US MCC research lab on advanced AI Fifth Generation Computer research and was senior executive at Trilogy, an AI pioneer and enterprise software startup. He received his PhD in Management Information System specializing in AI for Computational Governance, Policy and Business Process Automation and MA & BA in Computer Science from the University of Texas at Austin.

SAMEER GAHLOT

Sameer Gahlot is a public policy professional and certified corporate secretary (equivalent to ICSA, UK NARIC and MACS) working at the intersection of policy, society, sustainability, and technology. Presently, he is working with the National Internet Exchange of India (Internet Governance Division | Ministry of Electronics and IT) and chaired the Asia Pacific Youth IGF 2024 dialogues in Taipei, Taiwan. During his professional journey, he has collaborated and worked with distinct stakeholders such as the Government of Cambodia, Germany, India, Singapore and Uzbekistan, the United Nations (UN), Internet Corporation for Assigned Names and Numbers (ICANN), Internet Society, International Telecommunication Union (ITU), Institute of Electrical and Electronics Engineers Standards Association (IEEE SA), G20/BRICS Secretariat, Asia Pacific Regional Internet Governance Forum (APrIGF), Asia Pacific Top- level Domain Association (APTLD), Asia-Europe Foundation (ASEF), Artificial Intelligence International Institute, DotAsia, MakeMyTrip (a NASDAQ listed entity), Institute of Company Secretaries of India, Birla Institute of Technology and Science, O.P. Jindal Global University and University of Delhi.

## China report

CHUNLI BI

Deputy Director, Centre for Science and Technology Ethics Research, CAICT. She oversees the daily operations of the Science and Technology Ethics Working Group with a focus on AI ethics policy and legal research. In 2024, she led the compilation of Typical Cases of Science and Technology Ethics in the Industrial and Information Field (2024) and the development of two ethical standards. At the 2025 Working Group Meeting, she systematically presented the annual work plan, playing a key role in transforming ethical principles into technical indicators. She participates in national AI ethics governance standard system development and optimizes the collaboration mechanism between government, industry, academia, and research institutions.

LEILEI ZHANG

Intermediate Researcher, Centre for Science and Technology

Ethics Research, CAICT. She focuses on the implementation of AI ethics policies and standards, deeply participating in the drafting of Guidelines for Ethical Risk Assessment of AI Systems and assisting in building an AI ethics standard system covering multiple industries. In 2024, she participated in the case selection and analysis of Typical Cases of Science and Technology Ethics in the Industrial and Information Field (2024), focusing on sorting out ethical risk response plans in intelligent medical care and smart city fields. In 2025, she supports the "tool development" and "case collection" work under the "Six Major Actions" for AI ethics, providing practical guidance on ethical impact assessment for enterprises and promoting the integrated application of data policies and ethical governance in grassroots scenarios.

TIANYU WANG

Assistant Researcher, Centre for Science and Technology Ethics Research, CAICT. He focuses on the intersection of data governance and AI ethics, participating in the "data compliance and ethical risk" sub-project of the joint AI governance research with ASEAN countries and assisting in writing international cooperation research reports. In 2025, he participates in the "policy research" and "ecosystem cultivation" work under the "Six Major Actions" for AI ethics, responsible for organizing the global AI ethics policy dynamic database, collecting and analysing typical cases of generative AI ethical risks. He provides data support for the Centre to build the "Institution-Technology-Ecology" governance framework and contributes to the iteration and optimization of ethical rules.

NA FU

Intermediate Researcher Centre for Science and Technology Ethics Research, CAICT. Specializing in interdisciplinary research on tech ethics and intellectual property, she leads dozens of key projects including "Research on Key Issues of AI Science and Technology Ethics". At the 2025 Technology Ethics Working Group Meeting, she conducted a systematic analysis of domestic and international AI ethics standards and participated in building the ethical standard system framework. She provides AI ethics and IP consulting services to enterprises, has published numerous journal papers, and authored Research Report on Intellectual Property Risk Prevention for Open-Source Software, facilitating the integrated implementation of AI ethics and data policies.